\documentclass[preprint,floatfix,epsfig]{revtex4}

\usepackage{amsfonts} % collection is a set of miscellaneous TeX fonts
\usepackage{amsmath}
\usepackage{amssymb}
\usepackage{bm}% bold math
\usepackage[usenames, dvipsnames]{color} % use colored tex
\usepackage{dcolumn}
\usepackage{epic} % enhanced picture environment
\usepackage{epsfig}
\usepackage{pstool}
\usepackage{eucal} % The package changes the font that is selected by \mathcal
\usepackage{feynmp}    %Package for feynman diagrams.
\usepackage{graphicx}
\usepackage[breaklinks,colorlinks = true,linkcolor = red,urlcolor=cyan,citecolor=red]{hyperref}
\usepackage{mathrsfs}%\mathscr
\usepackage{soul}
\usepackage{subfigure}
\usepackage{tikz}
\usepackage{wrapfig} % to include figure with text wrapping around it
\usepackage{xy} % general drawing package useful for drawing complex commutative diagrams
\renewcommand{\v}[1]{\ensuremath{\boldsymbol{#1}}} % for vectors
\newcommand{\beq}{\begin{equation}}
\newcommand{\eeq}{\end{equation}}
\newcommand{\beqq}{\begin{equation*}}
\newcommand{\eeqq}{\end{equation*}}
\newcommand\beqa{\begin{eqnarray}}
\newcommand\eeqa{\end{eqnarray}}
\newcommand\beqaa{\begin{eqnarray*}}
\newcommand\eeqaa{\end{eqnarray*}}
\newcommand\bea{\begin{array}}
\newcommand\eea{\end{array}}
\newcommand\beal{\begin{align}}
\newcommand\eeal{\end{align}}

\newcommand{\ds}{\displaystyle\sum}

\newcommand{\de}{\delta}
\newcommand{\f}{\frac}

\newcommand{\al}{\alpha}
\newcommand{\be}{\beta}
\newcommand{\ga}{\gamma}

\newcommand{\td}{\tilde} 
\newcommand{\nn}{\nonumber \\}

\begin{document}
\title{Ab initio holography}

\author{Peter Lunts$^{1,2}$, Subhro Bhattacharjee$^{3}$, Jonah Miller$^{1,4}$, Erik Schnetter$^{1,4}$,  Yong Baek Kim$^{5}$, Sung-Sik Lee$^{1,2}$}

\address{
$^1$ Perimeter Institute for Theoretical Physics, Waterloo, Ontario, Canada N2L 2Y5.\\
$^2$ Department of Physics \& Astronomy, McMaster University, Hamilton, Ontario, Canada  L8S 4M1.\\
$^3$ Max-Planck-Institut f\"ur Physik komplexer Systeme, N\"othnitzer Str. 38, 01187 Dresden, Germany. \\
$^4$ Department of Physics, University of Guelph, Guelph, ON, Canada N1G 2W1. \\
$^5$ Department of Physics, University of Toronto, Toronto, Ontario, Canada M5S 1A7.\\
}

\email{
\\ plunts@pitp.ca
\\ subhro@pks.mpg.de
\\ jmiller@pitp.ca
\\ eschnetter@pitp.ca
\\ ybkim@physics.utoronto.ca
\\ slee@pitp.ca
}

\begin{abstract}

We apply the quantum renormalization group 
to construct a holographic dual 
for the U(N) vector model for complex bosons
defined on a lattice.
The bulk geometry becomes dynamical 
as the hopping amplitudes 
which determine connectivity of space 
are promoted to quantum variables.
In the large $N$ limit, 
the full bulk equations of motion
for the dynamical hopping fields 
are numerically solved for finite systems.
From finite size scaling, 
we show that different phases 
exhibit distinct geometric features
in the bulk.
In the insulating phase, 
the space gets fragmented into isolated islands 
deep inside the bulk, exhibiting {\it ultra-locality}.
In the superfluid phase,
the bulk exhibits a horizon 
beyond which the geometry becomes {\it non-local}.
Right at the horizon, 
the hopping fields
decay with a universal power-law 
in coordinate distance between sites,
while they decay
in slower power-laws with 
continuously varying exponents
inside the horizon. 
At the critical point, 
the bulk exhibits a {\it local} geometry 
whose characteristic length scale diverges
asymptotically in the IR limit.

\end{abstract}
\date{\today}
\maketitle
\tableofcontents
%%%%%%%%%%%%%%%%%%%%%%%%%%%%%%%%%%%%%%%%%%%%%%%%%%%%%%%%%%%%%%%%%%%%
\section{Introduction}

According to the AdS/CFT correspondence\cite{Maldacena:1997re,Witten:1998qj,Gubser:1998bc},
quantum field theories are dual to gravitational theories in a spacetime with one higher dimension.
The extra dimension in the bulk can be interpreted as a length scale in 
the renormalization group (RG) flow\cite{1998PhLB..442..152A,deBoer:1999xf,Skenderis:2002wp,Heemskerk:2010hk,2011JHEP...08..051F}.
However, the flow generated along the extra dimension in holography
is different from the conventional RG flow 
because the bulk theories are in general quantum theories.
In order to make the connection between RG and holography precise,
one has to introduce quantum RG (QRG)\cite{Lee2012,Lee:2012xba,Lee:2013dln}. 
According to QRG, general $D$-dimensional quantum field theories are 
equivalent to $(D+1)$-dimensional theories that include quantum gravity,
provided that the quantum field theories can be regularized
covariantly in curved spacetime\cite{Lee:2012xba,Lee:2013dln}.

Conventional renormalization group describes the flow
of the couplings (sources) generated by coarse graining
defined in the space of all sources allowed by symmetry. 
In QRG, only a subset of the sources is included.
The sources in the subset are promoted to quantum mechanical operators,
and a quantum Hamiltonian governs the evolution 
of the dynamical sources under scale transformations.
Quantum fluctuations in the RG flow
precisely capture the effect of other couplings
that are not explicitly included.
In the bulk, this amounts to the fact that
composite operators are generated after lowering the position of the UV boundary 
by integrating out bulk degrees of freedom\cite{Heemskerk:2010hk,2011JHEP...08..051F}. 
Intuitively, sources become quantum mechanical 
because high-energy modes 
act as fluctuating sources from the point of view of 
low-energy modes\cite{Lee2012,Lee:2012xba,Lee:2013dln,2013JHEP...01..030K}.

Despite the formal mapping between general quantum field theories and gravitational theories,
classical and local geometries are expected to arise 
only for a special set of quantum field theories\cite{Heemskerk:2009pn,2014JHEP...09..118H,Benjamin:2015hsa}.
In the presence of a large number of degrees of freedom,
the quantum fluctuations of the RG flow become weak, 
which allows one to use a classical description.
However, it is much more non-trivial to have locality in the bulk.
Generally, there exist infinitely many non-local operators,
and locality arises only when the sources for the 
non-local operators are suppressed.
In QRG, the condition for the emergence of local geometry
translates into stringent constraints 
on the beta functions of the quantum field theory\cite{2014MPLA...2950158N,2015arXiv150207049N}.
From the constraints, one can try to find quantum field theories that exhibit 
a local geometry in the bulk.
On the other hand, one can use varying
degrees of locality that emerge in the bulk
as a diagnostic that differentiates one phase from another. 
The bulk locality serves as a useful order parameter
for characterizing phases of matter using geometry,
which is one of the primary goals of applying holography to condensed matter systems\cite{2008JHEP...12..015H,Hartnoll:2009sz,2009arXiv0909.0518M,2011arXiv1108.1197S}.

In this paper, we apply QRG to a three-dimensional U(N) vector model regularized on a Euclidean lattice,
which can be viewed as a quantum Bose-Hubbard model of $N$ components in two space dimensions and an imaginary time. 
It supports an interacting field theory at the critical point 
between the insulating (gapped) phase 
and the superfluid (long-range ordered) phase.
The full bulk equations of motion derived from QRG 
are solved in the large $N$ limit on finite size lattices. 
From a finite size scaling analysis, 
we show that different phases indeed exhibit
different degrees of bulk (non-) locality in the thermodynamic limit.

Here is the outline of the paper.
In Sec. II, we start with the U(N) vector model
for $N$ complex bosons 
defined on a $D$-dimensional lattice.
In this model, the general single-trace operators
can be written as $\sum_{a=1}^N \phi^*_{ia} \phi_{ja}$,
where $\phi_{ia}$ is a complex boson with flavour $a$ 
defined at site $i$.
The single-trace operators involve only one flavour contraction
and  are bi-local in space\cite{Das:2003vw,Koch:2010cy,2014arXiv1411.3151M}.
The set of single-trace operators has a special status in that
all other singlet operators can be constructed as composites of single-trace operators.
They describe the kinetic term
that connects pairs of sites through hopping.
In the vector model, all single-trace operators are quadratic,
and it is not enough to include only 
single-trace operators to describe the phase transition 
from the insulating (gapped) phase to the superfluid (symmetry-broken) phase. 
Therefore, we also include a quartic double-trace operator.
We treat the single-trace hopping terms
as deformations to the on-site action
that includes the single-trace mass term 
and the double-trace interaction term.
In the conventional RG,
a series of coarse graining generates 
an infinite tower of multi-trace operators\cite{2013arXiv1303.6641P}.
In QRG, one only keeps track of single-trace deformations 
at the expense of promoting the sources for the single-trace operators
to quantum variables.
We derive a $(D+1)$-dimensional
bulk action whose degrees of freedom
are scale dependent bi-local hopping fields $( t_{ij}(z) )$
and their conjugate variables $( t^*_{ij}(z) )$,
where $z$ is the logarithmic length scale.
The bulk theory describes 
the quantum mechanical RG flow in the space of single-trace operators.
In the large $N$ limit, the path integral in the bulk
can be replaced by saddle point equations
whose solution determines the classical geometry in the bulk.   
In Sec. III A, an analytic solution to the bulk equations of motion
is found in the deep insulating phase 
by treating hoppings as small perturbations 
compared to the on-site terms.
Away from the deep insulating phase,  
we resort to numerical solutions.
Sec. III B and C outline the numerical scheme.
In Sec. IV, the numerical solutions obtained for  three dimensional lattices
with linear sizes $3 \leq L \leq 13$ are presented.
Then we extract the behaviour of the solution in the thermodynamic limit
from finite size scaling.

\begin{figure}[h]
\includegraphics[width=3.2in,height=3in]{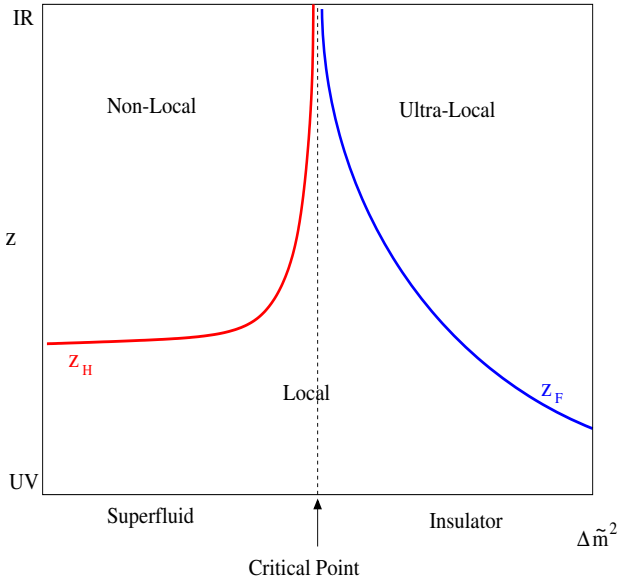}
\caption{
A schematic holographic phase diagram in the thermodynamic limit.
$\Delta \td m^2$ is the mass which tunes the system from the 
superfluid phase to the insulating phase.
$z$ is the extra dimension in the bulk,
which corresponds to a logarithmic length scale in the renormalization group.
In the insulating phase, the space gets fragmented beyond the 
 fragmentation scale, $z_F$.
In the superfluid phase, non-locality emerges beyond the horizon scale, $z_H$.
At the critical point, the bulk exhibits a local geometry.
}
\label{fig:holo_pd_L_inf}
\end{figure}

Fig. \ref{fig:holo_pd_L_inf},
which is the key result of the paper,  
summarizes the numerical solution in the thermodynamic limit, 
which is suggested from the finite size scaling analysis.
Here $\Delta \td m^2$ represents the on-site mass term,
which tunes the system away from the critical point 
either to the insulating phase 
($\Delta \td m^2 > 0$) 
or to the superfluid phase 
($\Delta \td m^2 < 0$) 
depending on its sign.
Although only the nearest neighbor hoppings are turned on 
at the UV boundary at $z=0$,
further neighbor hoppings are generated inside the bulk with $z>0$.
In the insulating phase, 
the hopping fields $t_{ij}(z)$ decay exponentially not only in $|i-j|$
but also in $z$.
As a result, the lattice is fragmented into decoupled sites in the large $z$ limit.
While the scale $z_F$ beyond which fragmentation occurs remains small in the deep insulating phase,
it diverges as the critical point is approached in the thermodynamic limit.
In the superfluid phase, on the other hand, a new scale emerges.
Instead of fragmentation, the system loses locality 
beyond a critical scale $z_H$, which is called the horizon. 
At the horizon, the hopping fields decay in a universal power-law in $|i-j|$.
Inside the horizon with $z>z_H$, 
the hopping fields
decay with a slower power-law with continuously varying exponents.
The horizon scale $z_H$ diverges as the critical point is approached
from the superfluid side.
At the critical point, the bulk exhibits a local geometry 
with a characteristic length scale that remains finite for finite $z$ 
and diverges only in the IR limit.
In summary, the insulating phase, the superfluid phase and the critical point exhibit
ultra-local,  non-local and local bulk geometries, respectively.
In this sense, {\it locality serves as a holographic order parameter
that characterizes different phases of matter}.
The goal of Sec. IV is to justify 
the holographic phase diagram in Fig. \ref{fig:holo_pd_L_inf}. 
For related works on the holographic description of the vector models, 
see Refs. 
\cite{
Das:2003vw,
Koch:2010cy,
Douglas:2010rc,
2013arXiv1303.6641P,
2014PhRvD..90h5003S,
Leigh:2014tza,
2015PhRvD..91b6002L,
2014arXiv1411.3151M}.

%%%%%%%%%%%%%%%%%%%%%%%%%%%%%%

%%%%%%%%%%%%%%%%%%%%%%%%%%%%%%%
\section{Holographic action for vector model on a lattice }

We consider a bosonic vector model in $D$-dimensional Euclidean space,
\beq
\mathcal{S}=\int d^Dx\left[\vert\nabla \v\phi\vert^2 + m^2 \vert \v\phi\vert^2+\frac{\lambda}{N}
( \vert\v \phi\vert^2 )^2 \right],
\label{eq_zero_action}
\eeq
where $\v\phi$ refers to the complex boson field with $N$ flavours, 
$\vert \v\phi\vert^2 \equiv \v \phi^* \cdot \v \phi$,
$m$ is the mass of the bosons, and  $\lambda$ is the quartic coupling.
We regularize the continuum theory on a $D$-dimensional hypercubic lattice, 
\beq
\mathcal{S}= S_0 + S_1,
\label{eq_original_discrete_action}
\eeq
where
\beqa
S_0 & = &  m^2\sum_i\left(\v{\phi}^*_i\cdot\v{\phi}_i\right)+\frac{\lambda}{N}\sum_i\left(\v{\phi}^*_i\cdot\v{\phi}_i\right)^2, 
\label{S0} \\
S_1 & = & -\sum_{ij} \tilde{t}_{ij}\left(\v{\phi}^*_i\cdot\v{\phi}_j\right)
+ \sum_{ijpq} \frac{\tilde{J}_{ijpq}}{N}\left(\v{\phi}^*_i\cdot\v{\phi}_j\right)\left(\v{\phi}^*_p\cdot\v{\phi}_q\right).
\label{S1} 
\eeqa
Here $i,j,p,q$ run over the $D$-dimensional Euclidean lattice for the $N$ component bosonic fields 
$\phi_{ia}$ with $a=1,\cdots,N$. 
$S_0$ represents on-site terms,
and $S_1$ includes the hopping term and general quartic interactions.
To guarantee the action is real, 
we impose 
$\tilde{t}_{ij} = \tilde{t}^*_{ji}$ and
$\tilde{J}_{ijpq} = \tilde{J}_{jiqp}^*$.
Here we consider the case where $\tilde{t}_{ij}$ is real.
We could make $\tilde{t}_{ij}$ complex by turning on a background gauge field.
The background gauge field in general breaks the time-reversal symmetry,
if we go to Minkowski space by Wick-rotating one of the Euclidean directions into real time.	
For general $\tilde{t}_{ij}$ and $\tilde{J}_{ijpq}$,
the model has the $U(N)$ symmetry.
The symmetry is enhanced to $O(2N)$ 
when $\td t_{ij}$'s are real
and $\tilde{J}_{ijpq} \propto \delta_{ij}\delta_{pq}$.
The action is proportional to $N$ in the large $N$ limit with fixed $m^2$, $\lambda$, $\td t_{ij}$ and $\tilde{J}_{ijpq}$.

There is a redundancy in the parameters of the action,
and the action is invariant under 
\beqa
m^2  \rightarrow m^2 + A,  && 
\td t_{ij} \rightarrow \td t_{ij} + A\delta_{ij}, \nn
\lambda \rightarrow \lambda + B, &&
\tilde{J}_{ijpq} \rightarrow  \tilde{J}_{ijpq} - B \delta_{ij} \delta_{ip} \delta_{iq}.
\eeqa
By shifting $\tilde t_{ii}$ and $\tilde{J}_{iiii}$,
$S_0$ could have been entirely absorbed into $S_1$.
Here $S_0$ is explicitly singled out because we will use $S_0$ as
`the reference theory\rq{} and treat $S_1$ as a deformation.
In other words, the renormalization group (RG) flow will be defined 
in terms of the flow of the parameters in $S_1$ 
which is defined with respect to the fixed $S_0$
with $m^2, \lambda > 0$.
It is emphasized that one can describe not only the insulating (gapped) phase 
but also the superfluid (symmetry broken) phase
because $\td t_{ij}$ can be arbitrarily large to support a Mexican hat potential
for $\v \phi$.
The freedom to choose different $m^2$ and $\lambda$ amounts to 
choosing different RG schemes, which does not affect physical observables.

We derive a holographic theory for the lattice action in Eq. (\ref{eq_original_discrete_action}) 
using the QRG scheme.
We start by writing the partition function:
\beqq
\mathcal{Z}=\int\mathcal{D}\v\phi\mathcal{D}\v\phi^*e^{-\mathcal{S}},
\eeqq
where the measure is given by
$\mathcal{D}\v\phi=\prod_{i=1}^{V}\prod_{a=1}^{N}d\phi_{ia}$,
$\mathcal{D}\v\phi^*=\prod_{i=1}^{V}\prod_{a=1}^{N}d\phi^*_{ia}$ and $V$ is the number of sites in the lattice.
The first step is to remove multi-trace operators in $S_1$ 
by promoting the sources 
for the single-trace operators into dynamical variables.
Using the representation of the Dirac-Delta function\cite{Lee2012} 
\beq
f(A)=\frac{N}{\pi}\int dt~dt^*~e^{-t(Nt^*-A)}~f( N t^*),
\label{eq_deltarep}
\eeq
we introduce a pair of complex conjugate link fields, $t^{(0)}_{ij}$ and $t^{*(0)}_{ij}$ 
for every ordered pair of sites $i,j$ to rewrite the partition function as
\beqq
\mathcal{Z} = \int \mathcal{D}\v \phi\mathcal{D}\v\phi^*
\mathcal{D}t^{(0)} \mathcal{D}t^{*(0)}
e^{-\mathcal{S}^{(0)} },
\eeqq
where 
$ \mathcal{D}t^{(0)} \mathcal{D}t^{*(0)} \equiv \Pi_{i,j} \left[ d t_{ij}^{(0)} d t_{ij}^{*(0)} \right]$
with $i,j$ running over all sites and
\beq
\mathcal{S}^{(0)} = S_0 + N\mathcal{S}_{UV}\left[t^{(0)}_{ij},t^{*(0)}_{ij}\right]
-\sum_{ij}t^{(0)}_{ij}\left(\v{\phi}^*_i\cdot\v{\phi}_j\right)
%%+m^2\sum_i\left(\v{\phi}^*_i\cdot\v{\phi}_i\right)
%%+\frac{\lambda}{N}\sum_i\left(\v{\phi}^*_i\cdot\v{\phi}_i\right)^2
\label{eq_S0prime}
\eeq
with
\beq
\mathcal{S}_{UV}\left[t^{(0)}_{ij},t^{*(0)}_{ij}\right]
=- \sum_{ij} \tilde{t}_{ij}t^{*(0)}_{ij}+
\sum_{ijpq} \tilde{J}_{ijpq}~t^{*(0)}_{ij}~t^{*(0)}_{pq}
+\sum_{ij}t^{(0)}_{ij}t^{*(0)}_{ij}.
\label{eq_SUV}
\eeq
Here and henceforth, multiplicative constants in the partition function are ignored.
In the new action, only single-trace deformations are present.
Although the double-trace operator in $S_0$ could have been removed as well,
we need to keep the on-site double-trace operator to make sure that 
the path integral for $\v{\phi}$ is well defined for any values 
of fluctuating sources $t_{ij}^{(0)}$.
The presence of the double-trace operator is also important 
because it makes the hopping fields $t^{(0)}_{ij}$ 
genuinely dynamical variables.

\begin{figure}[h]
%	\centering
\subfigure[]{\includegraphics[width=1.5in]{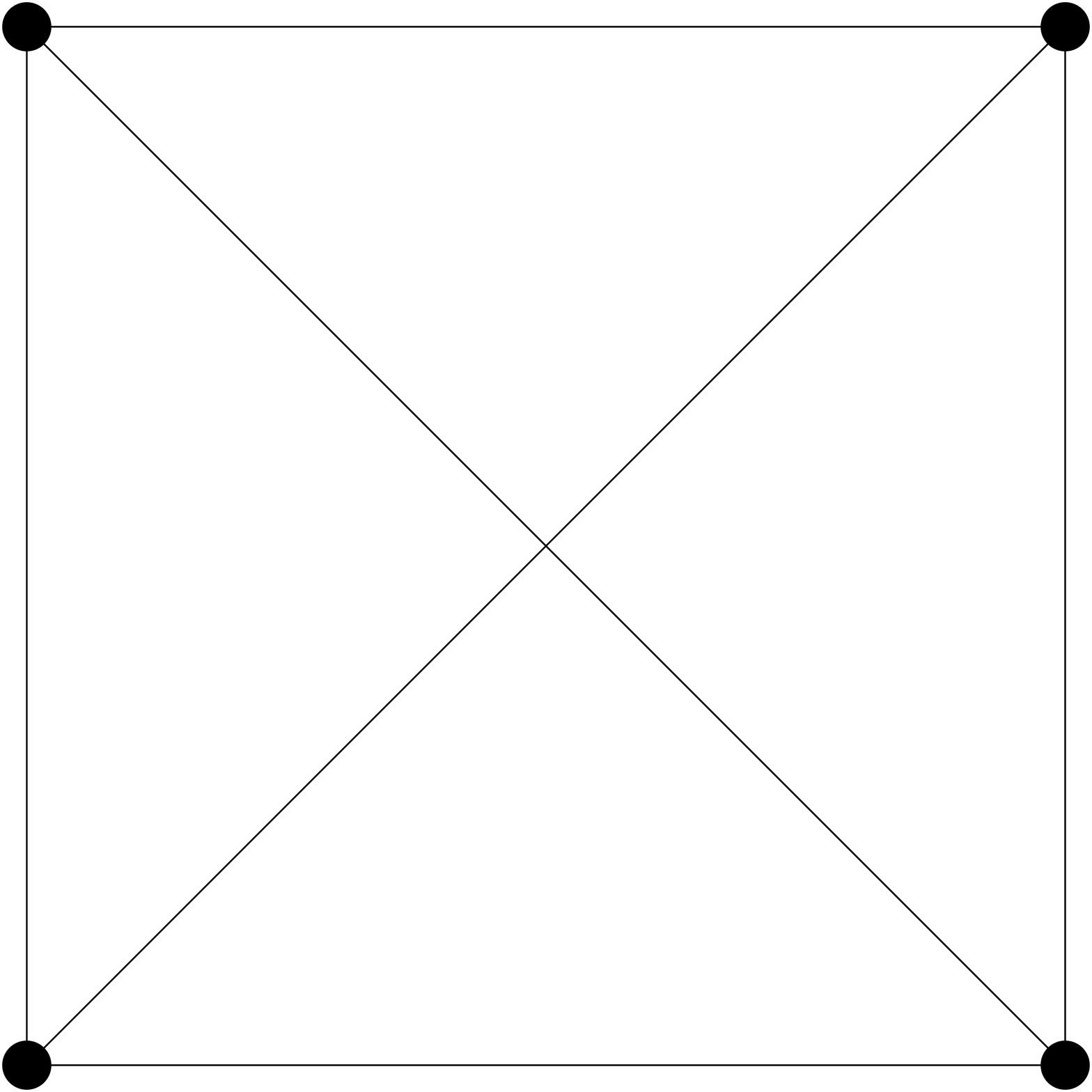}}
\hspace{3mm}
\subfigure[]{\includegraphics[width=1.7in]{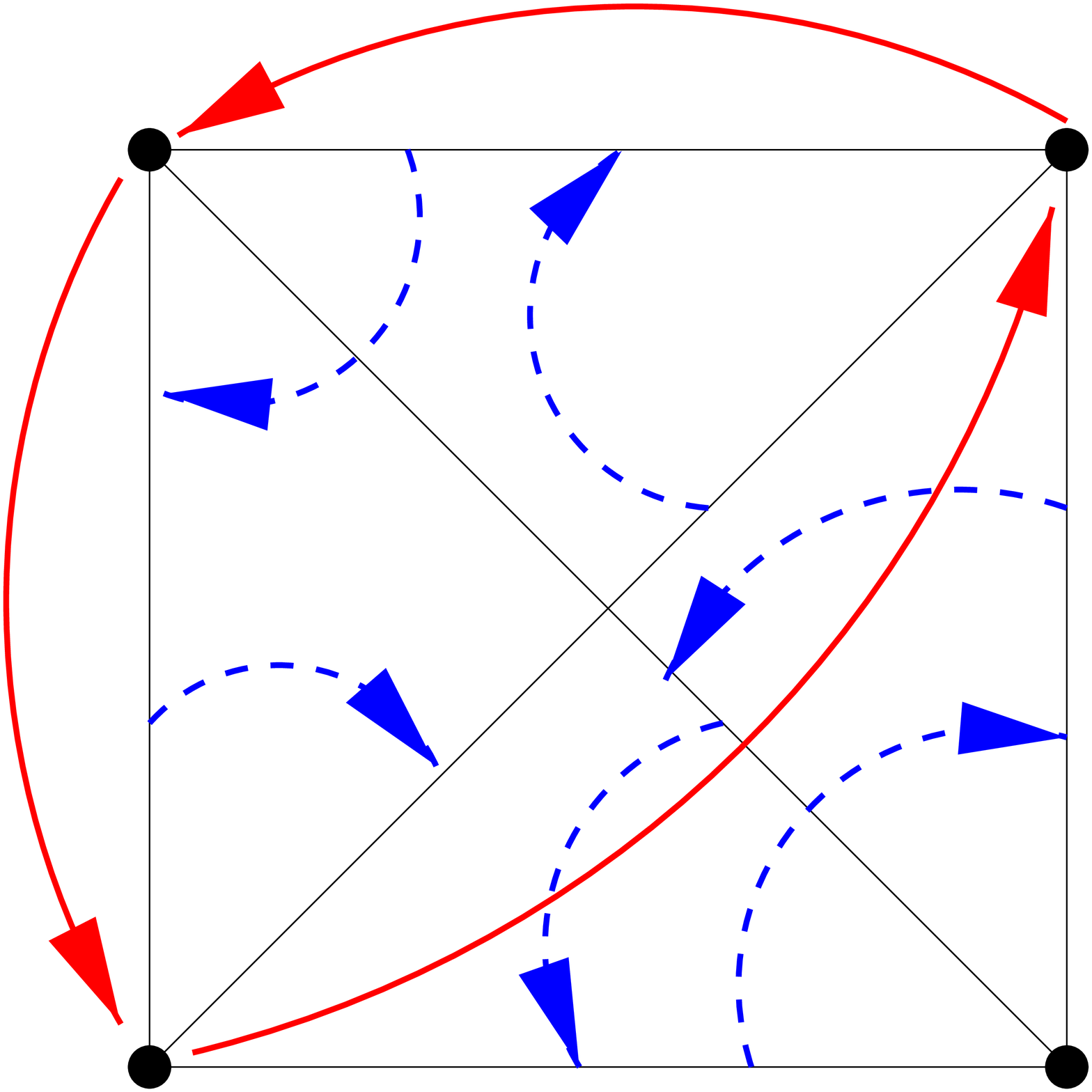}}
\hspace{10mm}
\subfigure[]{\includegraphics[width=1.5in]{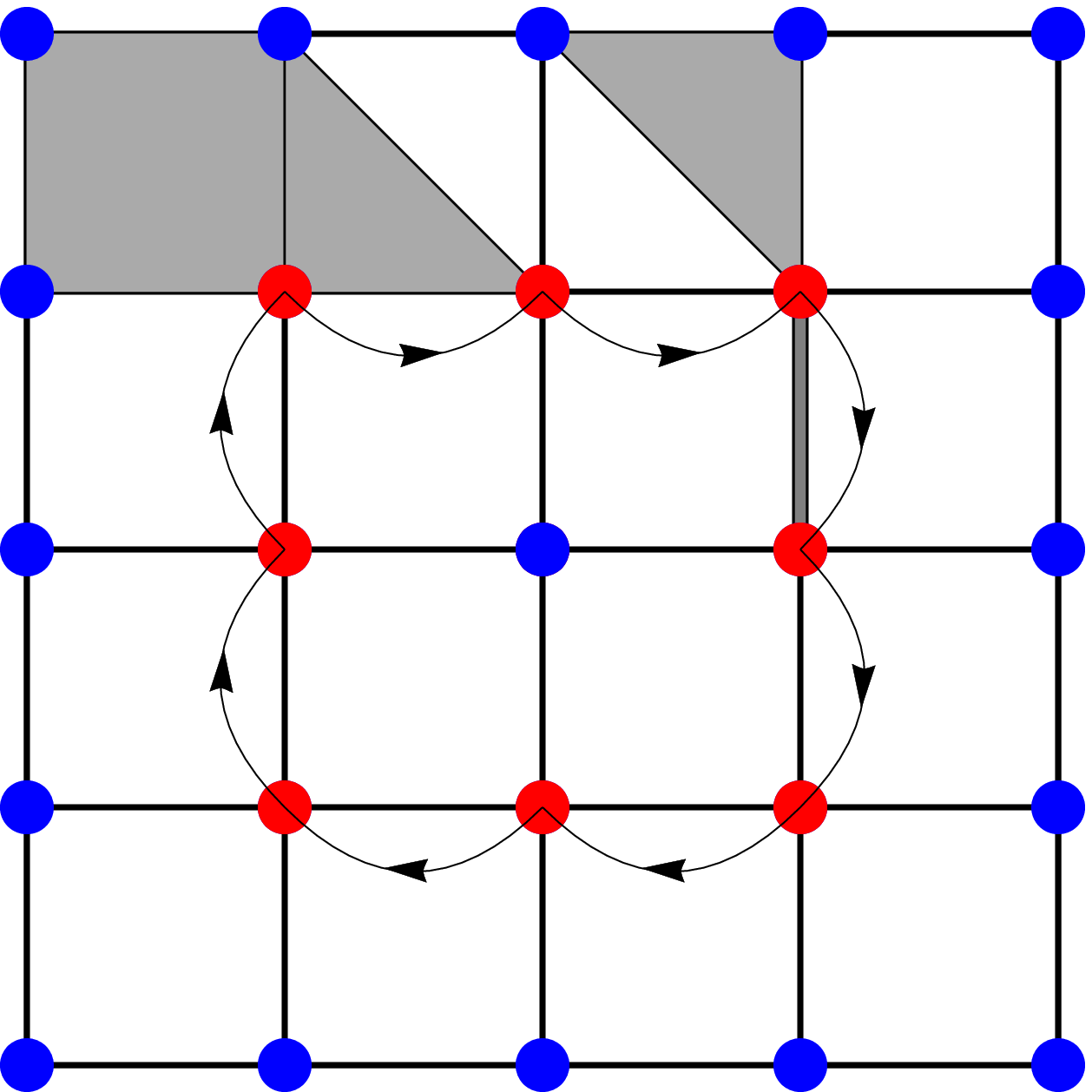}}
\caption{
(a) A lattice of four sites where every site is coupled with all other sites through dynamical hopping fields.
(b) Upon permutation among a subset of sites represented by solid (red) arrows,
the action remains invariant if
the dynamical hopping fields
are also permuted 
as is denoted by dashed (blue) arrows.
(c) An example of a permutation that doesn't preserve the local coordinate volume element. 
The blue lattice sites remain fixed, while the red lattice sites move according to the denoted arrows.
Under the transformation, some rectangles with coordinate volume $1$ are mapped to
regions with coordinate volume $1.5$ or $0.5$, as is indicated with shaded regions in the figure. 
}
\label{fig:permutation}
\end{figure}

Note that $\v{\phi}$ is coupled to dynamical sources $t^{(0)}_{ij}$
which connect all possible sites.
Therefore the new action is defined on a globally connected network 
rather than a fixed lattice.
The globally coupled network with dynamical hoppings
does not have a fixed lattice structure, and
$i,j$ should be regarded as indices for `events' rather than fixed coordinates.
Since there is no fixed lattice,
there is a freedom to relabel each event differently,
which corresponds to a discrete local coordinate transformation.
So we perform a discrete coordinate transformation,
\beq
\v{ \phi }_i \rightarrow \v{ \phi }_{ i+{\bf N}^{(0)}_i },
\label{eq:phi_shift}
\eeq
where site $i$ is displaced by a site-dependent D-dimensional shift vector, 
${\bf N}^{(0)}_i$\cite{Douglas:2010rc,Lee:2012xba,Lee:2013dln}.
Since the full set of events must map to itself,
the transformation should form cyclic permutations of the events.
This is illustrated in Figs. \ref{fig:permutation}(a),(b).
After the local coordinate transformation,
$S_0$ is manifestly invariant
and the hopping term is transformed to
\begin{align}
-\sum_{ij}t^{(0)}_{ij}(\v{\phi}^*_i\cdot\v{\phi}_j)
\rightarrow
-\sum_{ij} t^{'(0)}_{ij}(\v{\phi}^*_i\cdot\v{\phi}_j),
\label{S_22}
\end{align}
where the shifted hopping field is given by
\beq
t^{'(0)}_{ij} \equiv t^{(0)}_{i - {\bf N}^{(0)}_i, j - {\bf N}^{(0)}_j}.
\label{eq:t_shift}
\eeq
The point is that the change introduced by the local coordinate transformation
can be compensated by a transformation of the hopping fields
because $t^{(0)}_{ij}$ is {\it dynamical}.
Therefore, the dynamical hopping fields play the role of a dynamical metric in the continuum limit,
which guarantees invariance under 
a set of $D$-dimensional local coordinate transformations in the continuum limit.
The set of permutations generated by the shift includes
$D$-dimensional local coordinate transformations, 
which don't necessarily preserve the coordinate volume locally.
An example is given in Fig. \ref{fig:permutation}(c).
The fact that the hopping fields play the role of a metric is expected 
because the physical distance between two sites is determined 
from the strength of the hopping between them:
the larger the hopping between two sites, 
the shorter the physical distance.

Now we generate RG flow by coarse graining the system in real space.
The original field $\v{\phi}$ is split up into high- and low- energy fields,
where the low-energy field has a mass slightly larger than $m^2$. 
The missing fluctuations are carried away by the very massive high-energy fields. 
Then, the high-energy modes are integrated out 
to  generate quantum corrections for the low-energy modes\cite{Polchinski:1983gv,2003CEJPh...1....1P}.
For this, we introduce an auxiliary field $\v{\Phi}$ with an arbitrary mass $\mu$,
\begin{widetext}
\beqq
\mathcal{Z} = \int\mathcal{D}\v\phi\mathcal{D}\v\phi^*\mathcal{D}\v\Phi\mathcal{D}\v\Phi^*\mathcal{D}t^{(0)} \mathcal{D}t^{*(0)}e^{-\mathcal{S}^{(0)'} },
\eeqq
where
\beqq
\mathcal{S}^{(0)'} = N \mathcal{S}_{UV}\left[t^{(0)}_{ij},t^{*(0)}_{ij}\right]
-\sum_{ij}t^{'(0)}_{ij}\left(\v{\phi}^*_i\cdot\v{\phi}_j\right)+\sum_i\left[m^2\left(\v{\phi}^*_i\cdot\v{\phi}_i\right)+\mu^2\left(\v{\Phi}^*_i\cdot\v{\Phi}_i\right)\right]+\frac{\lambda}{N}\sum_i\left(\v{\phi}^*_i\cdot\v{\phi}_i\right)^2.
\eeqq
\end{widetext}
Now we rotate the physical field and the auxiliary field 
into high- and low-energy fields,
\begin{align}
\v\phi_i&=\v\phi_i'+\tilde{\v\phi}_i\nonumber\\
\v\Phi_i&=A_i \v\phi_i'+B_i \tilde{\v\phi_i},
\label{TR}
\end{align}
with $A_i =\frac{m^2}{\tilde{\mu}_i \mu}$, $B_i =-\frac{\tilde{\mu}_i }{\mu}$ and 
$\tilde{\mu_i}=\frac{m}{\sqrt{e^{2\alpha^{(1)}_i dz}-1}}$. 
$\alpha^{(1)}_i$ is the `lapse', 
which is a site-dependent constant 
that controls the local speed of coarse graining\cite{1991NuPhB.363..486O,Lee:2012xba,Lee:2013dln}.
Here $\v\phi'$ is the low-energy mode with mass $m^2e^{2\alpha_i^{(1)} dz}$, which is larger than $m^2$ by an infinitesimal amount $O(dz)$, and $\tilde{\v\phi}$ is the high-energy mode with  a large mass $\tilde{\mu_i}^2e^{2\alpha_i^{(1)} dz}$.
After rescaling the fields 
$\v\phi'_i \rightarrow e^{-\alpha_i^{(1)} dz}\v\phi'_i ,~\tilde{\v\phi}_i \rightarrow e^{-\alpha_i^{(1)} dz}\tilde{\v\phi}_i$ 
and renaming $\v\phi'_i \rightarrow \v\phi_i$, 
we rewrite the partition function as
\begin{widetext}
\beqq
\mathcal{Z} = \int\mathcal{D}\v\phi\mathcal{D}\v\phi^*
\mathcal{D}\tilde{\v\phi}\mathcal{D}\tilde{\v\phi}^*
\mathcal{D}t^{(0)} \mathcal{D}t^{*(0)}
e^{-\mathcal{S}^{(0)''} },
\eeqq
\end{widetext}
where
%{\bf\underline{(we now drop the $'$ from the low energy fields, {\em i.e.}, $\v\phi'\equiv\v\phi$)}}
\begin{align}
&\mathcal{S}^{(0)''} =N\mathcal{S}_{UV}\left[t^{(0)}_{ij},t^{*(0)}_{ij}\right]+\sum_i\left[m^2\left(\v{\phi}^*_i\cdot\v{\phi}_i\right)+\tilde{\mu_i}^2\left(\tilde{\v{\phi}}^*_i\cdot\tilde{\v{\phi}}_i\right)\right]\nonumber\\
&-\sum_{ij}t^{'(0)}_{ij}e^{-( \alpha_i^{(1)} + \alpha_j^{(1)} ) dz} \left[\left(\v{\phi}^*_i+\tilde{\v\phi}^*_i\right)\cdot\left(\v{\phi}_j+\tilde{\v\phi}_j\right)\right] \nonumber\\ 
& +\frac{\lambda}{N}
\sum_i
e^{-4\alpha_i^{(1)} dz}
\left[\left(\v{\phi}^*_i+\tilde{\v\phi}^*_i\right)\cdot\left(\v{\phi}_i+\tilde{\v\phi}_i\right)\right]^2.
\label{S_1}
\end{align}
Integrating out the high energy mode $\tilde{\v\phi}$, we obtain the renormalized action for the low energy mode,
\begin{align}
\tilde{\mathcal{S}}^{(0)''} =& 
N~\mathcal{S}_{UV}\left[t^{(0)}_{ij},t^{*(0)}_{ij}\right]
+ 2N dz \sum_i \alpha_i^{(1)} ~\left\{-\frac{1}{m^2} t^{'(0)}_{ii}\right\}\nonumber\\
&+2 dz \sum_i \alpha_i^{(1)} 
\left\{\frac{2\lambda\left(1+\f{1}{N}\right)}{m^2} 
\left({\v \phi}_i^*\cdot{\v \phi}_i\right)
-\frac{4\lambda^2}{m^2N^2} ({\v \phi}_i^*\cdot{\v \phi}_i)^3\right\}\nonumber\\
&+2 dz \sum_i \alpha_i^{(1)} \left\{\frac{2\lambda}{m^2N}\sum_{j} 
\left( t^{'(0)}_{ij}({\v \phi}_i^*\cdot{\v \phi}_j)
+ t^{'(0)}_{ji}({\v \phi}_j^*\cdot{\v \phi}_i) \right)
({\v \phi}_i^*\cdot{\v \phi}_i) \right\} \nonumber\\
&+2 dz \sum_i \alpha_i^{(1)} \left\{
-\frac{1}{m^2}\sum_{jk}t^{'(0)}_{ki}t^{'(0)}_{ij}({\v \phi}^*_k\cdot{\v \phi}_j)
-2\frac{\lambda}{N} \left(\v{\phi}^*_i\cdot\v{\phi}_i\right)^2
+\sum_{j} \frac{ 
  t^{'(0)}_{ij}\left(\v{\phi}^*_i\cdot\v{\phi}_j\right)
+ t^{'(0)}_{ji}\left(\v{\phi}^*_j\cdot\v{\phi}_i\right)
}{2}  
\right\}\nonumber\\
&-\sum_{ij}t^{'(0)}_{ij}\left(\v{\phi}^*_i\cdot\v{\phi}_j\right)+m^2\sum_i \left(\v{\phi}^*_i\cdot\v{\phi}_i\right)+\frac{\lambda}{N}\sum_i \left(\v{\phi}^*_i\cdot\v{\phi}_i\right)^2,
\label{S_1 plus delta S_1}
\end{align}
where quantum corrections are kept only to the linear order $O(dz)$ (see Appendix A for details).
In the renormalized action, multi-trace operators have been generated for the low energy field.
Under the conventional (classical) Wilsonian RG procedure,
one would apply the same coarse graining procedure to Eq. (\ref{S_1 plus delta S_1})
to generate even higher-trace deformations.
In QRG, we remove the multi-trace deformations 
before repeating the coarse graining.
This way, one can keep only single-trace deformations
at each step of coarse graining
by making the sources for the single-trace deformations fluctuating.
Just as before, we introduce new link fields $t^{(1)}_{ij}$ and $t^{*(1)}_{ij}$ 
to remove the multi-trace deformations,
\beqq
\mathcal{Z} = \int\mathcal{D}\v\phi\:\mathcal{D}\v\phi^*\:\mathcal{D}t^{(0)} \:\mathcal{D}t^{*(0)} \:\mathcal{D}t^{(1)}\:\mathcal{D}t^{*(1)} \:e^{-\mathcal{S}^{(1)} },
\eeqq
where
%\begin{widetext}
\begin{align}
\mathcal{S}^{(1)} =&N~\mathcal{S}_{UV}\left[t^{(0)}_{ij},t^{*(0)}_{ij}\right]
+N~ \sum_{ij}  (t^{(1)}_{ij}-t^{'(0)}_{ij})~t^{*(1)}_{ij} \nonumber\\
&+2N dz \sum_i \alpha_i^{(1)} ~\Biggl\{
-\frac{1}{m^2} t^{'(0)}_{ii}
+\frac{2\lambda\left(1+\f{1}{N}\right)}{m^2} t^{*(1)}_{ii}
-\frac{4\lambda^2}{m^2} \left(t^{*(1)}_{ii}\right)^3  \nonumber\\
& + \frac{2\lambda}{m^2} \sum_{j} (t^{'(0)}_{ij}t^{*(1)}_{ij} + t^{'(0)}_{ji}t^{*(1)}_{ji} ) t^{*(1)}_{ii} 
-\frac{1}{m^2}\sum_{jk}t^{'(0)}_{ki}t^{'(0)}_{ij}t^{*(1)}_{kj} \nn
& -2\lambda \left(t^{*(1)}_{ii}\right)^2
+\sum_{j} \frac{ t^{'(0)}_{ij} ~t^{*(1)}_{ij} +t^{'(0)}_{ji} ~t^{*(1)}_{ji} }{2} 
\Biggr\}\nonumber\\
&-\sum_{ij}t^{(1)}_{ij}(\v{\phi}^*_i\cdot\v{\phi}_j)+m^2\sum_i \left(\v{\phi}^*_i\cdot\v{\phi}_i\right)+\frac{\lambda}{N}\sum_i \left(\v{\phi}^*_i\cdot\v{\phi}_i\right)^2.
\label{S_2}
\end{align}

%%%%%%%%%

Now we can repeat the coarse graining procedure as before.
The coordinates of the $\v{\phi}_i$'s can be shifted  
by a new set of discrete translations, ${\bf N}^{(1)}_i$.
Then $\v{\phi}$ is split into low-energy and high-energy modes 
where the low energy mode at site $i$ has a mass $m^2 e^{\alpha_i^{(2)} dz}$.
After rescaling the fields, we integrating out the high energy modes 
to generate quantum corrections which include multi-trace operators for the low energy fields. 
We introduce another set of auxiliary fields $t^{(2)}_{ij}$ and $t^{*(2)}_{ij}$ to remove the multi-trace operators.
From here the pattern is  clear.
After doing this $\Gamma$ times, 
the partition function is written as
\beqq
\mathcal{Z} = \int 
\mathcal{D}\v\phi\:\mathcal{D}\v\phi^*
\prod_{l=0}^{\Gamma} 
\mathcal{D}t^{(l)}
\mathcal{D}t^{*(l)}
e^{-\mathcal{S}_{\Gamma}},
\eeqq
where
\begin{align}
\mathcal{S}_\Gamma =&N~\mathcal{S}_{UV}\left[t^{(0)}_{ij},t^{*(0)}_{ij}\right]
+N~\sum_{l=1}^\Gamma \Biggl[ 
\sum_{ij} (t^{(l)}_{ij}-t^{'(l-1)}_{ij}) ~t^{*(l)}_{ij} 
\nonumber\\
&+2 dz \sum_i \alpha_i^{(l)} ~\biggl\{
-\frac{1}{m^2} t^{'(l-1)}_{ii}
+\frac{2\lambda\left(1+\f{1}{N}\right)}{m^2} t^{*(l)}_{ii}
-\frac{4\lambda^2}{m^2} \left(t^{*(l)}_{ii}\right)^3  \nonumber\\
& + \frac{2\lambda}{m^2} \sum_{j} (t^{'(l-1)}_{ij}t^{*(l)}_{ij} + t^{'(l-1)}_{ji}t^{*(l)}_{ji} ) t^{*(l)}_{ii} 
-\frac{1}{m^2}\sum_{jk}t^{'(l-1)}_{ki}t^{'(l-1)}_{ij}t^{*(l)}_{kj} \nn
& -2\lambda \left(t^{*(l)}_{ii}\right)^2
+\sum_{j} \frac{ t^{'(l-1)}_{ij} ~t^{*(l)}_{ij} +t^{'(l-1)}_{ji} ~t^{*(l)}_{ji} }{2} 
\biggr\}
\Biggr]
\nonumber\\
&-\sum_{ij}t^{(\Gamma)}_{ij}(\v{\phi}^*_i\cdot\v{\phi}_j)
+m^2\sum_i \left(\v{\phi}^*_i\cdot\v{\phi}_i\right)
+\frac{\lambda}{N}\sum_i \left(\v{\phi}^*_i\cdot\v{\phi}_i\right)^2 \label{S_Gamma}
\end{align}
with $t^{'(l)}_{ij} \equiv t^{(l)}_{i-{\bf N}^{(l)}_i,j-{\bf N}^{(l)}_j}$.
Here 
$t^{(l)}_{ij}$, $t^{(l)*}_{ij}$ are the dynamical sources and their conjugate variables 
introduced at the $l$-th step of coarse graining.
The site-dependent lapse and shift 
$\{ \alpha_i^{(l)}, {\bf N}_i^{(l)} \}$
are chosen independently at each step of coarse graining.
Finally, we take the continuum limit in the RG direction 
by introducing $z=l dz$ in the limit $dz\rightarrow 0$ with $z^*\equiv \Gamma dz$ fixed .
In the limit where $z$ is continuous, 
$t_{ij}^{(l)}$,
$t_{ij}^{*(l)}$,
$\alpha_{i}^{(l)}$,
${\bf N}_{i}^{(l)}$
become 
$t_{ij}(z)$,
$t_{ij}^{*}(z)$,
$\alpha_{i}(z)$,
${\bf N}_{i}(z)$.
The partition function is written in terms of the path integral 
for the scale dependent hopping fields and their conjugate fields,
\beq
\mathcal{Z} 
=\int\mathcal{D}t\mathcal{D}t^{*}
~ e^{-N\left[\mathcal{S}_{UV}+\mathcal{S}_{bulk}+\mathcal{S}_{IR}\right]}.
\label{new_Z}
\eeq
Here $\mathcal{S}_{UV}$, defined in Eq. ($\ref{eq_SUV}$), 
is the UV boundary action for the dynamical source at $z=0$.
$S_{bulk}$ is the bulk action, 
\begin{align}
\mathcal{S}_{Bulk}=& ~ \int_0^{z^*} dz
\Biggl\{
\sum_{ij} t^{*}_{ij}(z) \left(
 \partial_z t_{ij}(z) 
 + \frac{t_{ij}(z) - t^{'}_{ij}(z)}{dz}
\right)  \nonumber\\
%%%%
+& ~\sum_i \alpha_i(z)  
\Biggl(-\frac{2}{m^2}t^{'}_{ii}(z)
+ ~ \frac{4\lambda\left(1+\f{1}{N}\right)}{m^2}t^{*}_{ii}(z)-4\lambda\left(t^{*}_{ii}(z)\right)^2-\frac{8\lambda^2}{m^2}\left(t^{*}_{ii}(z)\right)^3   \nonumber\\
%%%%
+& \sum_{j} [ t_{ij}^{'} (z)~t^{*}_{ij}(z) + t_{ji}^{'} (z)~t^{*}_{ji}(z) ]
+ \frac{4\lambda}{m^2} \sum_j [ t^{'}_{ij}(z)t^{*}_{ij}(z) + t^{'}_{ji}(z)t^{*}_{ji}(z) ] t^{*}_{ii}(z) \nn
& - \frac{2}{m^2} \sum_{jk} [ t^{'}_{ki}(z)t^{'}_{ij}(z) t^{*}_{kj}(z) ]
\Biggr) \Biggr\}
\label{eq:bulkaction}
\end{align}
with $t_{ij}^{'}(z) \equiv t_{ i-{\bf N}_i(z), j-{\bf N}_j(z)}(z)$.
$S_{IR}$ is an action,
\begin{align}
\mathcal{S}_{IR}[t_{ij}(z^*)]=&-\frac{1}{N}\ln\left[\int\mathcal{D}\v\phi\mathcal{D}\v\phi^*e^{-\left[-\sum_{ij}t_{ij}(z^*)(\v{\phi}^*_i\cdot\v{\phi}_j)+m^2\sum_i \left(\v{\phi}^*_i\cdot\v{\phi}_i\right)+\frac{\lambda}{N}\sum_i \left(\v{\phi}^*_i\cdot\v{\phi}_i\right)^2\right]}\right],
\label{S_IR}
\end{align}
which is defined at the IR boundary $z=z^*$.
$z^*$ is the scale at which we stop the coarse graining procedure,
which can be taken to be infinite.
%\end{widetext}

\begin{figure}[h]
%	\centering
\includegraphics[width=3in]{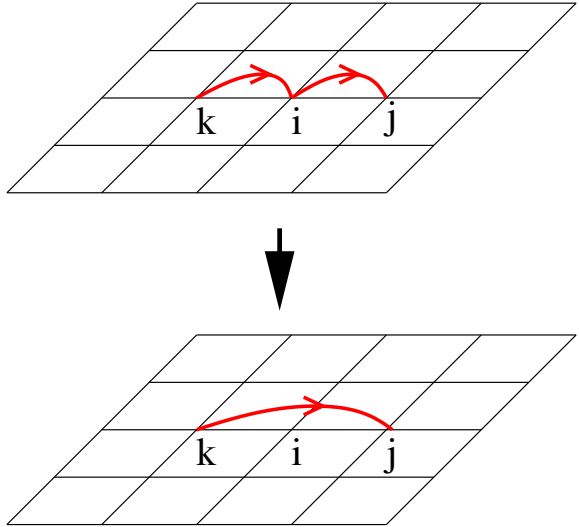}
\caption{
Two hopping fields $t_{ki}$ and $t_{ij}$ fuse 
to generate a longer range hopping between sites $k$ and $j$.
}
\label{fig:fusion}
\end{figure}

The new action is written in terms of the fields $t_{ij}(z)$ and $t^*_{ij}(z)$
which are bi-local within the original $D$-dimensional lattice
and local in the $z$ direction.
If one interprets $z$ as `time', 
the action describes a system of bi-local hopping fields under 
a Hamiltonian evolution in imaginary time.
$t_{ij}$ and $t_{ij}^*$
correspond to annihilation and creation operators respectively\cite{Lee2012},
and the last term in Eq. (\ref{eq:bulkaction}) describes the process
where two hoppings merge to become a longer-range hopping.
This is illustrated in Fig. \ref{fig:fusion}.
In the path integral language,
all RG paths are summed over 
for the single-trace operators.
The weight for each RG path is determined by the bulk action, $\mathcal{S}_{Bulk}$.

In the bulk, the lapse and shift vector parameterize local coordinate transformations.
The partition function does not depend on  $\{ \alpha_i(z), {\bf N}_i(z) \}$ because 
different choices of lapse and shift merely correspond to different renormalization group schemes \cite{Lee:2012xba,Lee:2013dln}.
As a result, the local coordinate transformation in the bulk is a gauge symmetry.
Unlike in the continuum, 
the action contains terms that are non-linear in the shift.
This is because the discrete shift is a finite transformation.
When $t_{ij}(z)$ vary slowly in $i,j$,
one can take a continuum limit along the $D$-dimensional directions 
by writing
\beq
t_{ij}(z)  \rightarrow t({\bf r}_1, {\bf r}_2, z), 
\eeq
where ${\bf r}_1 = a {\bf r}_i$, ${\bf r}_2 = a {\bf r}_j$
in the limit that the lattice spacing $a$ is small.
Then, the shift can be performed infinitesimally,
$ {\bf N}_i(z) \rightarrow {\bf n} ( {\bf r}, z) dz$,
and one can drop the terms that are non-linear in $dz$.
In this case, 
the action is linear in ${\bf n}({\bf r},z)$ and $\alpha({\bf r},z)$,
which generate $(D+1)$-dimensional coordinate transformations in the bulk.
Also, in the continuum, 
the bi-local fields can be represented 
by an infinite set of fields with arbitrarily large spin.
Here we don't take the continuum limit and proceed with the lattice action.

In the absence of the quartic term ($\lambda = 0$), 
the theory has a larger symmetry generated by\cite{Leigh:2014tza,
2015PhRvD..91b6002L}
\beqaa
\v{\phi}_i  \rightarrow  \sum_j V_{ij} \v{\phi}_j, \nn
t_{ij}^{}   \rightarrow  \sum_{i^{'} j^{'}} V_{i i^{'}}  t_{i^{'} j^{'}}^{} V^*_{j j^{'}},
\eeqaa
where $V_{ij}$ is a unitary matrix in the space of all sites in the lattice.
The enlarged symmetry is the source of the higher spin gauge symmetry for the free theory with $\lambda=0$.
In the interacting theory with $\lambda \neq 0$, 
the symmetry is broken down to the local coordinate transformation.
One might keep the higher spin symmetry in the bulk action even when $\lambda \neq 0$ 
by absorbing the quartic term into the  UV boundary action $S_{UV}$.
In that case, there is no $\lambda$ in the bulk,
and the quartic term is implemented only through the alternative boundary condition determined by $S_{UV}$\cite{2002PhLB..550..213K}.
However, this can be done only in the strict large $N$ limit 
where the fluctuations of the dynamical hopping field induced by $S_{UV}$ are negligible.
What is less satisfying in this description for the interacting theory
is the fact that $1/N$ corrections are singular 
because the path integral for $\v{\phi}$ is ill-defined
without the quartic term in $S_0$ for fluctuating hopping fields.
Therefore we choose to include the quartic term in the bulk for general $N$, 
which breaks the higher spin symmetry.

Because the fields $t_{ij}, t^*_{ij}$ are singlets under $U(N)$ rotations,
the action in Eq. (\ref{new_Z}) is proportional to $N$ in the large $N$ limit.
Therefore, the partition function can be obtained by the saddle point approximation in the large $N$ limit.
To that effect, all we need to solve are the equations of motion, which we focus on from now on. 
Although $t_{ij}$ and $t^*_{ij}$ are complex conjugates of each other in Eq. (\ref{new_Z}), their saddle points configurations do not have to satisfy that condition. 
This is because the amplitude and phase of $t_{ij}$ can take complex values at the saddle point. 
Therefore, we treat  $t_{ij}$ and $t^*_{ij}$ 
as independent complex fields in the saddle point equation.

In the gauge with $\alpha_i(z)=1$ and ${\bf N}_i(z)=0$,
the equations of motion for the bulk fields are given by
\begin{widetext}
\begin{align}
\partial_z t_{ij}&=-2 \left\{\frac{2\lambda\,\delta_{ij}}{m^2} - \delta_{ij}\left[4\lambda + \frac{12\lambda^2}{m^2}t^*_{ii} \right]t^*_{ii} + \f{2\lambda\, \delta_{ij}}{m^2} \sum_k \left(t_{ik}t^*_{ik} + t_{ki}t^*_{ki} \right) \right. \nonumber\\
&\qquad \qquad \qquad \qquad \qquad +\left. \left[1+\frac{2\lambda}{m^2}\left(t^*_{ii}+t^*_{jj}\right)\right]t_{ij}-\frac{1}{m^2}\sum_{k}t_{ik}t_{kj}\right\}, \nonumber\\
\partial_z t^*_{ij}&=2 \left\{-\frac{\delta_{ij}}{m^2}+\left[1+\frac{2\lambda}{m^2}\left(t^*_{ii}+t^*_{jj}\right)\right]t^*_{ij}-\frac{1}{m^2}\sum_k\left(t^*_{ik}t_{jk}+t_{ki}t^*_{kj}\right)\right\}
\label{EOMs},
\end{align}
\end{widetext}
subject to two boundary conditions
\begin{align}
&t_{ij}(0)-\tilde{t}_{ij} + 2 \sum_{pq} \tilde{J}_{ijpq}t^{*}_{pq}(0)=0,
\label{eq_UV_BC}\\
&t^{*}_{ij}(z^*)+\frac{\partial\mathcal{S}_{IR}}{\partial t_{ij}(z^*)}=0,
\end{align}
which are derived by extremizing the boundary actions at $z=0$ and $z=z^*$, respectively. 
In the equations of motion, sub-leading terms are dropped in the large $N$ limit. 
From the IR boundary action, one can readily write the IR boundary condition as
\begin{align}
t^{*}_{ij}(z=z^*)=\frac{1}{N}\langle({\v \phi}^*_i\cdot{\v \phi}_j)\rangle_{\mathcal{S}_{IR}},
\label{eq_general_IR_BC}
\end{align}
where $< O >_{\mathcal{S}_{IR}}$ denotes the expectation value of $O$ evaluated with respect to $\mathcal{S}_{IR}$
at $z=z^*$.
This relationship must hold anywhere inside the bulk, since the RG process can be stopped at any point. 
Therefore, the on-shell value of $t^{*}_{ij}(z)$ coincides with the 2-point correlation functions at scale $z$,
which are completely determined by $t_{ij}(z)$.

In the end, one has to solve the first order differential equations for $t_{ij}$ and $t_{ij}^*$ 
with one set of boundary conditions imposed at the UV boundary and the other imposed at the IR boundary.
It is the IR boundary condition that imposes the constraints that the expectation values of operators ($t_{ij}^*$) have to satisfy for a given set of sources ($t_{ij}$).
In particular, the vacuum expectation value has to satisfy the Ward identity associated with the discrete coordinate transformation,
 \begin{align}
 \langle{\v \phi}^*_i\cdot{\v \phi}_j\rangle_{t} = \langle{\v \phi^{}}^{*}_{i+\textbf{N}_i}\cdot{\v \phi^{}}_{j+\textbf{N}_j}\rangle_{t^{'}},
 \end{align}
 where $t^{'}$ is defined in Eq. (\ref{eq:t_shift}). 
 Since this equation has to be satisfied at all $z$,
it becomes a dynamical constraint,  $t^*_{ij}(t(z)) = t^{*}_{i+\textbf{N}_i,j+\textbf{N}_j}(t^{'}(z))$,
which is the discrete version of the energy-momentum conservation imposed by ${\bf N}_i(z)$.
This condition is automatically implied by the IR boundary condition
even in the fixed gauge with ${\bf N}_i(z)=0$.
It is noted that $t_{ij}^*$ cannot be arbitrary at the IR boundary, 
but it has to be the actual vacuum expectation value
computed from the IR boundary action, $S_{IR}$.

%%%%%%%%%%%%%%%%%%%%%%%%%%%%%%%%%%%%%%%%%%%%%%%%%%%%%%%%%%%%%%%%%%%%%%%%%%%%%

\section{Solutions to saddle point equations}

In this section, we examine the solution to the equations of motion  
both analytically and numerically.
%Our goal is to gain holographic insight into the   insulating phase, the superfluid phase 
%and the critical point that separates the former two phases in the large $N$ limit, 
%by examining the behaviors of $t_{ij}(z)$, $t^*_{ij}(z)$ in the bulk.
We assume that translational symmetry is present, in which case  $t_{ij}$ and $t^*_{ij}$ depend only on $|i-j| \equiv |{\bf r}_i-{\bf r}_j|$. Also, since we are maintaining the full $SO(D)$ and $D$-dimensional inversion symmetries at each scale $z$, all $t_{ij}(z)$ and $t^*_{ij}(z)$ are real.
We focus on the three dimensional cubic lattice of linear size $L$  with periodic boundary conditions.

To uniquely determine the solutions we need to impose the boundary conditions in Eqs. (\ref{eq_UV_BC}) and (\ref{eq_general_IR_BC}). 
At the UV boundary, Eq. (\ref{eq_UV_BC}) imposes the Dirichlet boundary condition 
for $t_{ij}(0) = \td t_{ij}$ in the absence of $\tilde{J}_{ijpq}$.
To impose the IR boundary condition in Eq. (\ref{eq_general_IR_BC}) we compute $\mathcal{S}_{IR}[t_{ij}(z^*)]$ 
by introducing a Hubbard-Stratonovich field $\sigma$ which satisfies the saddle-point equation in the large $N$ limit 
(see Appendix B), 
\beq
\sigma = \lambda~\frac{1}{V}\sum_{k\in V} \frac{1}{m^2+2\sigma-t_k(z^*)}
\label{eq_self_constist_momentspace_discrete},
\eeq
where $t_k(z) = \sum_j e^{-i {\bf k} \cdot ( {\bf r}_{j} - {\bf r}_i ) } t_{ij}(z)$.
Then the IR boundary condition becomes
\beq
t^{*}_{k}(z^*)=\frac{1}{N}\langle({\v \phi}^*_k\cdot{\v \phi}_{-k})\rangle_{\mathcal{S}_{IR}}=
\frac{1}{m^2+2\sigma-t_k(z^*)}.
\label{eq_tstar_IR_general}
\eeq

\subsection{Analytic solutions in the deep insulating phase }

For  $m^2 \gg \tilde{t}_{ij}$ and $m^4 \gg \lambda$, 
the quadratic on-site action dominates.
All correlation functions decay exponentially, resulting in the   insulating phase.
In this case, one can easily obtain an analytic solution to the equations of motion 
as a perturbation series in $\f{\tilde{t}_{ij}}{m^2}$ and $\f{\lambda}{m^4}$. 
To order $1/m^6$, the solutions are
\begin{align}
t^*_{ij}&=\f{\delta_{ij}}{m^2}+\f{\tilde{t}_{ij}}{m^4}e^{-2 z} +\f{e^{-2 z}}{m^6}\left(\sum_{k}\tilde{t}_{ik}\tilde{t}_{kj} - 2\lambda \delta_{ij}\right),\\
t_{ij}&=\tilde{t}_{ij}e^{-2 z}+\f{1}{m^2}\left[\left(1- e^{-2z}\right)\left(2\lambda\delta_{ij}+\sum_{k}\tilde{t}_{ik}\tilde{t}_{kj}\:e^{-2 z}\right)\right]\nonumber\\
&~~~+\f{1}{m^4}
\left[\sum_k\sum_{k'}\tilde{t}_{ik}\tilde{t}_{kk'}\tilde{t}_{k'j}\left(e^{-6 z} - 2\: e^{-4 z} + e^{-2 z}\right) 
+4 \lambda \tilde{t}_{ij}\left(e^{-4 z} - e^{-2 z}\right)\right]\nonumber\\
&~~~+\f{1}{m^6}(e^{-2z} - 1)\left[4 e^{-2z}\lambda \left(\tilde{t}_{ii}\tilde{t}_{ij} - \lambda \delta_{ij}\right)  + 6\lambda \left(e^{-2z} - e^{-4z}\right) \sum_{k} \tilde{t}_{ik}\tilde{t}_{kj} \right. \nonumber\\
&~~~ \left.  - \left(e^{-6z} - 2e^{-4z} +e^{-2z}\right) \sum_{k}\sum_{k'}\sum_{k''} \tilde{t}_{ik}\tilde{t}_{kk'} \tilde{t}_{k'k''}\tilde{t}_{k''j}
\right].
\label{analytical_pert_solution}
\end{align}
Both $t_{ij}(z)$ and $t^*_{ij}(z)$ decay exponentially as one moves towards the IR boundary. 
This is expected since bosons are localized at low energies.
%%%%%%%%%%%NEW
It is noted that $t_{ij}^*$ for $i \neq j$ decays as $e^{-2z}$ with increasing $z$.
This feature holds generally even when $t_{ij}$ is not small.
This is due to the fact that  the transformation in Eq. (\ref{TR}) followed by the rescaling by $e^{-dz}$ implies
${\v \phi}_i(z+dz) = e^{-dz}  {\v \phi}_i(z)$ up to the contribution from the auxiliary field 
introduced at each step of coarse graining which has zero correlation length.
This implies that 
\begin{align}
< {\v \phi}^*_i(z+dz)\cdot{\v \phi}_j(z+dz) >
= e^{-2 dz} <{\v \phi}^*_i(z)\cdot{\v \phi}_j(z)>
\label{2zdecay}
\end{align}
for $i \neq j$.

%%%%%%%%%%%%

In the insulating phase a large simplification can be made for the IR boundary condition. 
Since both $t_{ij}(z)$ and $t_{ij}^*(z)$ decay exponentially in $z$ except for $i=j$, 
we can approximate them as 
$t_{ij}(z^*) \approx \f{2 \lambda \delta_{ij}}{m^2}$ and $t^*_{ij}(z^*) \approx \f{\delta_{ij}}{m^2}$ 
for a sufficiently large $z^*$.
In this case, the IR boundary condition reduces to a single-site problem,
%large enough, we can incorporate this into our boundary conditions. With this assumption Eq. (\ref{eq_tstar_IR_general}) becomes an on-site problem in real space,
%Eq. (\ref{analytical_pert_solution})) that 
\beq
t^{*}_{ij}(z^*)= \frac{\de_{ij}}{m^2+2\sigma-t_{ii}(z^*)},
\label{eq_tstar_IR_on_site}
\eeq
where $\sigma$ is given by 
%\beq
$\sigma = \f14 \left(t_{ii}(z^*) - m^2 + \sqrt{(t_{ii}(z^*) - m^2)^2 + 8\lambda}\right)$,
%\label{sigma_on_site}
%\eeq
which is the solution of the on-site saddle point equation
%\beq
$\sigma = \frac{\lambda}{m^2+2\sigma - t_{ii}(z^*)}$.
%\eeq

%%%%%%%%%%%%%%%
\subsection{Numerical solutions}

Since it is hard to solve the full equations of motion analytically in general, 
we resort to numerical solutions for the superfluid phase 
and the critical point.
We solve the coupled differential equations using a spectral method. 
The details of the numerical method can be found in Appendix C. 
Numerical solutions are obtained for the cubic lattice with linear sizes $3 \leq L \leq 13$. 
We set $\lambda = 1$, $m^2 = 25$, and $\tilde{t}_{ij} = 0$ 
except for the on-site and nearest neighbor hoppings.
The nearest neighbor hoppings are set to be $1$,
and the on-site `hopping' $\td t_{ii}$ is tuned to change the physical bare mass,
\beq
\tilde{m}^2 \equiv m^2 -\tilde{t}_{ii}
\eeq
to drive the phase transition
from the insulating phase to the superfluid phase.
The physical mass $\td m^2$ can be either positive or negative
whereas the mass $m^2$ of the reference action in Eq. (\ref{S0}) is fixed to be positive.
The reason for choosing a positive $m^2$ 
is to make sure that only short-distance modes 
are integrated out in the coarse graining procedure. 
Although there is no genuine symmetry breaking phase transition in finite systems,
features of the superfluid phase and the critical point can be still identified
from the behaviors at intermediate length scales before the finite size effect takes over.
Henceforth, we will keep using the terminology ``superfluid'' 
to refer to systems in parameter regimes
which become real symmetry broken states 
in the thermodynamic limit.
Although only the nearest neighbor hopping is turned on at the UV boundary,
all further neighbor hoppings are generated inside the bulk. 
From now on, we will use the notation 
$t_{(l,m,n)}$, $t^*_{(l,m,n)}$
to denote the specific fields 
$t_{ij}, t^*_{ij}$ with ${\bf r}_j - {\bf r}_i = (l,m,n)$ in units of the lattice spacing.
For example, the three nearest neighbor hopping fields are represented by 
$t_{(0,0,1)}(z)$, $t_{(0,1,0)}(z)$, $t_{(1,0,0)}(z)$.

%%%%%%%%%
\subsection{Universal IR boundary condition in finite systems}

%%%%%%%%%
\begin{figure}[h]
	\centering
\subfigure[]{\includegraphics[width=3in]{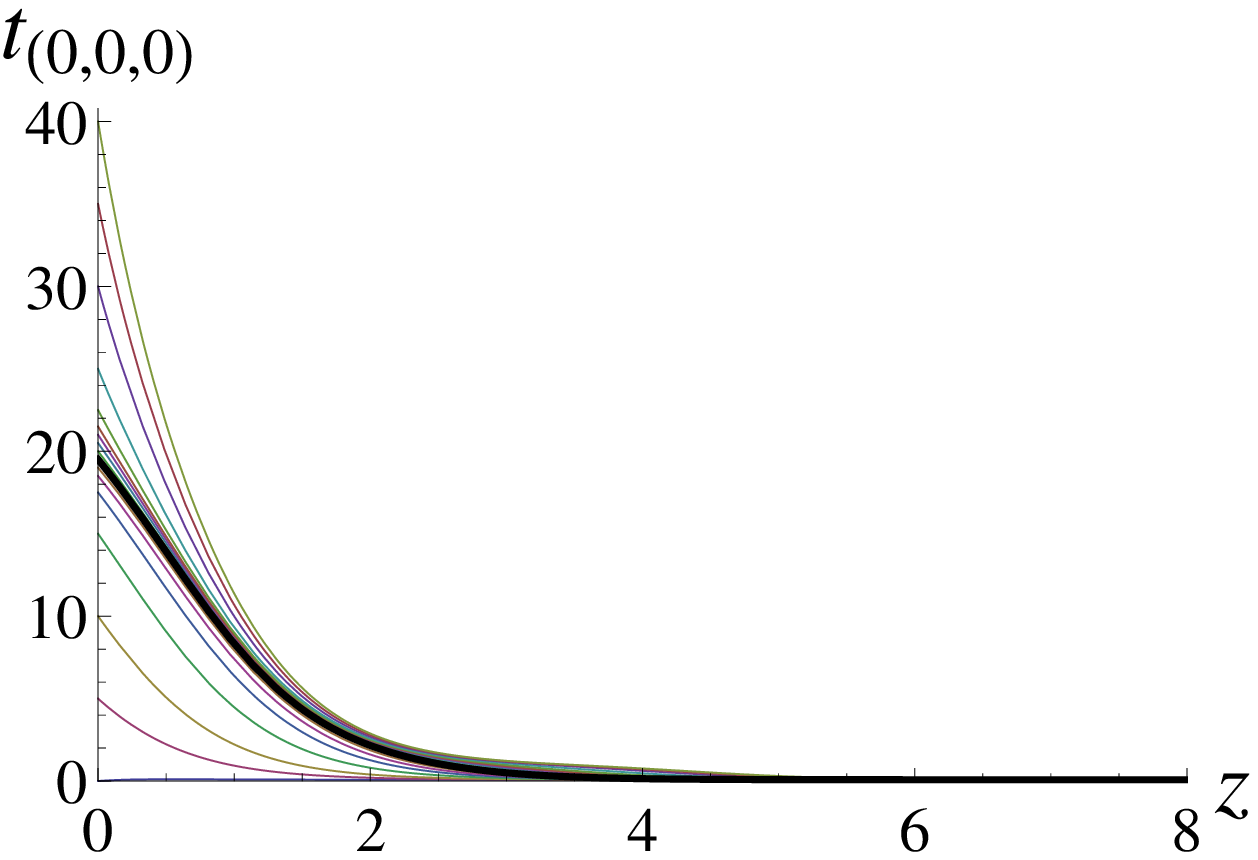}}
\subfigure[]{\includegraphics[width=3in]{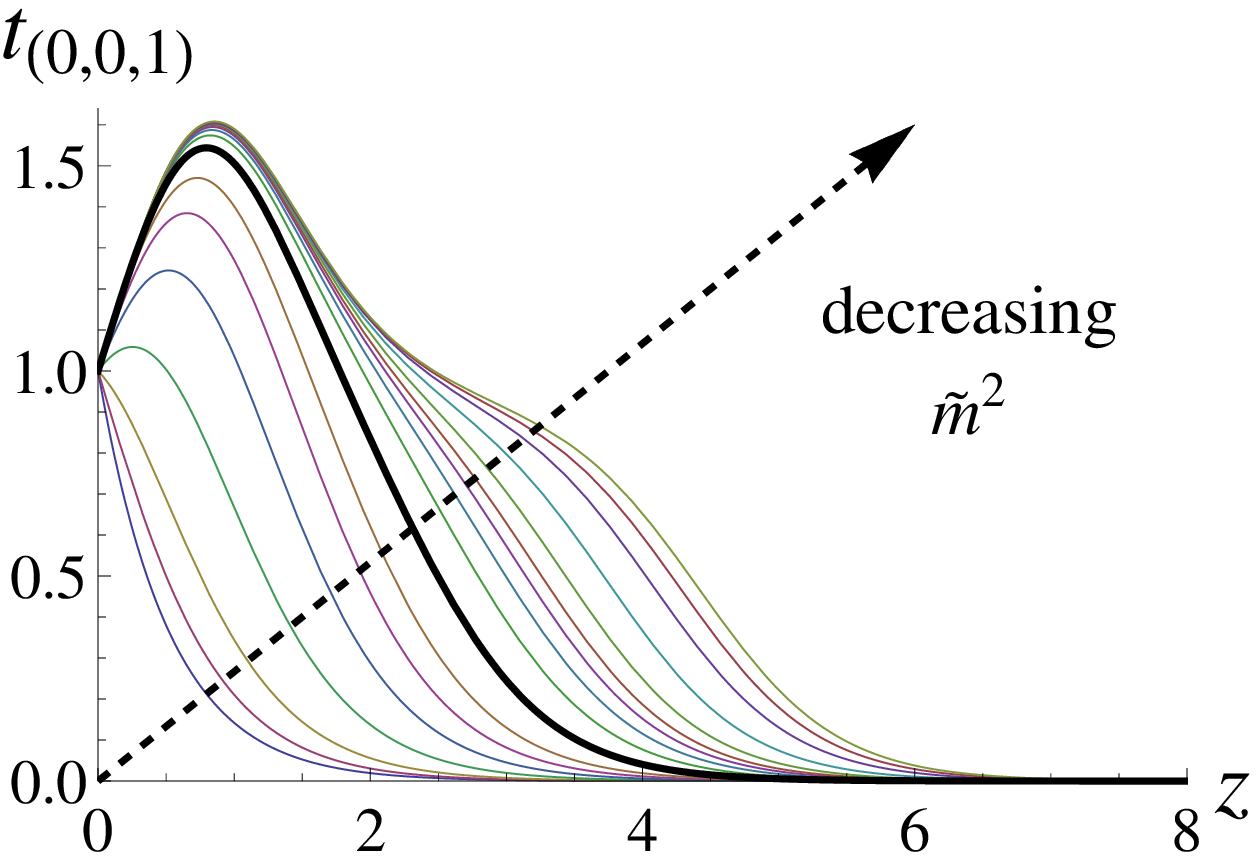}}\\
\subfigure[]{\includegraphics[width=3in]{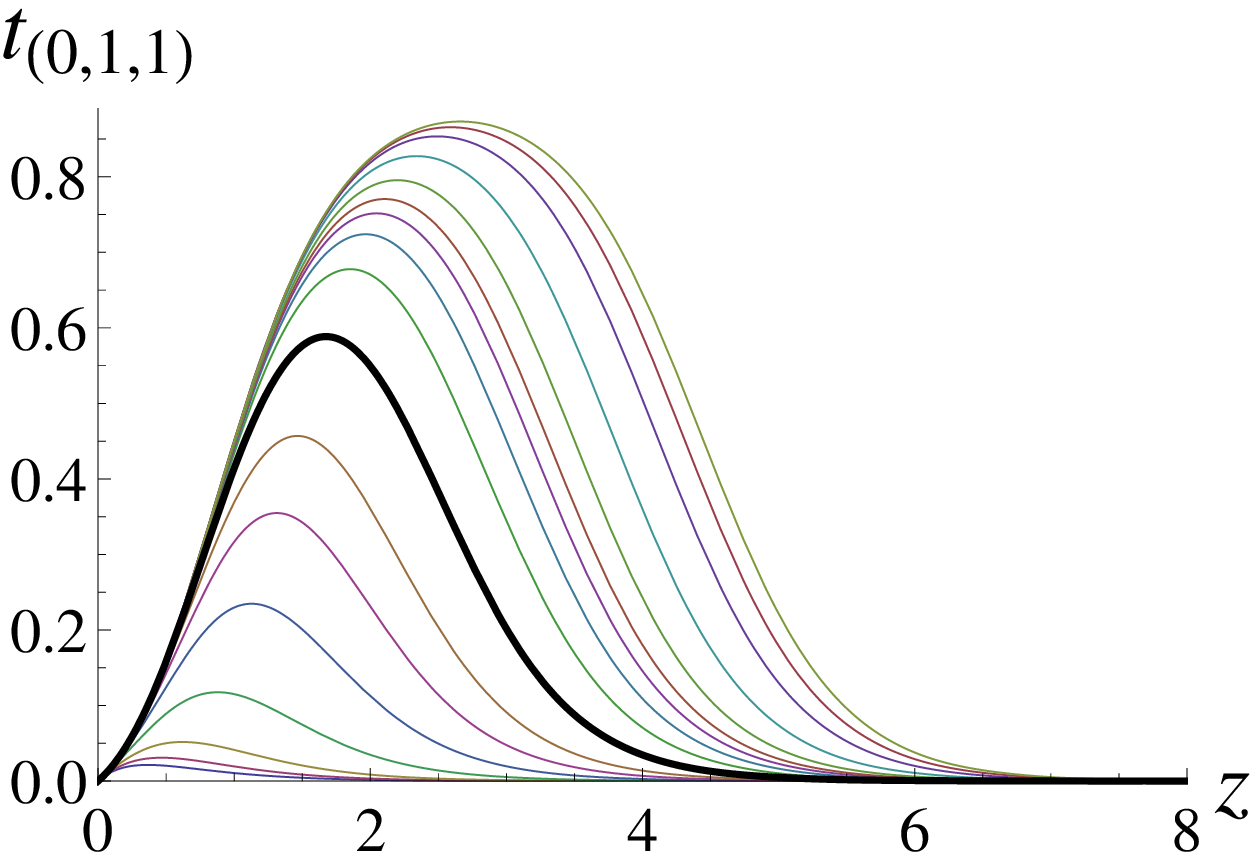}}
\subfigure[]{\includegraphics[width=3in]{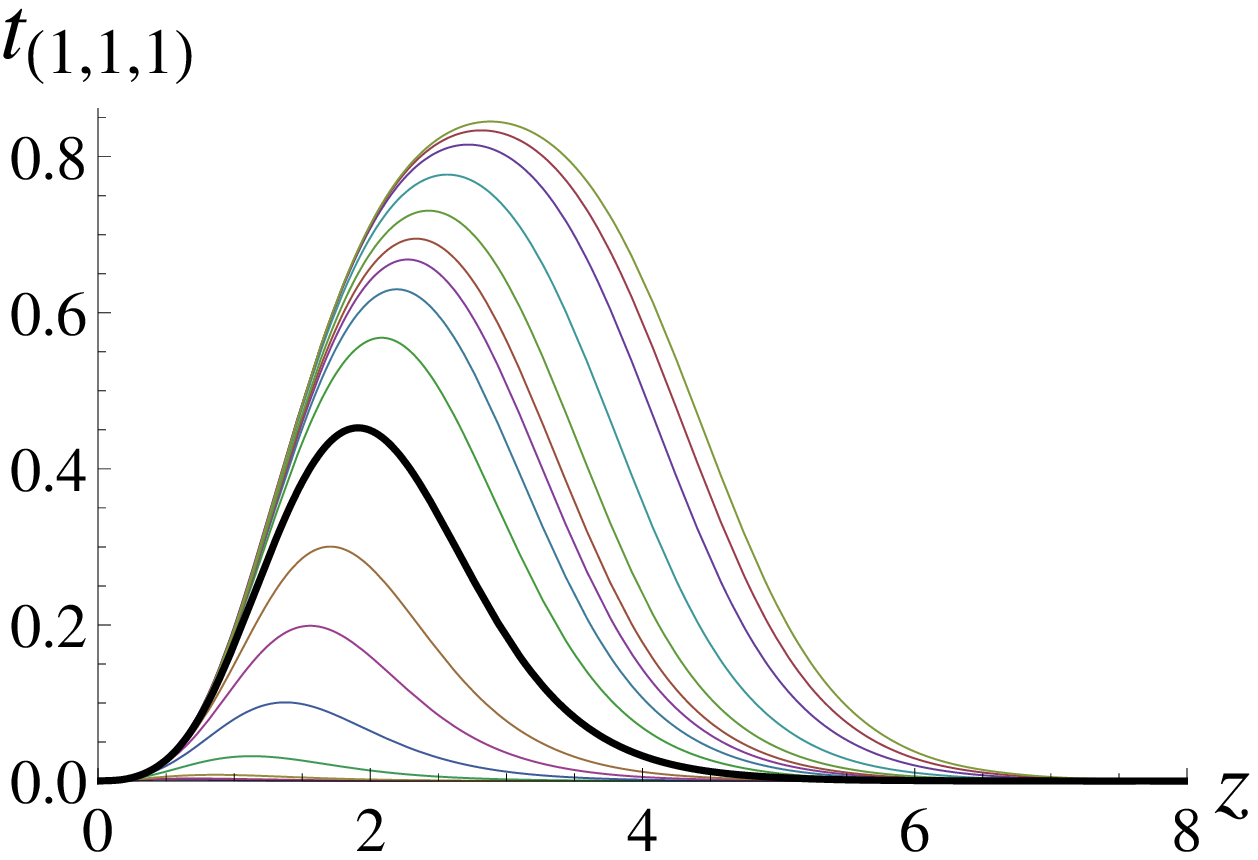}}
\caption{
The scale dependent dynamical hopping fields $t_{(l,m,n)}(z)$ 
for $ (l,m,n) \equiv {\bf r}_j - {\bf r}_i = (0,0,0), (0,0,1), (0,1,1), (1,1,1)$
for $L=3$ and $z^*=8$. 
Different curves in each panel are at different values of $\tilde{m}^2 = $  $25$, $20$, $15$, $10$, $7.5$, $6.5$, $6$, $5.49454$, $5$, $4.5$, $4$, $3.5$, $2.5$, $0$, $-5$, $-10$, $-15$. 
The hopping fields increase monotonically with decreasing $\tilde{m}^2$, as is denoted by the arrow in (b).
The thick line is for $\tilde{m}^2 = 5.49454$ which is the critical point in the thermodynamic limit.
The curves below (above) the thick lines are in the insulating phase (superfluid phase).
Signatures of the phase transition will be discussed in Sec. IV in detail.
}
\label{3x3x3 t fields GBC zstar 8}
\end{figure}
%%%%%%%%%
%%%%%%%%%
\begin{figure}[h]
	\centering
\subfigure[]{\includegraphics[width=3in]{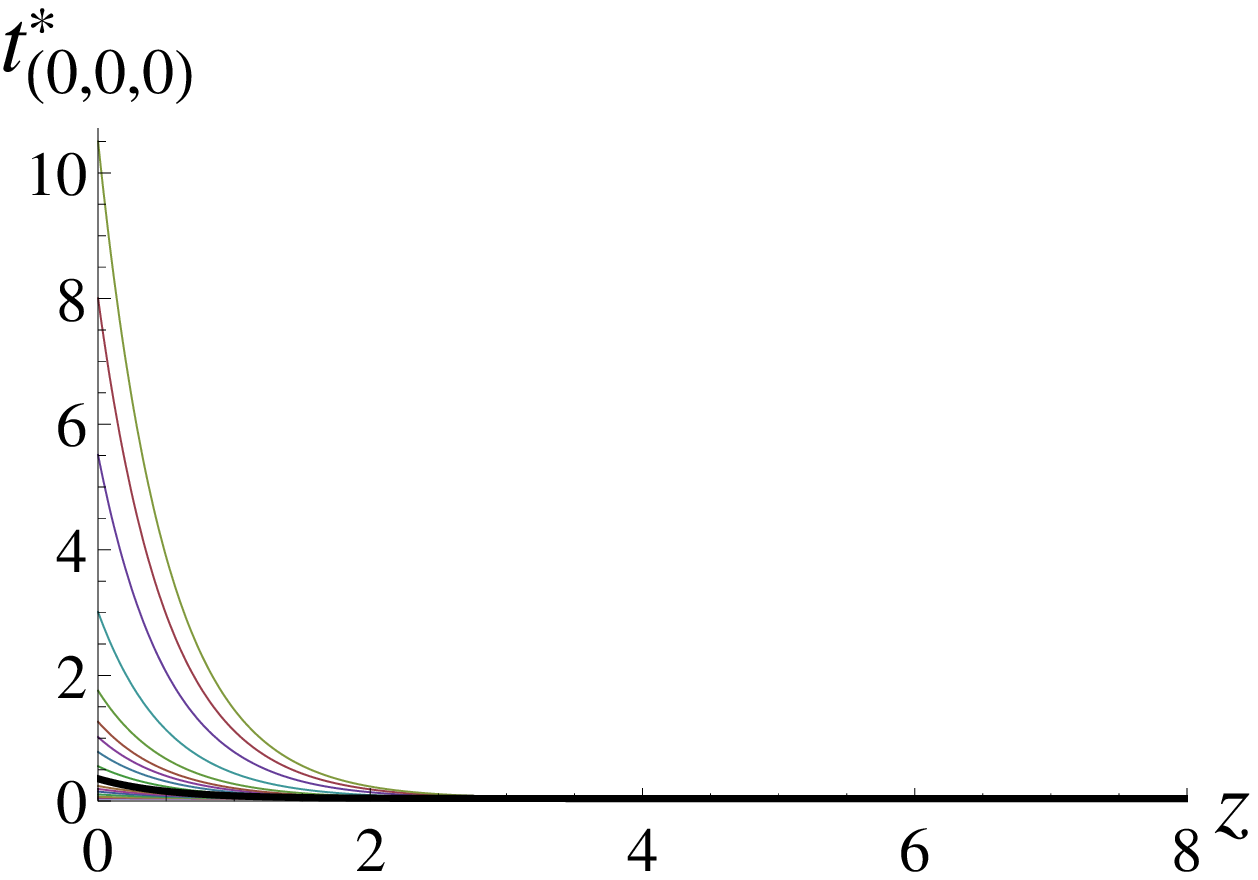}}
\subfigure[]{\includegraphics[width=3in]{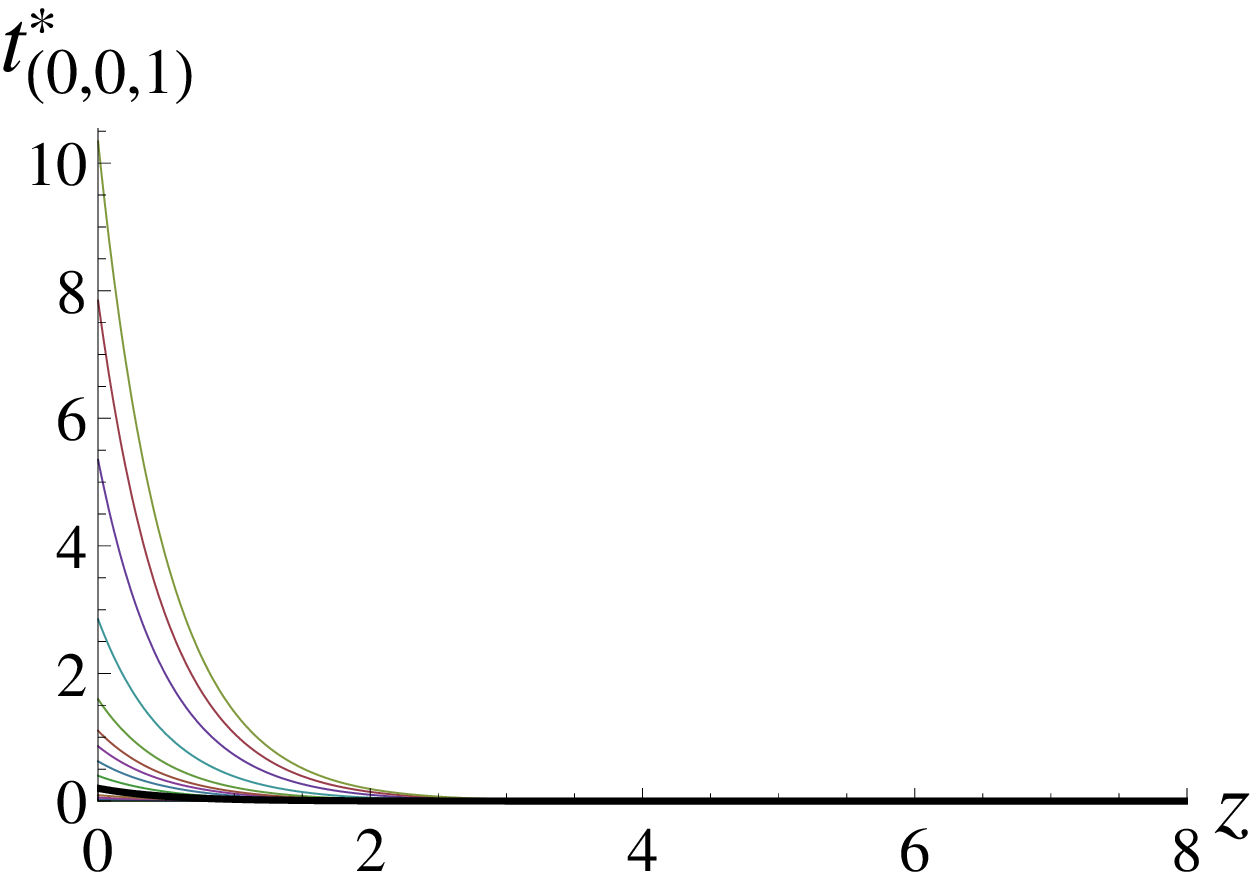}}\\
\subfigure[]{\includegraphics[width=3in]{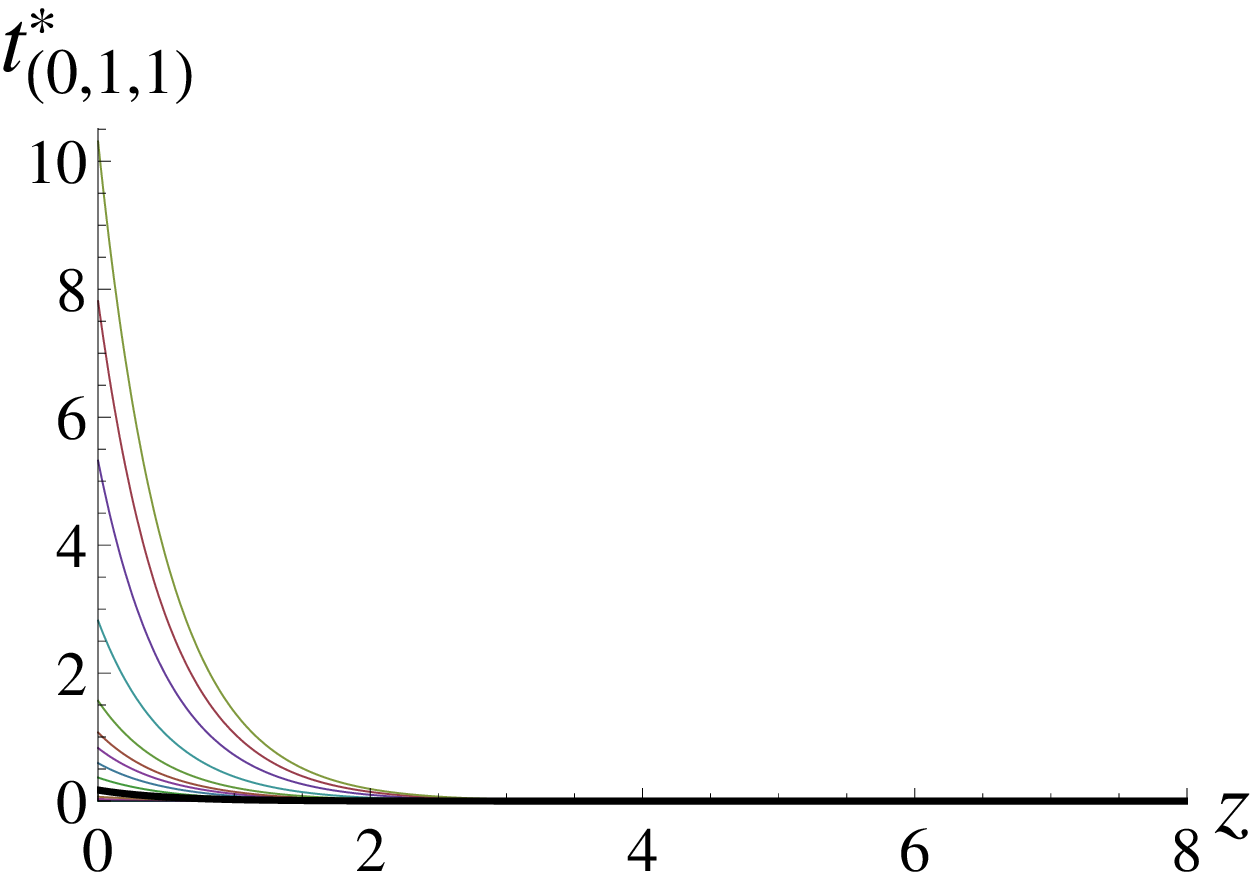}}
\subfigure[]{\includegraphics[width=3in]{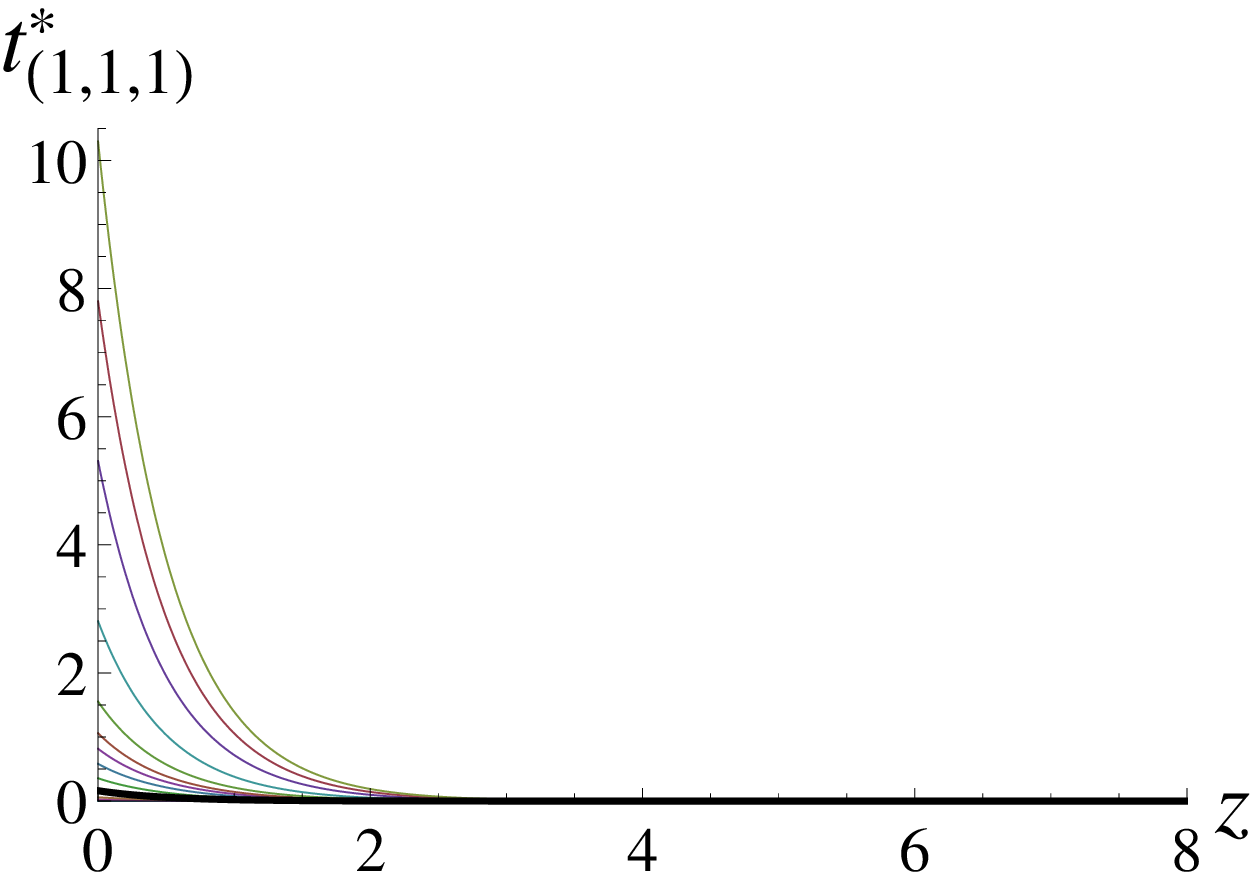}}
\caption{
The scale dependent correlation functions, $t^*_{(l,m,n)}(z)$ 
for $ (l,m,n) \equiv {\bf r}_j - {\bf r}_i = (0,0,0), (0,0,1), (0,1,1), (1,1,1)$
for $L=3$ and $z^*=8$. 
The same values of $\td m^2$ are chosen as in Fig. \ref{3x3x3 t fields GBC zstar 8},
and the magnitudes of the correlation functions increase monotonically with decreasing $\tilde{m}^2$.
}
\label{3x3x3 tstar fields GBC zstar 8}
\end{figure}
%%%%%%%%%

In Figs. \ref{3x3x3 t fields GBC zstar 8} and \ref{3x3x3 tstar fields GBC zstar 8},
we plot $t_{ij}(z)$ and $t^*_{ij}(z)$ for $L=3$ with $z^*=8$.    
For each $(l,m,n)$, we display multiple curves at  $25 \geq \tilde{m}^2 \geq -15$
that cover from the deep   insulating phase to the deep superfluid phase 
across the critical point, 
which is at $\tilde{m_c}^2 = 5.49454$
in the thermodynamic limit. 
Since the curves look very similar away from critical point and they start to change rapidly only in its vicinity, 
we choose the spacing of $\tilde{m}^2$ to be smaller near the critical point and larger away from it.
Here, we notice something interesting. 
No matter what the values of $\tilde{m}^2$ are, all $t_{ij}(z)$ and $t^*_{ij}(z)$ with $i\neq j$ approach zero in the large $z$ limit. 
In other words, the system flows to the on-site problem in the deep IR not only 
in the insulating phase but also in the deep superfluid state.
The universal insulating (on-site) behavior in the deep IR limit is due to the finite size effect. 
For $\tilde{m}^2 < \tilde{m}^2_c$, 
$t_{ij}$ increases with increasing $z$ for small $z$,
which is consistent with the expected RG flow in the superfluid phase 
where the hopping terms become more important at larger length scales.
However, the RG length scale eventually becomes greater than the system size. 
Then the finite size effect becomes important, 
and the system ``flows" back to the insulating behavior in the deep IR. 
This has to do with the fact that all phases are adiabatically
connected to the   insulating phase in finite systems.

At first, the finite size effect appears to be detrimental to our goal 
of characterizing different phases of matter 
in terms of different IR geometries.
However, one can still study the superfluid phase and the critical point
by focusing on the behavior of $t_{ij}, t^*_{ij}$ in the intermediate range of $z$ 
before the finite size effect becomes dominant. 
Moreover, we can use the finite size effect to our advantage.
The universal insulating behavior in the IR limit allows us to
use the single-site IR boundary condition in Eq. (\ref{eq_tstar_IR_on_site}), 
which is much simpler than
the full IR boundary condition in Eq. (\ref{eq_tstar_IR_general}).
This has a significant implication.
A practical difficulty in applying QRG to strongly coupled field theories
is that one has to impose the IR boundary condition dynamically.
This is done either by extremizing the bulk action in the $z^* \rightarrow \infty$ limit
or imposing Eq. (\ref{eq_general_IR_BC}) at a finite $z^*$.
If one uses the latter scheme, one has to know $S_{IR}$.
Although this does not pose any problem for the present  
vector model which is exactly solvable in the large $N$ limit,
this is in general a difficult task for strongly coupled theories  (such as matrix models). 
However, if one can always impose the insulating boundary condition for finite systems
one only needs to know the solution to the on-site problem, which is much easier.
This allows one to find solutions for finite systems
without knowing the IR behavior of the infinite system a priori. 
Then, one can extract the behaviors in the thermodynamic limit
through finite size scaling.
This is the strategy we will employ in the rest of the paper.

In order to guarantee that systems with finite sizes flow to the insulating phase,
one has to choose  $z^*$ to be sufficiently large.
The smaller $\td m^2$ is, and the larger $L$ is, $z^*$ has to be larger to ensure that
the system flows to the deep insulating state driven by the finite size effect at the IR boundary.
This is because it takes longer RG `time' before the finite size effect takes over
in deeper superfluid phases and larger lattices. 
We observed that for all values of $\tilde{m}^2$ 
we took within the range $25 \geq \tilde{m}^2 \geq -15$ 
and all lattice sizes $3 \leq L \leq 13$,  $z^* = 8$ is sufficient. 
For larger $L$, it is expected that $z^*$ should grow logarithmically in $L$ because $z$ corresponds to a logarithmic length scale.
Although computational cost increases with increasing $z^*$, 
it is still much cheaper to use the 
on-site  boundary condition with a logarithmically larger $z^*$
than using the full boundary condition whose computational cost increases much faster.
%%at least with a power-law in $L$ (in this case $O(L^6)$).
In Appendix D, we show that the solutions obtained from the on-site boundary condition
are indeed identical to the ones obtained with the full boundary condition.
From now on, we will use the on-site  boundary condition which allows us to 
obtain full numerical solutions for larger lattices. 
The plots of the hopping fields for larger lattices can be found in Appendix D.

\section{Analysis}

In this section, we present the main results of the paper.
We first establish the presence of the critical point 
that divides the insulating phase
and the superfluid phase
based on the correlation functions 
measured at the UV boundary.
Then we examine how different phases exhibit
distinct geometric features in the bulk.

\subsection{UV Boundary}

\begin{figure}[h]
	\centering
\includegraphics[width=3.5in]{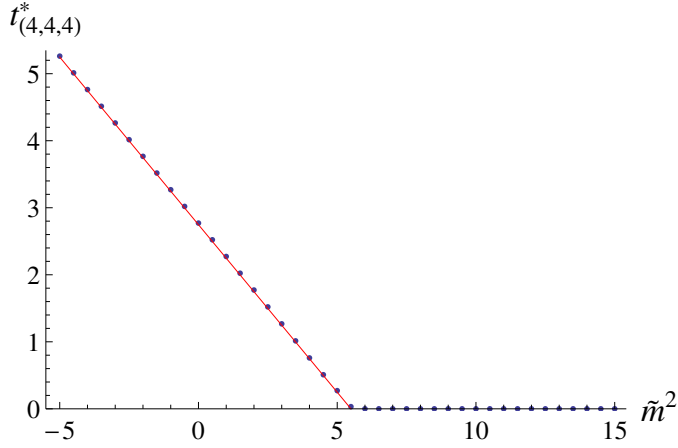}
\caption{
The correlation function at the UV boundary
$t^*_{ij}(z=0)=\frac{1}{N}\langle({\v \phi}^*_i\cdot{\v \phi}_j)\rangle_{UV}$ 
between farthest sites $(l,m,n) \equiv {\bf r}_j - {\bf r}_i =(4,4,4)$ 
plotted as a function of $\td m^2$ for $L=9$.
Although there is no true singularity due to finite size effect,
the correlation function exhibits a rather sharp kink at the critical point,
which is located at $\tilde{m}^2 \approx 5.5$ in the thermodynamic limit.
The straight line is 
$\frac{1}{N}\langle({\v \phi}^*_i\cdot{\v \phi}_j)\rangle_{UV} = -\f{\Delta \tilde{m}^2}{2\lambda}$,
which is the expected value in the thermodynamic limit.
}
\label{UV long distance tstar}
\end{figure}

The correlation functions, $\frac{1}{N}\langle({\v \phi}^*_i\cdot{\v \phi}_j)\rangle$ 
are given by $t^*_{ij}(z=0)$,
which are measured at the UV boundary. 
In the thermodynamic limit, 
$\lim_{r \rightarrow \infty} t^*_{i,i+r}(0)$ is the order parameter for the phase transition
between the insulating (symmetric) phase and the superfluid (symmetry broken) phase.
Although there is no real phase transition in finite lattices, 
there exists a rather sharp crossover which becomes the phase transition in the thermodynamic limit.
To identify the `would-be' phase transition in finite systems,
we plot $t^*_{ij}(0)$  for the largest possible $|i - j|$ as a function of $\tilde{m}^2$ 
for $L=9$ in Fig. \ref{UV long distance tstar}.
Although it is not strictly zero even in the insulating phase due to finite size effect,
there is a sharp crossover around $\tilde{m}^2 \approx 5.5$.
The critical point in the thermodynamic limit is at $\td m^2_c = 5.49454$
(see Appendix E).
From now on, we will use the notation
\beq
\Delta \td m^2 \equiv \td m^2 - \td m_c^2
\eeq
to denote the deviation of the mass away from the thermodynamic critical point.

\begin{figure}[h]
	\centering
\subfigure[]{\includegraphics[width=3.5in]{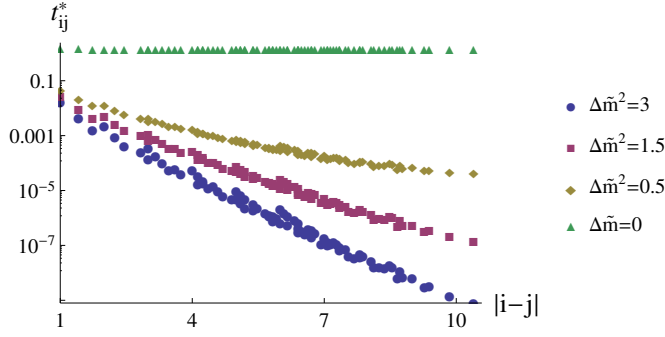}}
\hspace{15mm}
\subfigure[]{\includegraphics[width=3.5in]{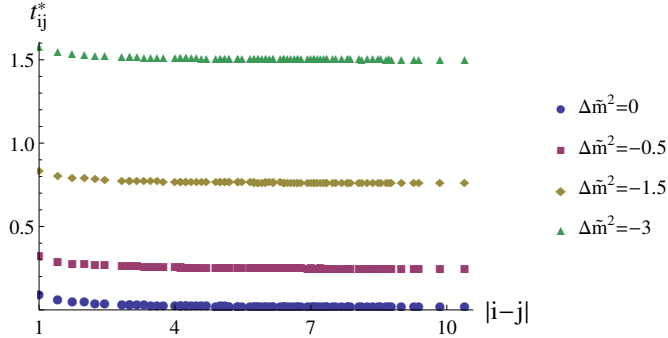}}
%%\subfigure[]{\includegraphics[width=2.5in]{corr_func_sub_dm_SF}}
\caption{
The correlation functions $t_{ij}^*$ at the UV boundary ($z=0$) as a function of $|i-j|$ at different $\Delta \tilde{m}^2$ for $L=13$. 
(a) 
Logarithmic plot of $t_{ij}^*$ at $\Delta \td m^2 \geq 0$.  
The exponential decay becomes slower as the critical point is approached from the insulating phase.
(b) 
$t_{ij}^*$ at $\Delta \td m^2 \leq 0$ in the linear scale. 
In the superfluid phase, 
the correlation functions approach non-zero values in the large $|i-j|$ limit,
exhibiting the off-diagonal long-range order.
}
\label{chi_r}
\end{figure}

In Fig. \ref{chi_r} 
we plot the full correlation functions $t_{ij}^*(z=0)$ as functions of $|i-j|$ for various values of $\Delta \tilde{m}^2$.
In the   insulating phase, the correlation functions decay exponentially as expected.
The correlation length tends to diverge as the critical point is approached, 
as is shown in Fig. \ref{chi_r} (a).
In the superfluid phase, the correlation functions approach
non-zero values in the large distance limit, 
exhibiting a long range order.
As the critical point is approached from the superfluid side, 
the long range order disappears continuously as is shown in Fig. \ref{chi_r} (b).

\begin{figure}[h]
	\centering
\subfigure[]{\includegraphics[width=3in]{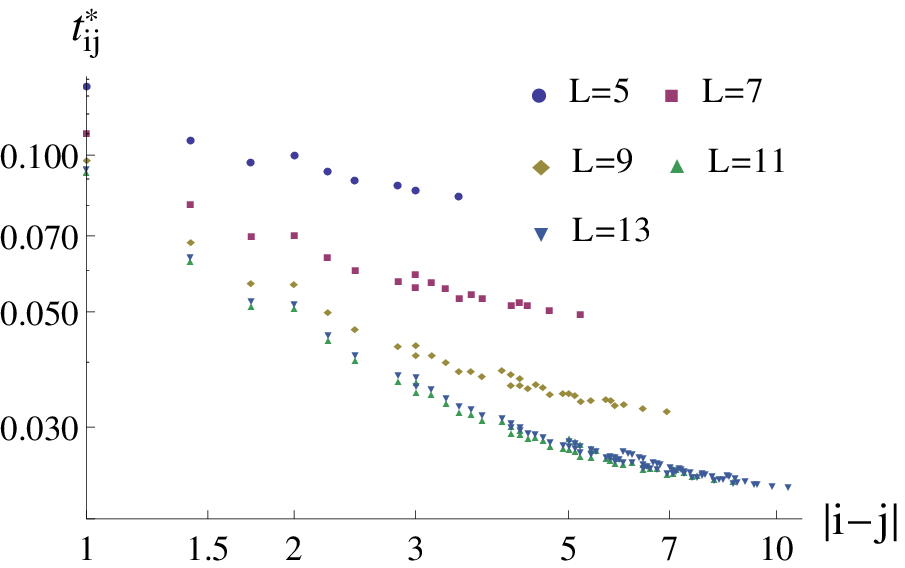}}
\subfigure[]{\includegraphics[width=3in]
{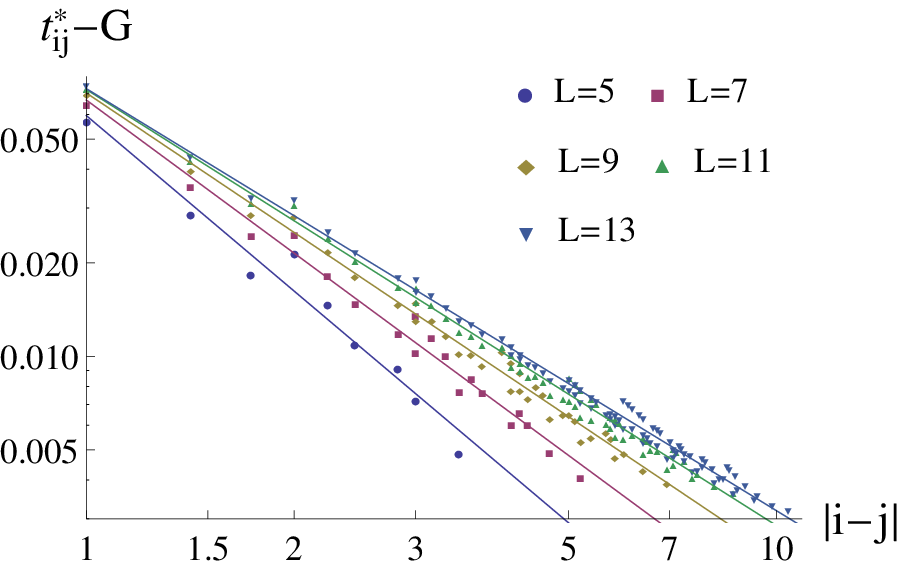}}
%\subfigure[]{\includegraphics[width=3in]{corr_func_minus_const_G_all_L_at_CP}}
\subfigure[]{\includegraphics[width=3in]{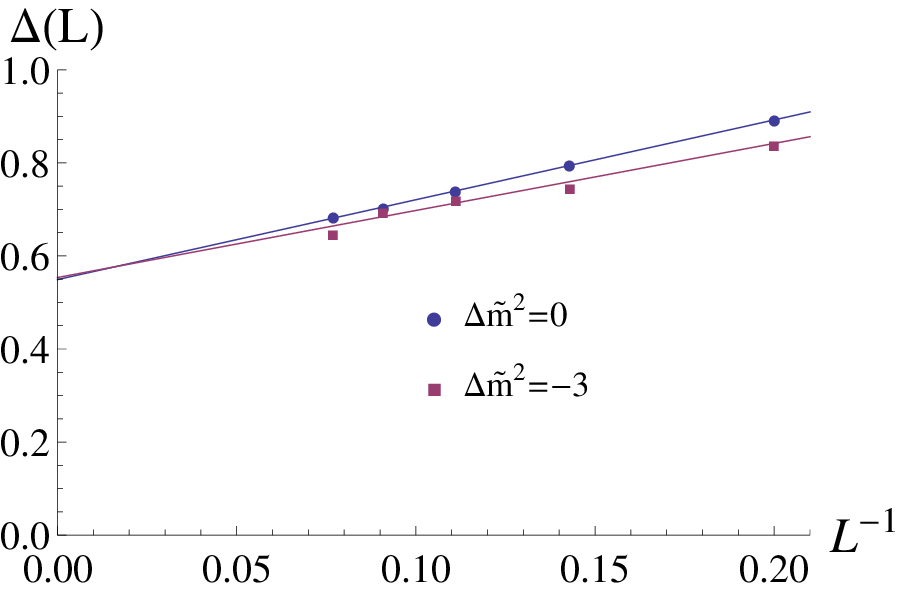}}
\subfigure[]{\includegraphics[width=3in]{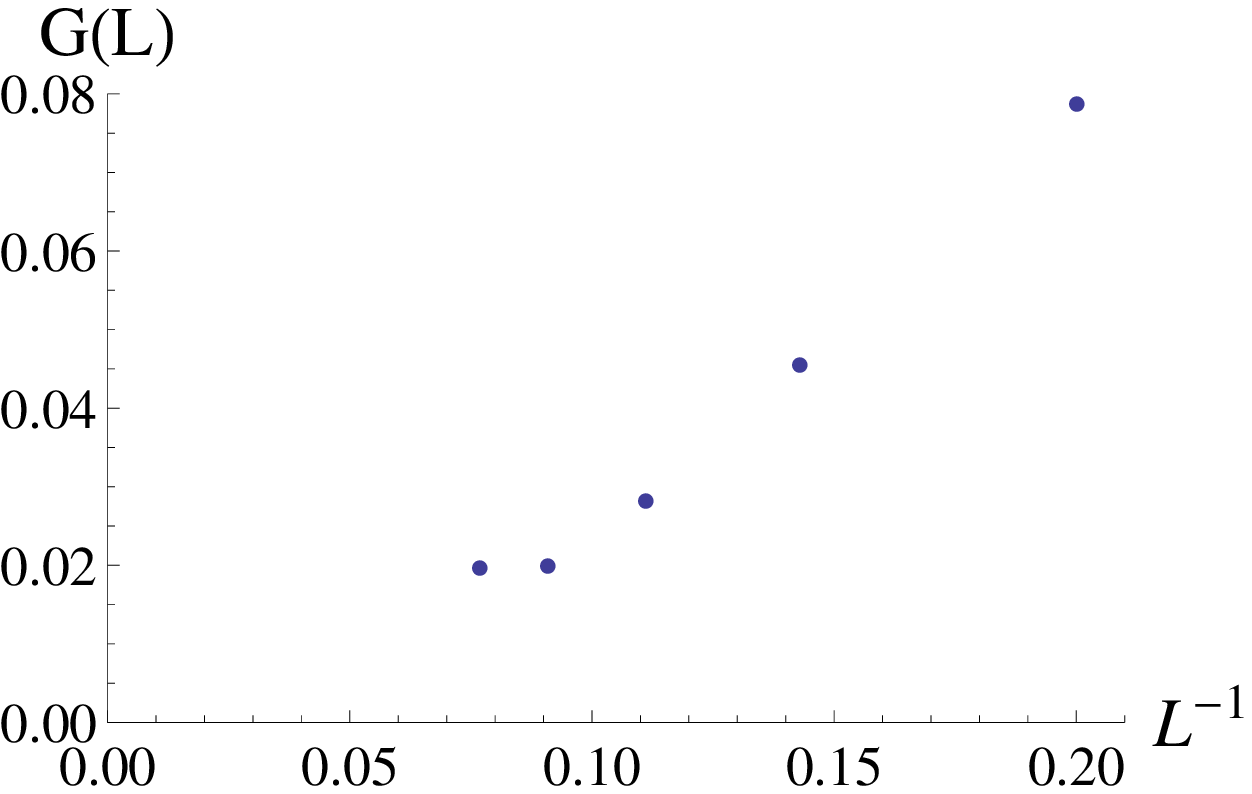}}
\caption{
The UV boundary ($z=0$) correlation functions $t_{ij}^*$ at $\Delta \td m^2 = 0$  for $L=5,7,9,11,13$.
(a) Log-log plot of the correlation functions. 
(b) Log-log plot of the correlation functions after constant pieces $G(L)$ are subtracted out. 
(c) The exponent $\Delta(L)$ 
defined in Eq. (\ref{eq:32}) 
as a function of $L^{-1}$ at $\Delta \td m^2 = 0$ and $\Delta \td m^2 = -3$ 
(the line with $\Delta \td m^2 = -3$ will be discussed at the end of Sec. IV B. 2). 
In the thermodynamic limit the exponent approaches $\Delta(\infty) \approx 0.55$.
(d) The offsets $G(L)$ 
defined in Eq. (\ref{eq:32}) 
as a function of $L^{-1}$. 
}
\label{chi_r_2}
\end{figure}
%%%%%%%%%%%

At the critical point, we expect that the correlation function decays in a power-law,
\beq
\frac{1}{N}\langle{\v \phi}^*(r)\cdot{\v \phi(0)}\rangle \sim \f{1}{r^{2 \Delta}}
\label{r scaling of corr fun}
\eeq      
with $\Delta=1/2$ for $D=3$.
In Fig. \ref{chi_r_2} (a), the correlation function is shown at $\Delta \td m^2 = 0$. 
The curvature in the log-log plot indicates that the correlation function actually decays slower than an algebraic decay. 
This is due to the finite size effect
which overestimates the correlations due to the 
periodic boundary condition. 
The finite size effect effectively pushes the system at $\Delta \td m^2=0$ 
to the superfluid side at finite $L$. 
As will be shown in Sec. IV B. 3, the correlation function decays algebraically 
at a positive value of $\Delta \td m^2$ 
that goes to zero only in the infinite $L$ limit.
To take this into account, we include a constant piece and 
fit the correlation function with 
\beq
t^*_{ij} = G(L) + \f{B(L)}{|i-j|^{2 \Delta(L)}}, 
\label{eq:32}
\eeq
as is shown in Fig. \ref{chi_r_2} (b). 
In the thermodynamic limit, 
the exponent $\Delta(L)$ approaches $0.55$  
and  $G(L)$ tends to vanish as is shown in Figs. \ref{chi_r_2} (c) and (d).
We expect that the $10 \%$ error in the critical exponent 
can be made smaller by increasing the number of basis points 
that are included to parameterize 
the radial profile of $t_{ij}(z)$ and $t_{ij}^*(z)$ 
in the numerical calculation.
For details, see Appendix C.

%%%%%%%%%%%
\subsection{Bulk}

Now we set out to understand the different phases 
from the behaviors of the bulk fields in the $(D+1)$-dimensional space.

\subsubsection{Insulating phase}

%%%%%%%%%%%%%
\begin{figure}[h]
\subfigure[]{\includegraphics[width=3in]{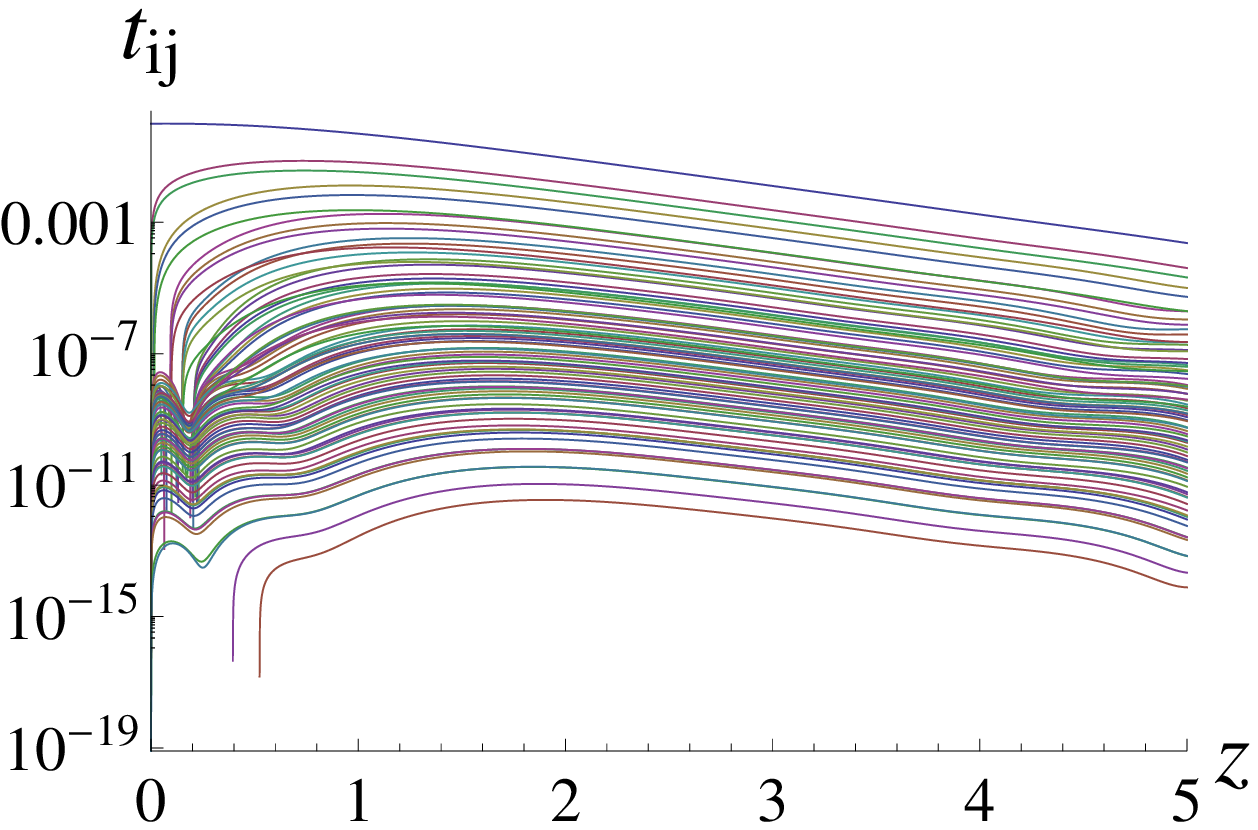}}
\subfigure[]{\includegraphics[width=3in]{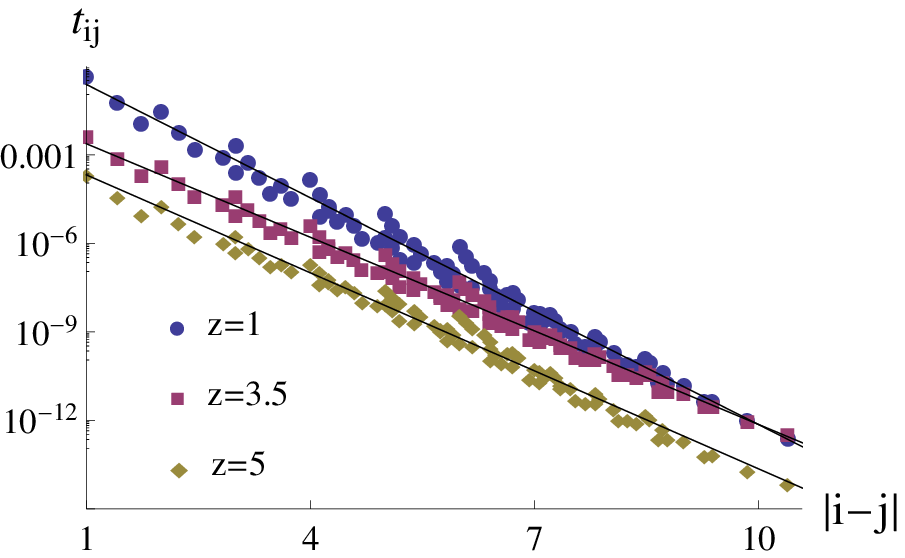}}
\caption{
(a) Logarithmic plot of $t_{ij}$ as a function of $z$  in the insulating phase at $\Delta \tilde{m}^2 = 6$ for $L=13$. 
The curves from the top to the bottom represent the hopping fields with increasing $|i-j|$ from the nearest to the farthest neighbor hoppings.
(b) Logarithmic plot of $t_{ij}(z)$ as a function of $|i-j|$ at different values of $z$. 
The fits with straight lines represent exponential decay. 
}
\label{fig:projected log plot}
\end{figure}

\begin{figure}[h]
	\includegraphics[width=3.5in]{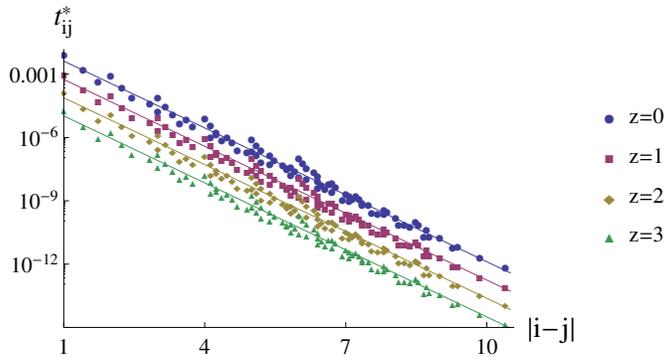}
\caption{
Logarithmic plot of $t^*_{ij}$ as a function of $|i-j|$ in the insulating phase at $\Delta \tilde{m}^2 = 6$ for $L=13$. 
The overall magnitude decreases with $z$.
The fits with straight lines represent an exponential decay
with the same exponent for all $z$.
}
\label{fig:log tstar bulk}
\end{figure}

\begin{figure}[h]
\subfigure[\;$\Delta \tilde{m}^2 = 1.5$]{\includegraphics[width=3in]{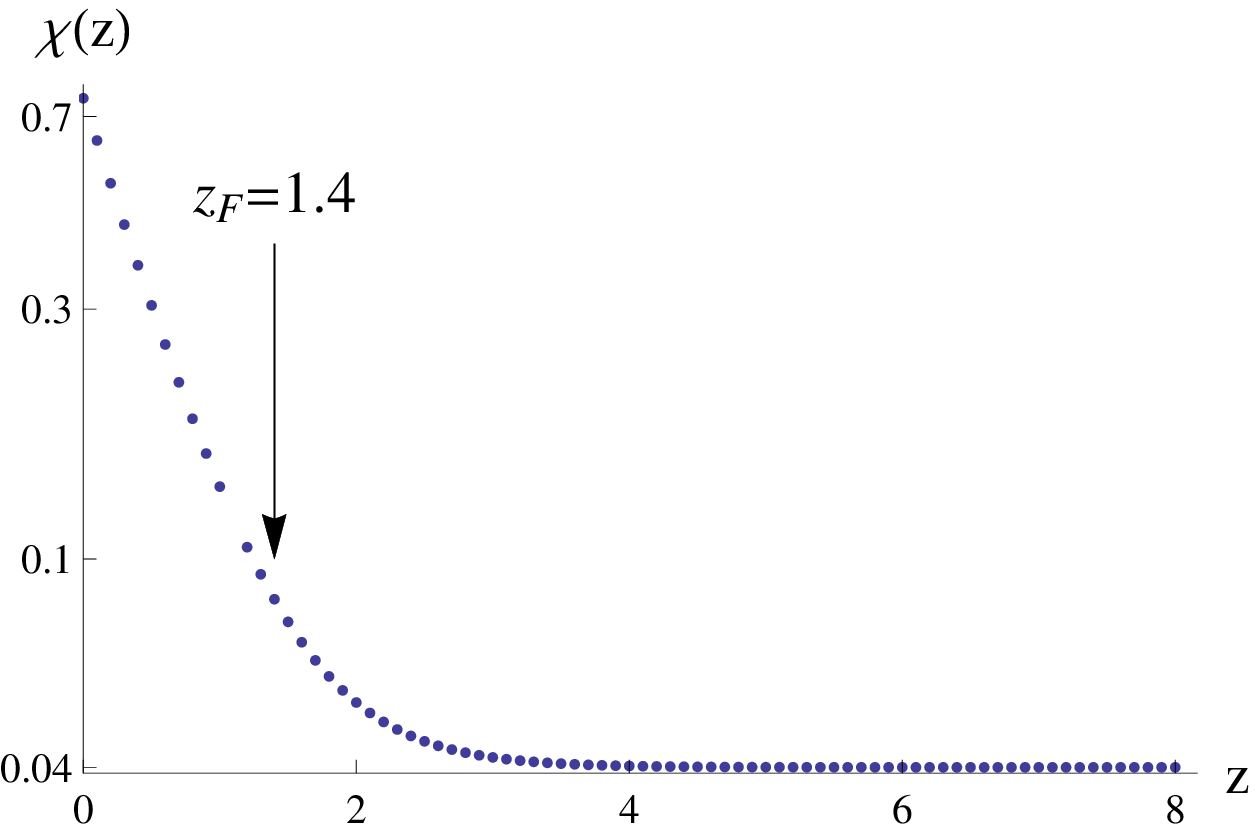}}
\subfigure[\;$\Delta \tilde{m}^2 = 0.5$]{\includegraphics[width=3in]{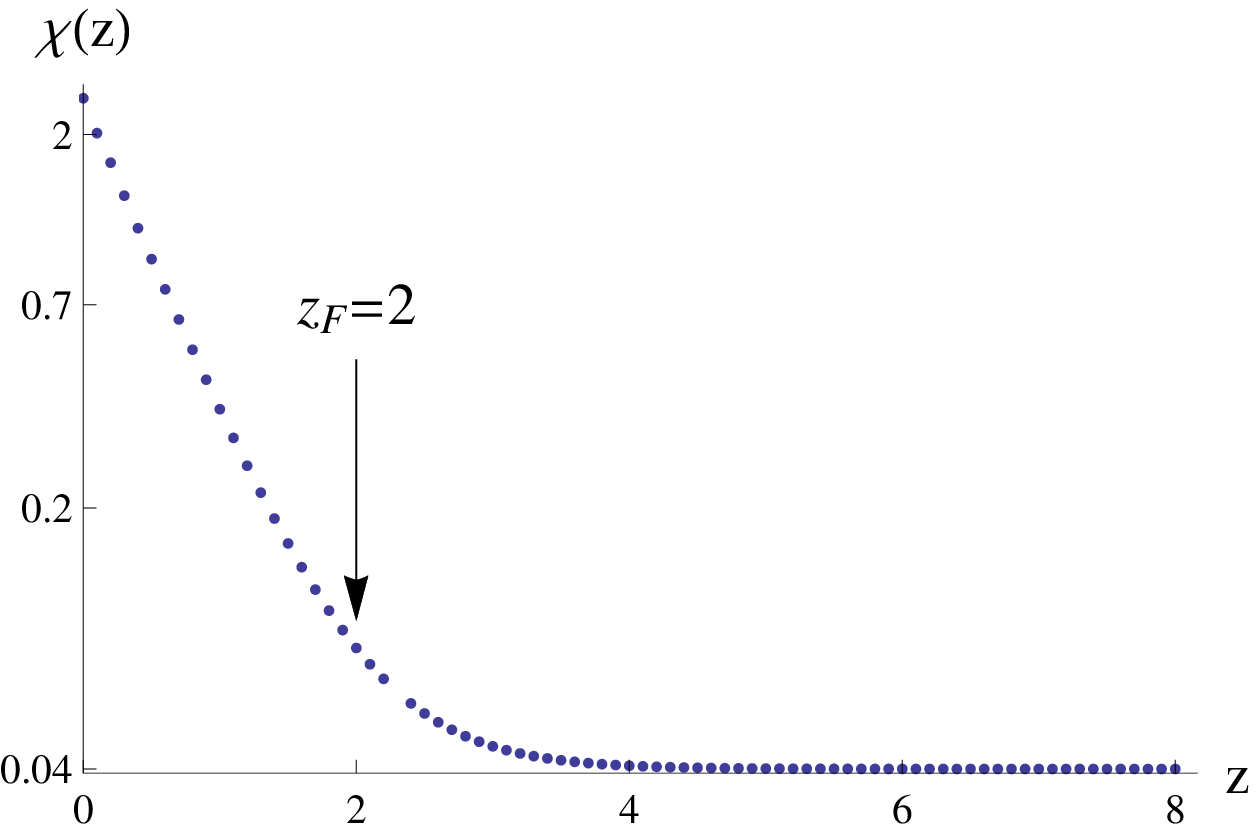}}
\caption{
Logarithmic plot of the susceptibility $\chi$ as a function of $z$ 
in the insulating phase with $\Delta \tilde{m}^2 = 1.5$(a) and $0.5$(b) for $L=13$.
Fragmentation scale is the crossover scale beyond which $\chi(z)$ becomes independent of $z$ and
becomes a constant whose value is determined by the on-site action $S_0$ without hopping.
For concreteness, we define $z_F$ to be the scale at which the second derivative of the logarithm of $\chi(z)$ is maximum. }
\label{fig: K MI phase}
\end{figure}

%%%%%%%%%%%%%
\begin{figure}[h]
\includegraphics[width=4in]{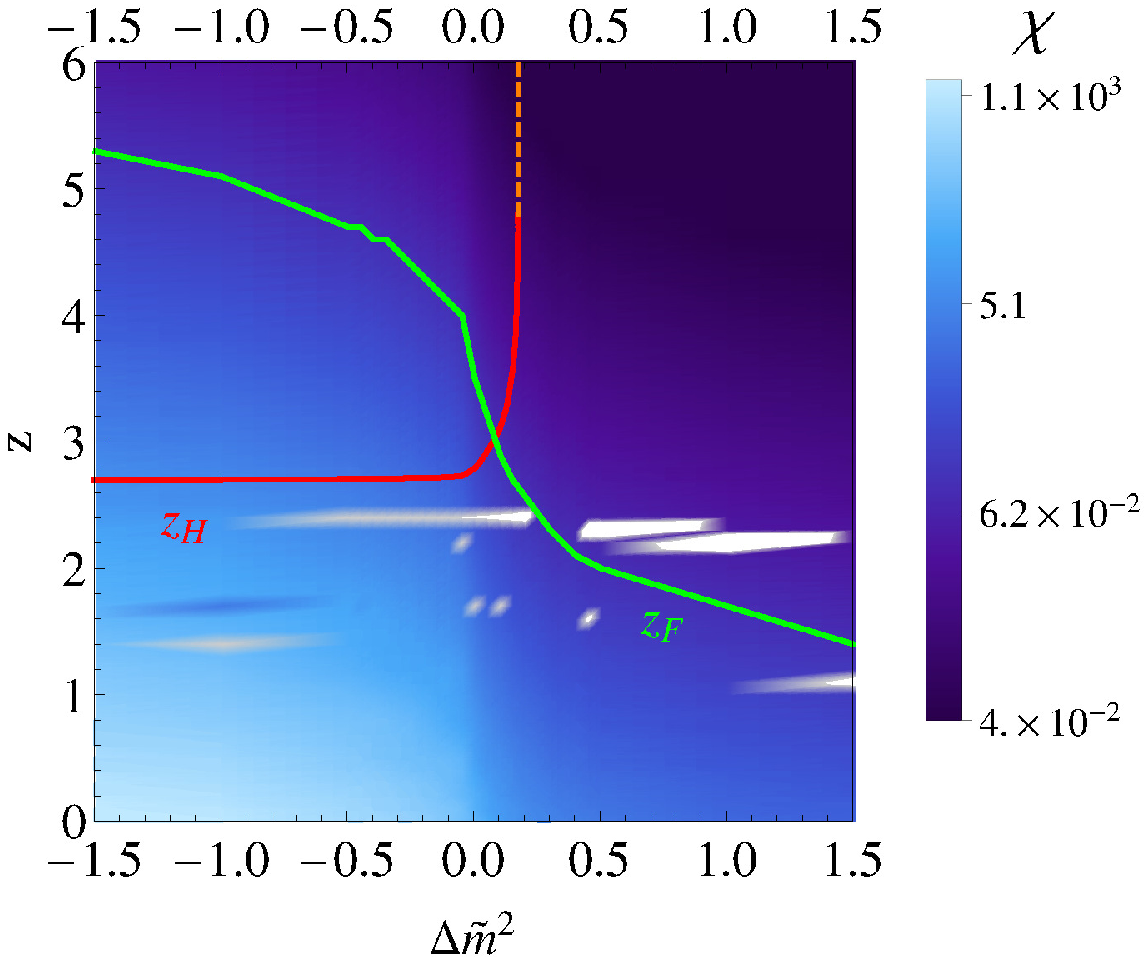}
\caption{
Holographic phase diagrams for $L=13$. 
The color contours represent
the value of the susceptibility
defined in Eq. (\ref{tijstarz}).
The darker the color is, the smaller the susceptibility is.
$z_F$ is the fragmentation scale, 
and $z_H$ represents the horizon (see text in Secs. IV B 1 and 2).
}
\label{fig:holo_pd_L_7_9_11}
\end{figure}

In Fig. \ref{fig:projected log plot}, the scale dependent hopping fields are plotted 
in a deep insulating state at 
%$\Delta \tilde{m}^2 \approx 6$ 
$\Delta \tilde{m}^2 = 6$ 
for $L=13$. 
Although all further neighbor hoppings are generated in the bulk,
they remain exponentially small in $|i-j|$ at all $z$.
Furthermore, all hopping fields decay exponentially 
in $z$ in the IR region.
Due to the exponential decay both in $|i-j|$ and $z$,
$t_{ij}(z)$ is significant only for small $|i-j|$ and $z$.
As the connectivity of space becomes weaker with increasing $z$,
the correlation functions $t_{ij}^*(z)$ become
smaller as is shown in Fig. \ref{fig:log tstar bulk}.
As $z$ increases, the overall magnitude of $t_{ij}^*(z)$ decreases
exponentially while the rate of decay in $|i-j|$ is unchanged. 
In the large $z$ limit, the space completely loses its connectivity,
and gets fragmented into isolated islands.
In the RG language, the IR fixed point is described by the decoupled sites. 
One can quantify the scale of fragmentation in terms of the susceptibility 
\beq
\chi(z) = 
\sum_{j} t_{ij}^*(z),
\label{tijstarz}
\eeq
which is proportional to $\xi^2$ in the insulating phase, where $\xi$ is the correlation length.
As $z$ increases, the susceptibility decreases because the hopping amplitudes become smaller in the bulk.
In the large $z$ limit, $\chi$ saturates to the value determined by the on-site mass term.
This is shown in Fig. \ref{fig: K MI phase}.
Although fragmentation is not a sharp transition,
we can choose a convenient criterion for definiteness: 
we define the fragmentation scale $z_F$ to be the scale at which $\f{d^2 ln\chi(z)}{dz^2}$ is maximal.    
It  represents a scale around which
a crossover occurs from a connected lattice to fragmented space.
In Fig. \ref{fig: K MI phase}, $z_F$ is indicated in the insulating phase,
which shows that $z_F$ increases with decreasing $\td m^2$, as expected.
Fragmentation occurs at longer distance scales
when the system is closer to the critical point.
The background color 
in Fig. \ref{fig:holo_pd_L_7_9_11} 
represents the value
of the susceptibility in the bulk for different values of $\Delta \td m^2$.
The curve denoted by $z_F$ represents the fragmentation scale.
The curve $z_H$ and other features in the superfluid phase ($\Delta \td m^2 < 0$) of Fig. \ref{fig:holo_pd_L_7_9_11} 
will be discussed in the following section.

%%%%%%%%%%%%%
\begin{figure}[h]
\includegraphics[width=3in]{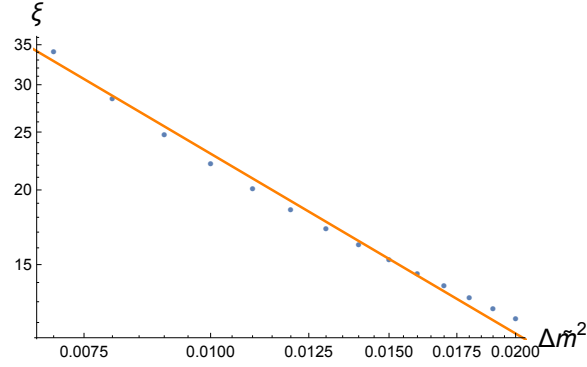}
\caption{
The correlation length $\xi$ plotted as a function of $\Delta \tilde{m}^2$ on a log-log scale. 
The data is obtained from solving the UV boundary theory with lattice size $L=151$. 
In the region where $1 \ll \xi \ll L/2$, the scaling is close to the one expected in the thermodynamic limit, 
$\xi \sim (\Delta \tilde{m}^2)^{-1}$, which is shown as a straight line.  
}
\label{fig:xi vs delta UV boundary}
\end{figure}
%%%%%%%%%%%%%

Near the critical point, the fragmentation scale $z_F$ can be used to extract the critical exponent $\nu$, which dictates how the correlation length depends on the distance to the critical point, $\xi \sim (\Delta \tilde{m}^2)^{-\nu}$. 
The susceptibility is given by $\chi \sim \xi^2$ at the UV boundary.
Inside the bulk, it falls off exponentially in $z$ as $\chi(z) \sim \xi^2 e^{-2z}$ 
because of Eq. (\ref{2zdecay}) before saturating to a fixed value of $\f{1}{m^2}$.
Therefore, the fragmentation scale is given by $z_F \sim \ln \xi \sim -\nu \ln (\Delta \tilde{m}^2)$.
In Fig. \ref{fig:xi vs delta UV boundary} we show $\xi$ as a function of  $(\Delta \tilde{m}^2)$, which shows $\nu \approx 1$, the expected value for the interacting model we are studying in the large $N$ limit.
Therefore the fragmentation scale should behave as  $z_F \sim -\ln (\Delta \tilde{m}^2)$ in the thermodynamic limit.

\subsubsection{Superfluid phase}

%%%%%%%%%%%%%
\begin{figure}[h]
\subfigure{\includegraphics[width=3in]{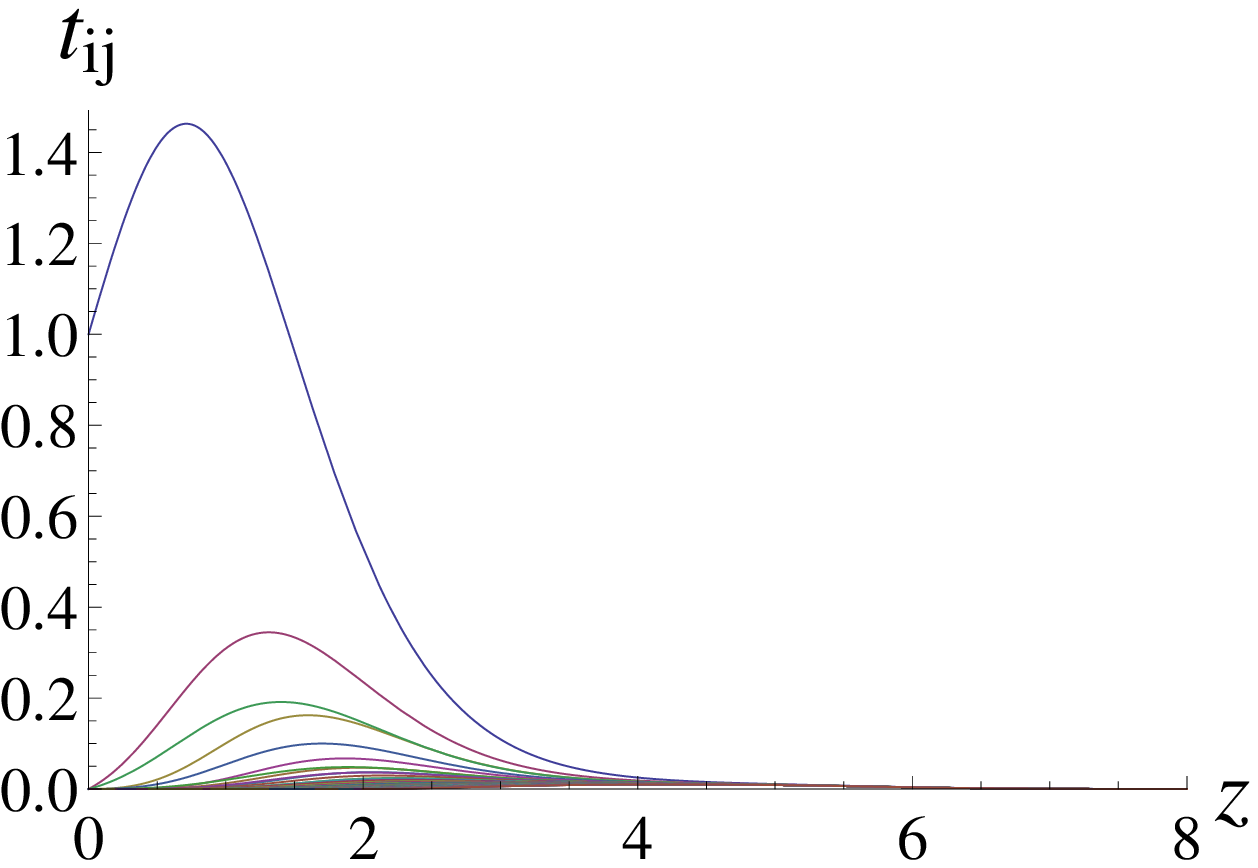}}
\subfigure{\includegraphics[width=3in]{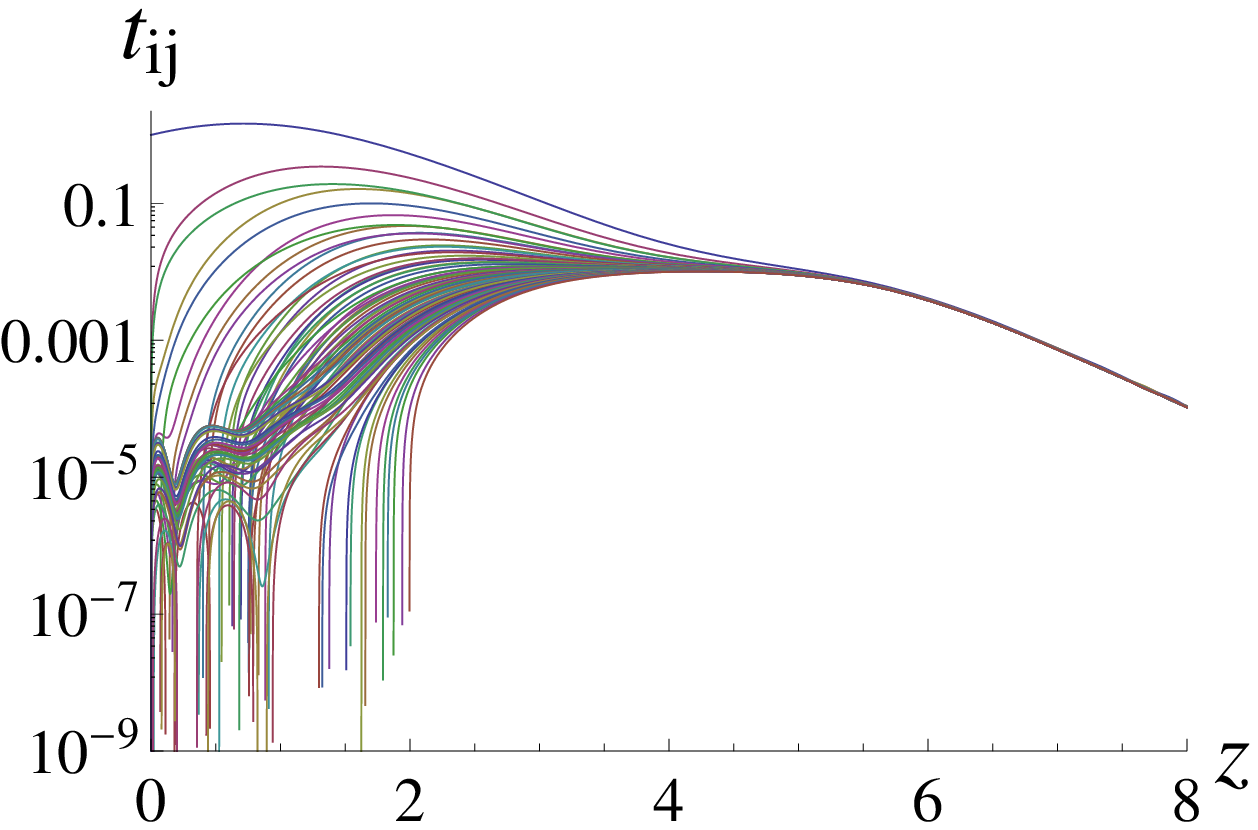}}
\caption{
Hopping fields $t_{ij}$ in the superfluid phase at $\Delta \tilde{m}^2 = -3$
for $L=13$ plotted in the linear scale (a) and in the logarithmic scale (b). 
The curves from the top to the bottom represent the hopping fields with increasing $|i-j|$ from the nearest to the farthest neighbor hoppings.
}
\label{all fields SF separate}
\end{figure}

\begin{figure}[h]
\subfigure[]{\includegraphics[width=3in]{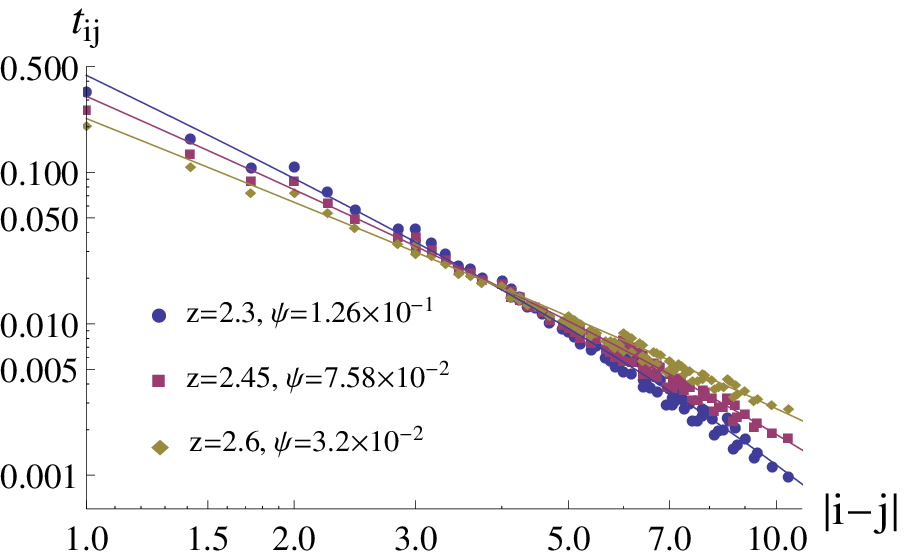}
\label{log t bulkc}
}
\subfigure[]{
\includegraphics[width=3in]{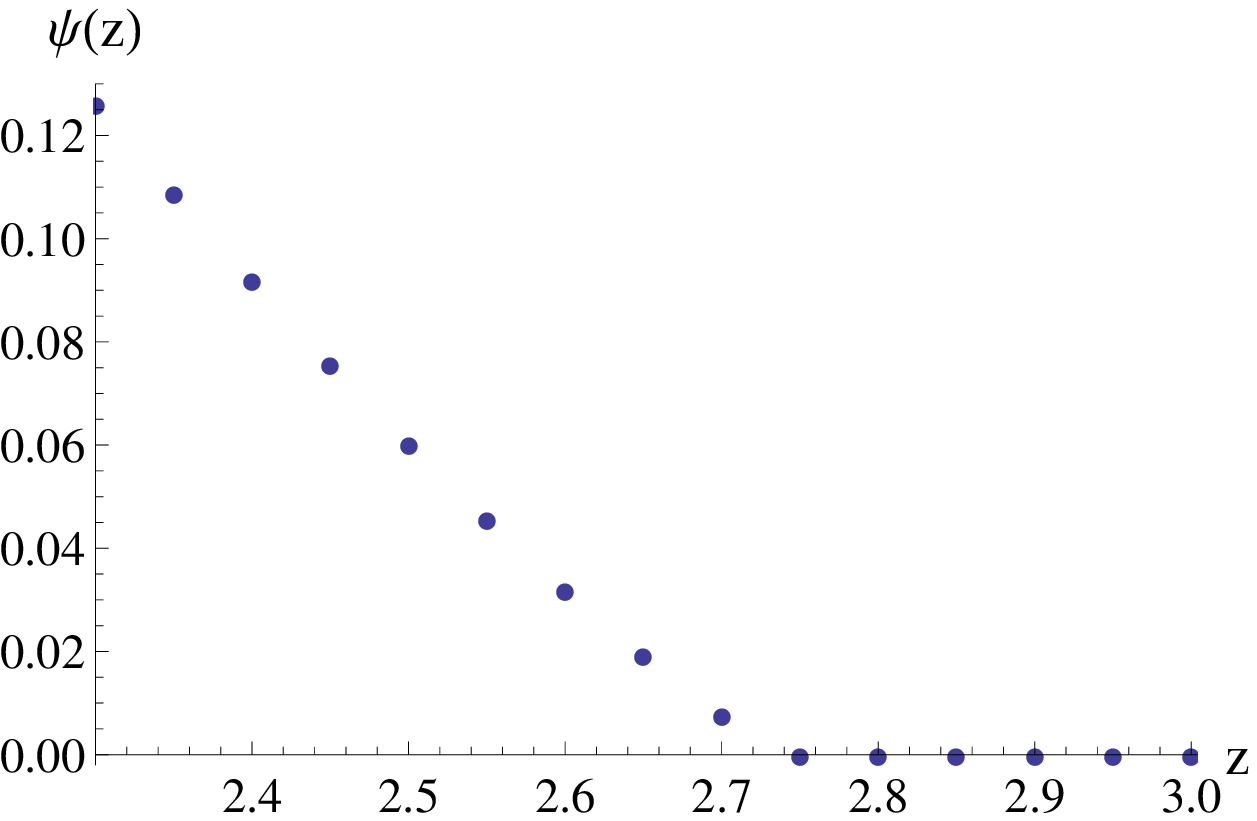}
\label{fig:psi_below_zH}
}
\caption{
(a) The log-log plot of the hopping fields $t_{ij}$ as functions of $|i-j|$ in the superfluid phase at $\Delta \td m^2 =  -3$  for $L=13$.
From the UV boundary to a critical scale $z_H \approx 2.8$, 
$t_{ij}$ decays exponentially. 
The solid curves represent fits of the form in Eq. (\ref {eq:expalg}) with $\psi(z) > 0$.
(b) The rate of the exponential decay $\psi(z)$ from (a) plotted as a function of $z$. 
$\psi(z)$ continuously vanishes at a critical scale $z_H \approx 2.8$.
}
\end{figure}

In Fig. \ref{all fields SF separate}, we display the hopping fields 
inside the superfluid phase at 
$\Delta \tilde{m}^2 = -3$
%%Let us use the convention 2.99=3
%%$\Delta \tilde{m}^2 = -2.99454$
for $L=13$.
Near the UV boundary, $t_{ij}(z)$ decays exponentially in $|i-j|$ as is the case in the insulating phase.
This has to be true even in the superfluid phase because 
only nearest neighbor hoppings are turned on at the UV boundary, 
and further neighbor hoppings are generated gradually in the bulk.
The exponential decay of the hopping fields 
is well captured by 
\beq
t_{ij}(z) \sim \frac{ e^{-\psi(z) |i-j|} }{ |i-j|^{\kappa(z)}}
\label{eq:expalg}
\eeq
with $\psi(z) > 0$ near the UV  boundary
as is shown in Fig. \ref{log t bulkc}.
However, the rate of the exponential decay, $\psi(z)$ becomes smaller as $z$ increases,
and vanishes at a critical scale $z_H$. 
This is shown in Fig. \ref{fig:psi_below_zH}.

\begin{figure}[h]
\subfigure[]{
\includegraphics[width=3in]{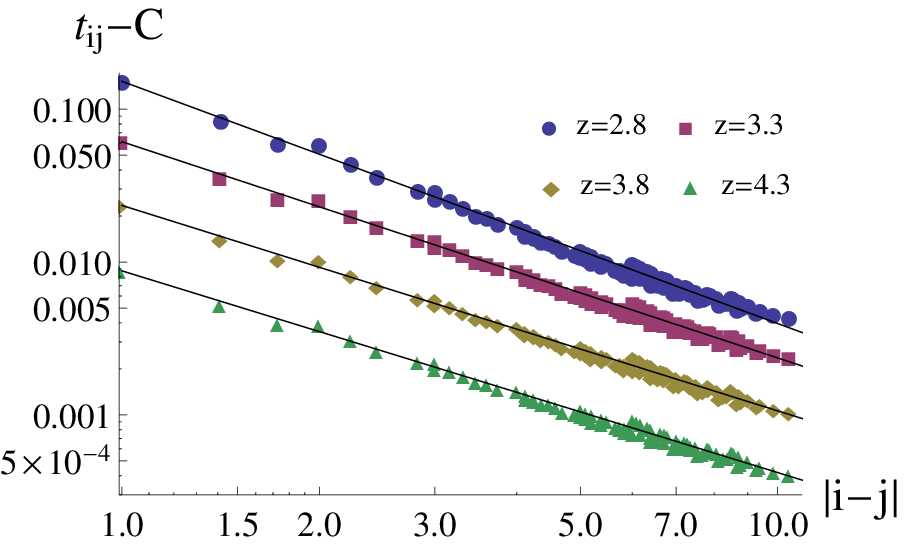}
\label{log t bulkd}
}
\subfigure[]{\includegraphics[width=3in]{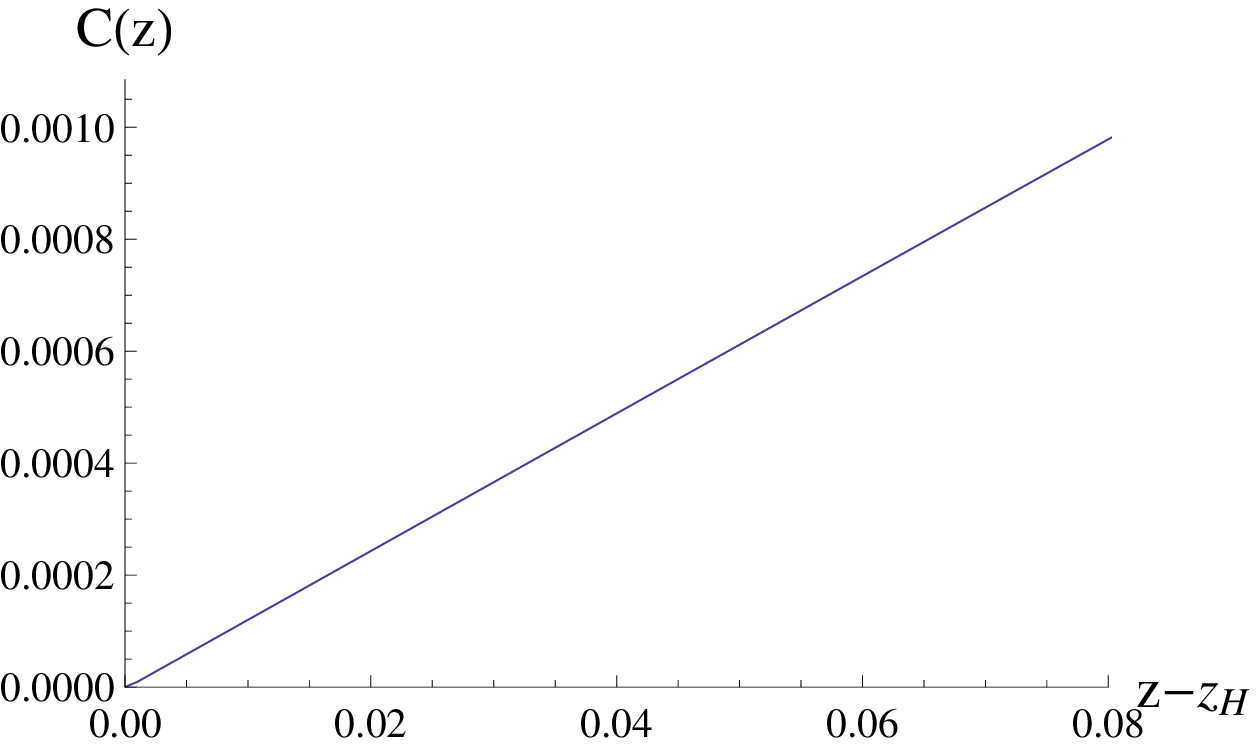}
%\subfigure[]{\includegraphics[width=3in]{C_of_z_onset_L791113_dm_minus_3}
\label{fig:c_exponent}
}
\caption{
(a) The hopping fields $t_{ij}$ at $\Delta \td m^2 = -3$ after a constant off-set is subtracted out in the bulk with $z \geq z_H = 2.8$ for $L=13$.
The straight lines represent pure power-law decays.
(b) The off-set, $C(z)$ in Eq. (\ref{tijCB}) plotted as a function of $z-z_H$ for $L=13$.
}
\label{}
\end{figure}

For $z \geq z_H$,  the hopping fields no longer decay exponentially 
but decay algebraically with a constant off-set, 
\beq
t_{ij}(z) = C(z) + \frac{B(z)}{|i-j|^{\kappa(z)}},
\label{tijCB}
\eeq
where the constant off-set turns on 
continuously as
\beq
C(z) \sim | z - z_H |^{\beta_H}
\label{eq:C}
\eeq
with an exponent $\beta_H \approx  1$ for $z \geq z_H$.
After the off-set is subtracted out the hopping fields exhibit pure power-law decays
as is shown in Fig. \ref{log t bulkd}.
The $z$ dependence of the off-set near $z_H$  is displayed in Fig. \ref{fig:c_exponent}.

\begin{figure}[h]
\subfigure[]{
\includegraphics[width=3in]{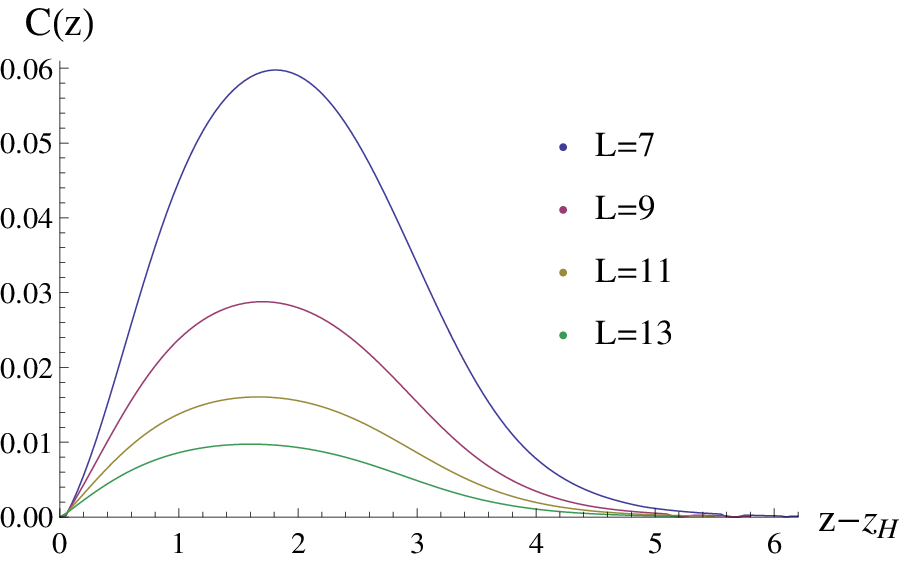}
\label{fig:C_profile_all_L}
}
\subfigure[]{
\includegraphics[width=3in]{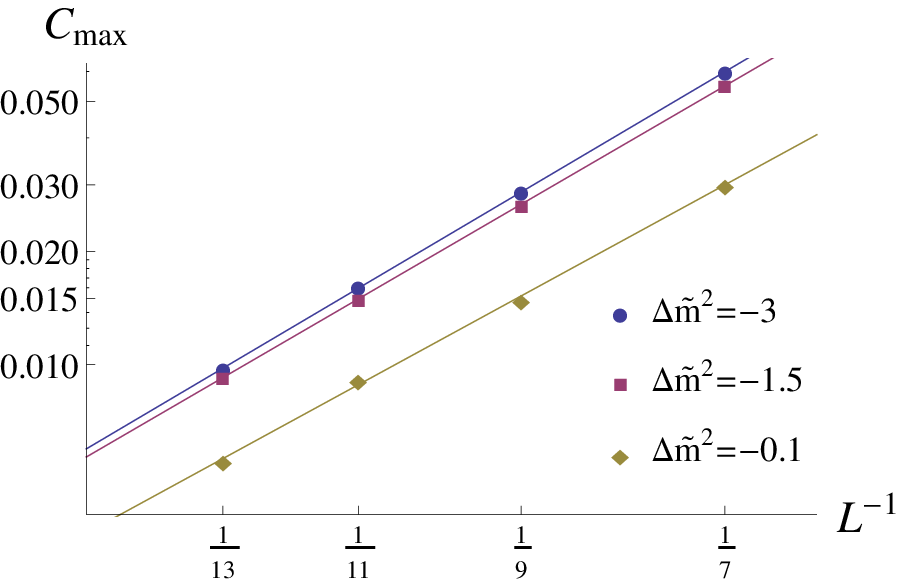}
\label{fig:d_max}
}
\caption{
(a) The off-set $C(z)$ at $\Delta \td m^2 = -3$ for $L=7,9,11,13$. 
(b) The log-log plot of the maximum value of $C(z)$ as a function of $L^{-1}$ at $\Delta \tilde{m}^2 = -0.1, -1.5, -3$. 
The lines represent fits, $C_{max} \sim \frac{1}{L^{r(\Delta \tilde{m}^2)}}$ with 
$r(-0.1) \approx 2.7$, $r(-1.5) \approx 2.88$ and $r(-3) \approx 2.93$,
which suggests that $C(z)$ vanishes at all $\Delta \td m^2$ 
in the thermodynamic limit.
}
\end{figure}

The scale $z_H$ represents a `horizon' 
at and beyond which the system loses locality
as sites are connected with each other 
through non-local hoppings.
As a result,
the `coordinate distance' $|i-j|$
is no longer a good measure of the `physical distance' 
that dictates the actual correlations. 
Because of the non-local hopping, the physical distance between two sites effectively shrinks to zero. 
If we were studying a system with a finite temperature with a compact Euclidean time direction 
in such a way that its length $\beta$ would remain finite as $L \rightarrow \infty$, we would first observe this loss of locality only in the Euclidean time direction. 
In that case, $z_H$ corresponds to the usual horizon at which only the thermal circle shrinks to zero. 
In the present case, the $L \rightarrow \infty$ limit corresponds to the zero temperature limit, 
where all directions are equivalent.
As a result, the size of space shrinks to zero in all directions at the horizon.
It will be interesting to understand what this generalized horizon corresponds to in Minkowski space.

If the off-set $C$ was indeed nonzero in the thermodynamic limit,
it would imply that any two sites remain
coupled with a non-vanishing hopping 
no matter how far they are.
However, Fig. \ref{fig:C_profile_all_L} shows that $C$
systematically decreases as the system size increases. 
In order to understand the behavior of $C$ in the thermodynamic limit,
we need to do a finite size scaling analysis. 
Fig.  \ref{fig:d_max} shows that
the maximum value of $C$ 
at fixed $\Delta \tilde{m}^2$ 
decays as  $C_{max} \sim \frac{1}{L^{r}}$ with 
$r = 2.81 \pm 0.12$ for $-3 \leq \Delta \td m^2 \leq -0.1$
as the system size $L$ increases. 
This implies that $C$ goes to zero,  
and  the hopping fields decay in a purely algebraic manner for $z \geq z_H$
in the thermodynamic limit.

\begin{figure}[h]
\subfigure[]{
\includegraphics[width=3in]{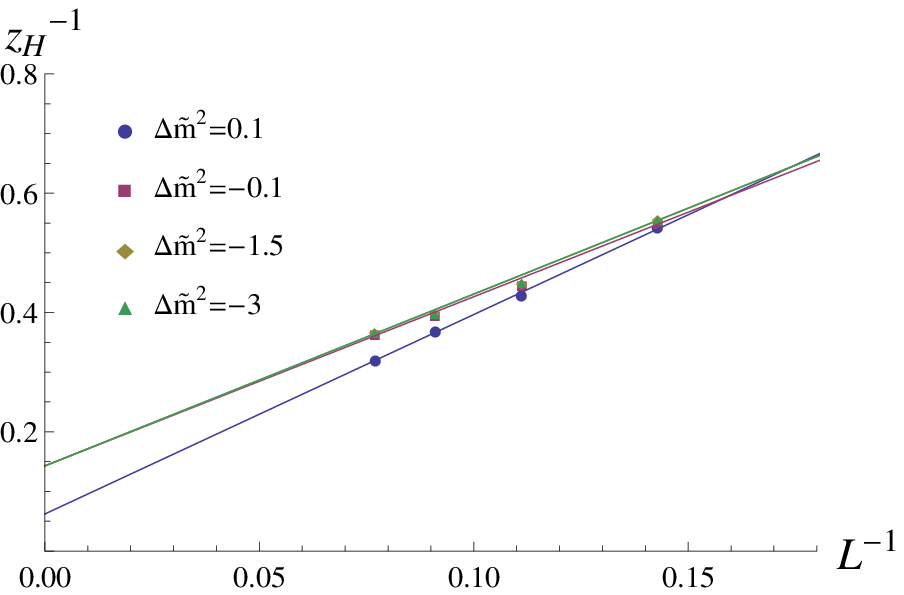}
\label{fig:zH of L in SF}
}
\subfigure[]{
\includegraphics[width=3in]{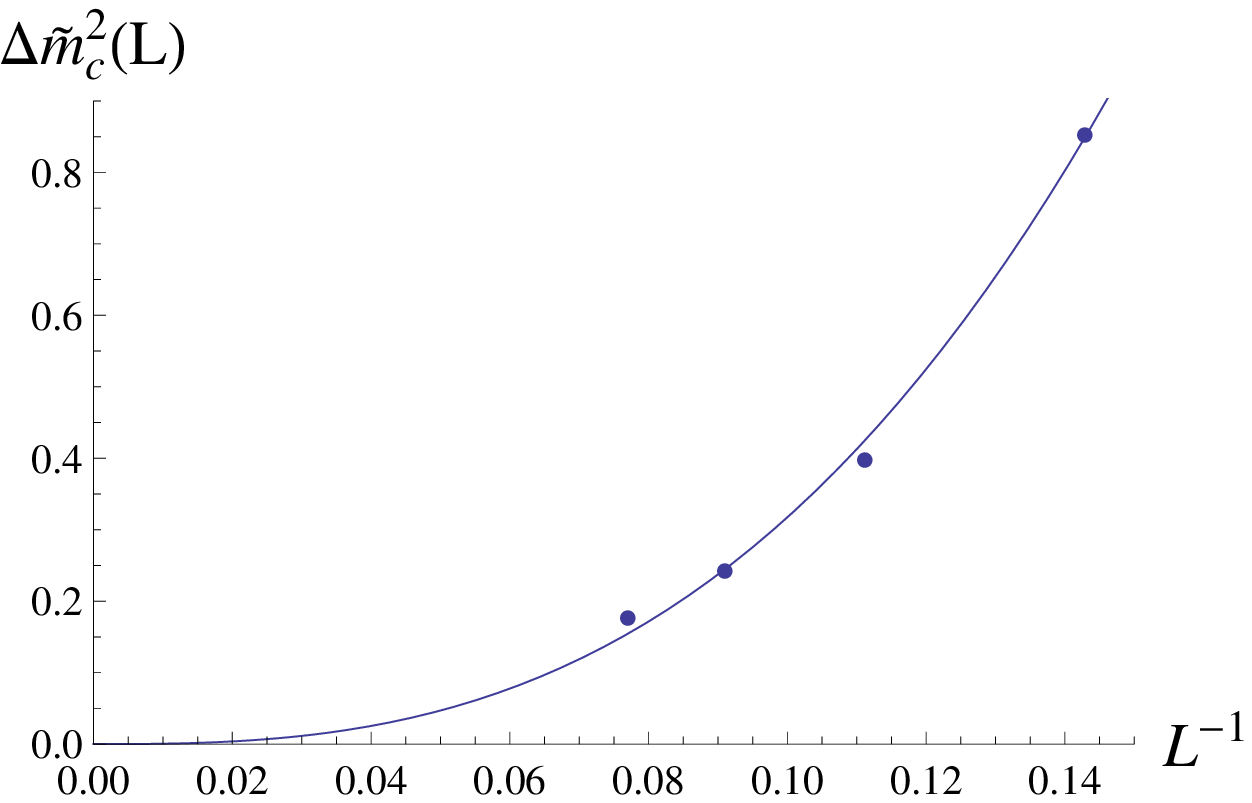}
\label{fig: delta_sub_m as function of L}
}
\caption{
(a) The inverse of the horizon scale $z_H^{-1}$ as a function of $L^{-1}$ with  $\Delta \td m^2 =  -3, -1.5, -0.1, 0.1$.
The nonzero intercept in the large $L$ limit implies that the horizon arises at a finite scale in the thermodynamic limit.
(b) The critical mass determined from the location of horizon in the bulk, 
$\Delta \td m_c^2(L)$, plotted as a function of $L^{-1}$. 
The line is a fit to an algebraic function of the form $f(L) = A L^{-D}$ with $A \approx 181$ and $D \approx 2.76$.
}
\end{figure}
Although the off-set $C$ vanishes in the thermodynamic limit,
the location of the horizon $z_H$ remains finite,
as is shown in Fig. \ref{fig:zH of L in SF}.
In the deep superfluid phase, $z_H$ is more or less independent of $\Delta \td m^2$,
but $z_H$ sharply increases as the critical point is approached from the superfluid side.
This is displayed in Fig. \ref{fig:holo_pd_L_7_9_11}.
For $L=13$, 
the horizon scale diverges at $\Delta \td m^2 = 0.2$,
and the horizon no longer exists for $\Delta \td m^2 > 0.2$. 
From the location of the horizon in the bulk, 
we define $\Delta \td m_c^2(L)$ to be the critical mass for finite lattices, 
e.g., $\Delta \td m_c^2(13) = 0.2$.
Although the critical point is at $\Delta \td m^2 = 0$  in the thermodynamic limit,
$\Delta \td m_c^2(L)$ is different from zero for finite $L$
due to the finite size effect.
A finite size scaling in Fig. \ref{fig: delta_sub_m as function of L} 
suggests that $\Delta \td m_c^2(L)$ indeed goes to zero in the thermodynamic limit.

\begin{figure}[h]
\subfigure[]{\includegraphics[width=3in]{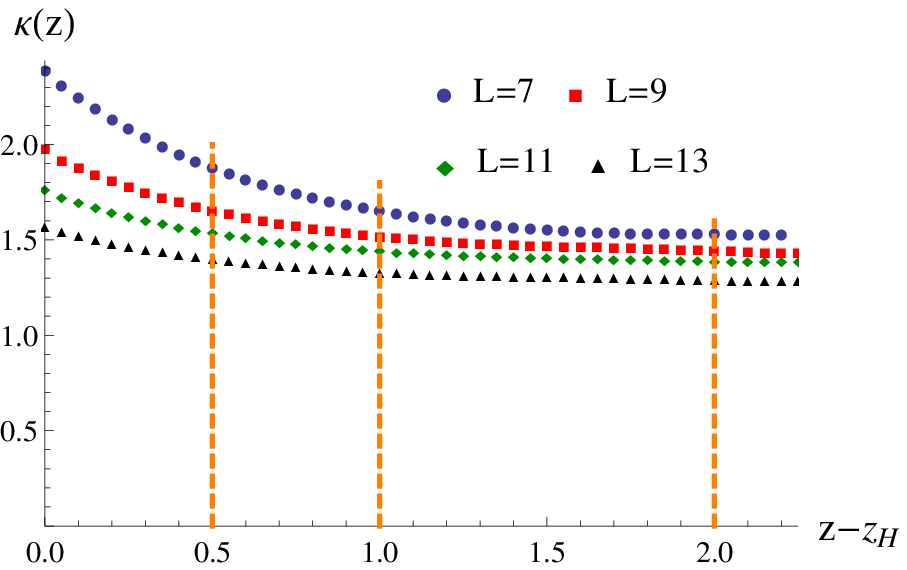}
\label{fig:kappa_profile}
}
\subfigure[]{\includegraphics[width=3in]{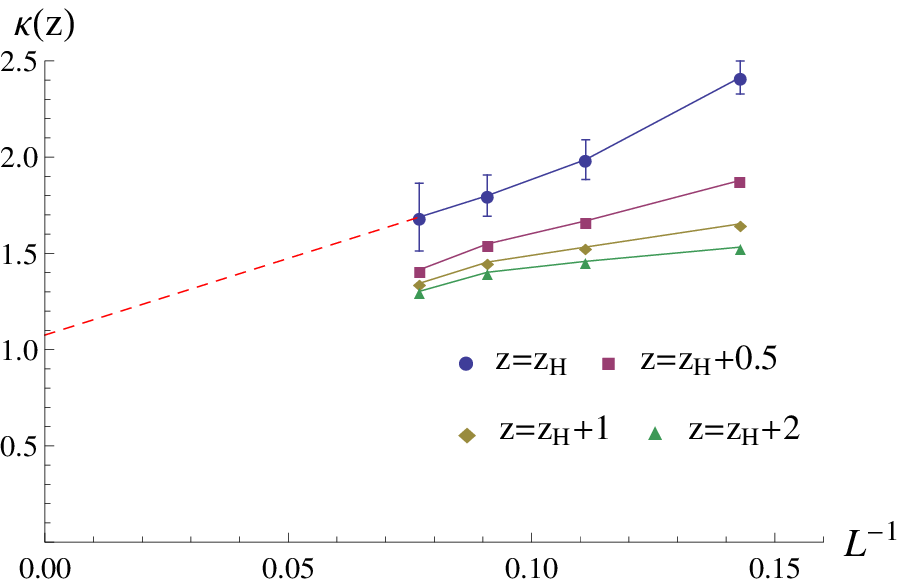}
\label{fig:kappa_L}
}
\caption{
(a) $\kappa(z)$ from Eq. (\ref{tijCB}) as a function of $z$ in the superfluid phase at  $\Delta \tilde{m}^2 = -3$.
It continuously decreases with increasing $z$ inside the horizon.
%The dotted orange line denotes the scale $z=z_H + 2$ at which we look at the $L$-scaling of $\kappa(z)$ inside the bulk.  
(b) $L$-dependence of $\kappa(z)$ at and inside the horizon at $\Delta \td m^2 = -3$.
}
\end{figure}

In the absence of the off-set in the thermodynamic limit,
what distinguishes the horizon at $z=z_H$
from the region inside the horizon with $z > z_H$
is the exponent $\kappa$ with which
the hopping fields decay.
Fig. \ref{fig:kappa_profile} shows that 
$\kappa$ inside the horizon is smaller than $\kappa_H$
defined at the horizon. 
The exponent $\kappa$ tends to approach nonzero values for all z 
in the thermodynamic limit
as is shown in Fig. \ref{fig:kappa_L}.

% % 
% % The location of the horizon is largely independent of $\td m^2$
% % in the deep superfluid side,
% % and rapidly increases as $\td m^2$ approaches 
% % $\td m_c^2(L) \approx \td m_c^2 + \delta_m(L)$,
% % where $\td m_c^2$ is the critical point in the thermodynamic limit. $\delta_m(L)$ is shown to approach zero as $L \rightarrow \infty$ in Fig. \ref{fig: delta_sub_m as function of L}.
% % As $L$ increases, $\td m_c^2(L)$ converges to $\td m_c^2$.

\begin{figure}[h]
\includegraphics[width=3.5in]{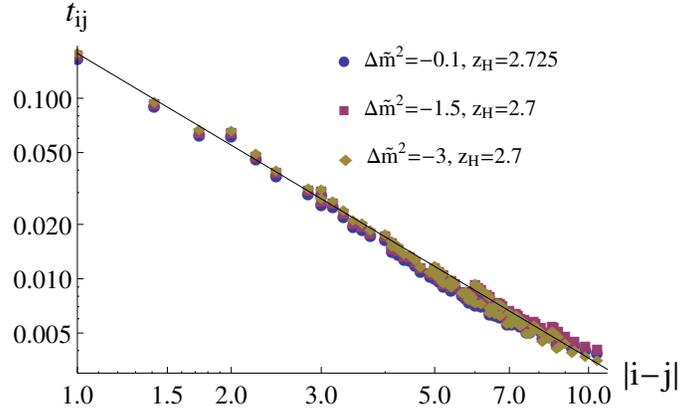}
\caption{
The log-log plot of the dynamical hopping fields $t_{ij}$ as functions of $|i-j|$ at the horizon, $z_H$,
at different values of $\Delta \tilde{m}^2$ in the superfluid phase. 
The slope of the decay is the same at all values of $\Delta \tilde{m}^2$, 
which indicates the universality of $\kappa$ along $z_H$.
}
\label{fig: t along zH in SF phase}
\end{figure}

Remarkably, the exponent $\kappa_H$ at the horizon
is independent of $\Delta \td m^2$ 
as is shown in Fig. \ref{fig: t along zH in SF phase}.
This suggests that the horizon is characterized
by a universal exponent associated with the decay
of the hopping fields.
The location where the decay of the hopping fields exhibits the universal exponent 
precisely coincides with the location where
 $C$ turns on and $\psi$ vanishes in finite lattices.
It is interesting to note that the hopping fields
exhibit universal `critical behavior' that is akin to 
a continuous phase transition. 
This should not be confused with 
the usual phase transition 
where the correlation length diverges 
as $\Delta \td m^2$ is tuned from
the insulating phase to the superfluid phase.
The `transition' that happens at $z_H$ is associated with
the divergence of the length scale for the renormalized hopping fields
within the superfluid phase.

As the `short-distance' modes are integrated out,
the effective action is renormalized by further neighbor hoppings.
In the insulating phase, 
the further neighbor hopping fields remain exponentially small,
and the renormalized action
remains local at all scales.
In the superfluid phase, 
the system can not keep the locality 
as further neighbor hoppings are proliferated
 beyond the critical scale $z_H$.
The presence/absence of the horizon 
can be used as a `holographic order parameter'
that distinguishes the superfluid/insulating phases.

The non-locality in the bulk could have been avoided 
if one had allowed for spontaneous symmetry breaking
by turning on an infinitesimally small symmetry breaking field
before taking the thermodynamic limit.
In that description, one starts with a new vacuum with broken symmetry,
and fluctuations around the new vacuum are described by the Goldstone modes.
Here the spontaneous symmetry breaking is not allowed
because the large volume limit is taken in the absence of 
the symmetry breaking field.
The emergence of the non-local geometry is a sign
that the system cannot maintain both locality
and the full symmetry in the superfluid phase.

\begin{figure}[h!]
\subfigure[\;$\Delta \td m^2 = -0.5$]{\includegraphics[width=3in]{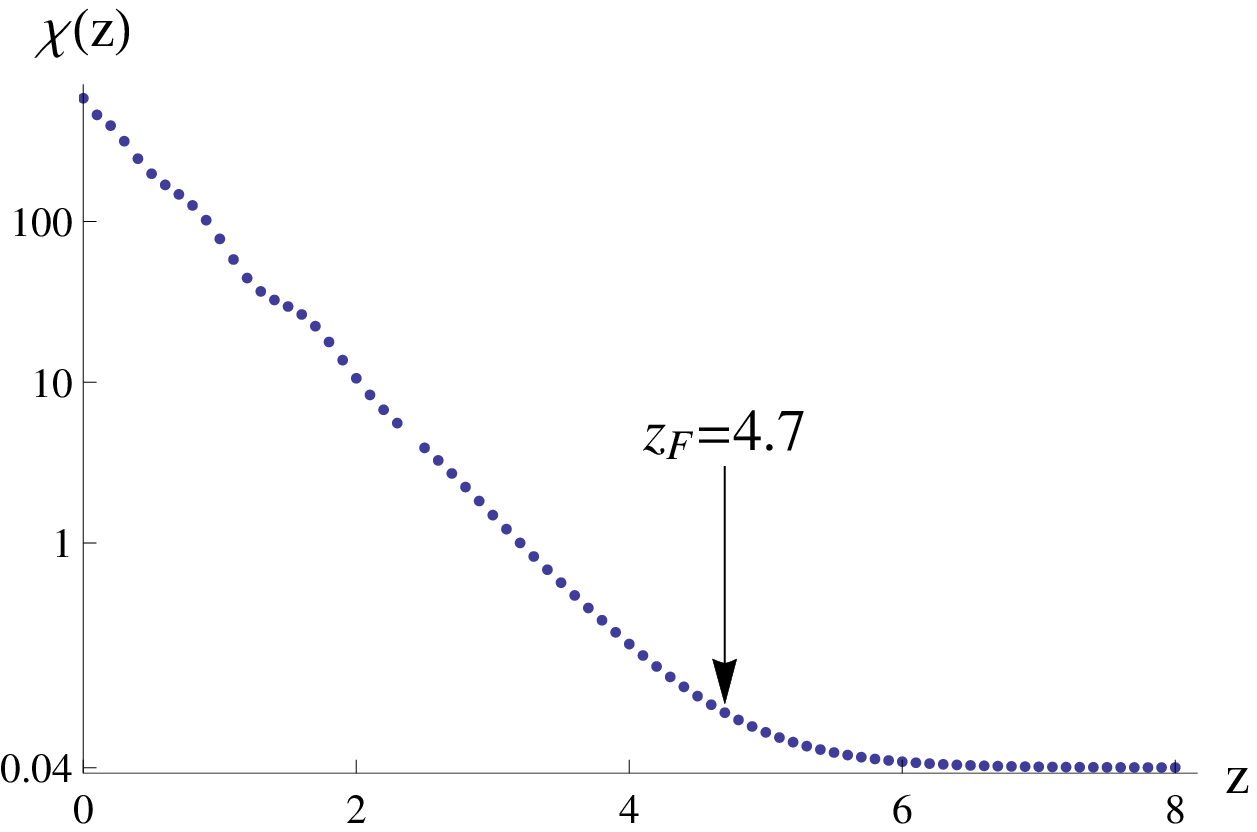}}
\subfigure[\;$\Delta \td m^2 = -1.5$]{\includegraphics[width=3in]{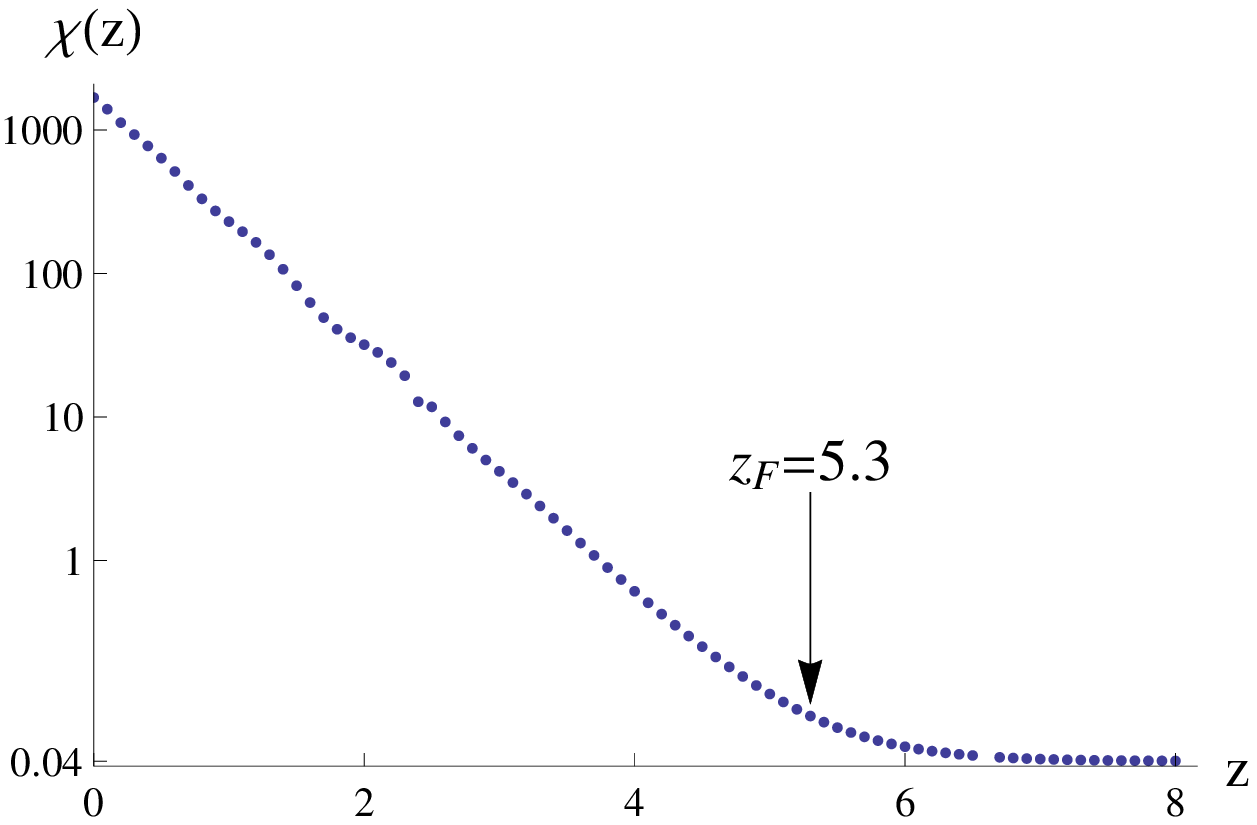}}
\caption{The $z$-dependence of $\chi$ in the superfluid phase at $\Delta \tilde{m}^2 = -0.5$(a) and $-1.5$(b) for $L=13$.}
\label{fig: K SF phase}
\end{figure}
\begin{figure}[h!]
\includegraphics[width=3.5in]{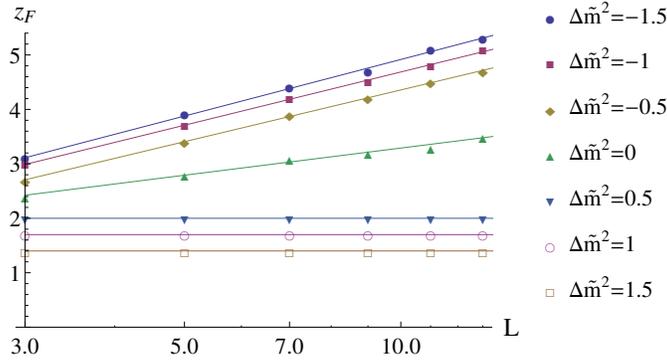}
\caption{
$L$-dependence of $z_F$ in the insulating phase and the superfluid phase.
The $L$ axis is shown on a logarithmic scale.
The non-zero slopes imply that $z_F$ diverges logarithmically in $L$ in the superfluid phase. 
On the contrary, $z_F$ remains finite in the thermodynamic limit in the insulating phase.}
\label{fig:zF_L}
\end{figure}

As $z$ increases further,  $t_{ij}(z)$ eventually decays exponentially in $z$
while remaining flat in $|i-j|$.
This leads to fragmentation in the deep IR region even in the superfluid phase.
Fig. \ref{fig: K SF phase} shows that the susceptibility $\chi$ becomes
negligible beyond a fragmentation scale $z_F$.
The fragmentation scale $z_F$ for $L=13$ is displayed along with 
the horizon scale $z_H$ in Fig. \ref{fig:holo_pd_L_7_9_11}.
However, the fragmentation in the superfluid phase is a finite size effect 
unlike in the insulating phase.
This can be seen from  Fig. \ref{fig:zF_L},
where  $z_F$ increases logarithmically in $L$ in the superfluid phase.
This is in contrast to the insulating phase where
$z_F$ is independent of $L$. 
In the thermodynamic limit,  $z_F = \infty$ in the superfluid phase 
while $z_F$ stays finite in the insulating phase. 
In this sense, the superfluid phase is characterized by  
the algebraically non-local geometry in the IR limit.

\begin{figure}[h!]
\includegraphics[width=3.5in]{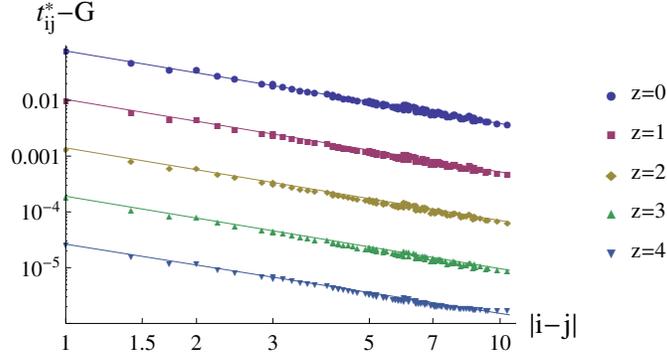}
%\subfigure[\;$\Delta \tilde{m}^2 = -3$]{\includegraphics[width=3in]{tStar_minus_G_L13_decay_diff_z_SF_dm_neg_3}}
%\subfigure[\;$\Delta \tilde{m}^2 = 0$]{\includegraphics[width=3in]{tStar_minus_G_L13_decay_diff_z_CP}}
\caption{
Log-log plots of $t^*_{ij}$ as a function of $|i-j|$ at different $z$ below and above the horizon, $z_H$, in the superfluid phase
at $\Delta \tilde{m}^2 = -3$. }
\label{fig: log log tstar minus C bulk}
\end{figure}
In Fig. \ref{fig: log log tstar minus C bulk}, the correlation functions $t^*_{ij}(z)$ are plotted in the bulk in the superfluid phase.
$t^*_{ij}(z)$ at each $z$ is fitted to Eq. (\ref{eq:32}).
%%$t^*_{ij}(z) = G(z) + \f{B(z)}{|i-j|^{2 \Delta(z)}}$.
The constant piece describes the long-range order,
and the power-law decay originates from the Goldstone mode in the superfluid phase.
The exponent $\Delta(z)$ remains constant as a function of $z$. 
In the thermodynamic limit, the exponent approaches $\Delta(L=\infty) \approx 0.55$ (Fig. \ref{chi_r_2} (c)).
The correlation function in the bulk is insensitive to the 
divergence in the length scale of the hopping field 
across the horizon, which is at $z_H=2.8$ at $\Delta \tilde{m}^2 = -3$.

\subsubsection{Critical point}

\begin{figure}[h]
%%\subfigure[]{\includegraphics[width=3in]{t_L13_loglog_CP}}
\subfigure[]{\includegraphics[width=3in]{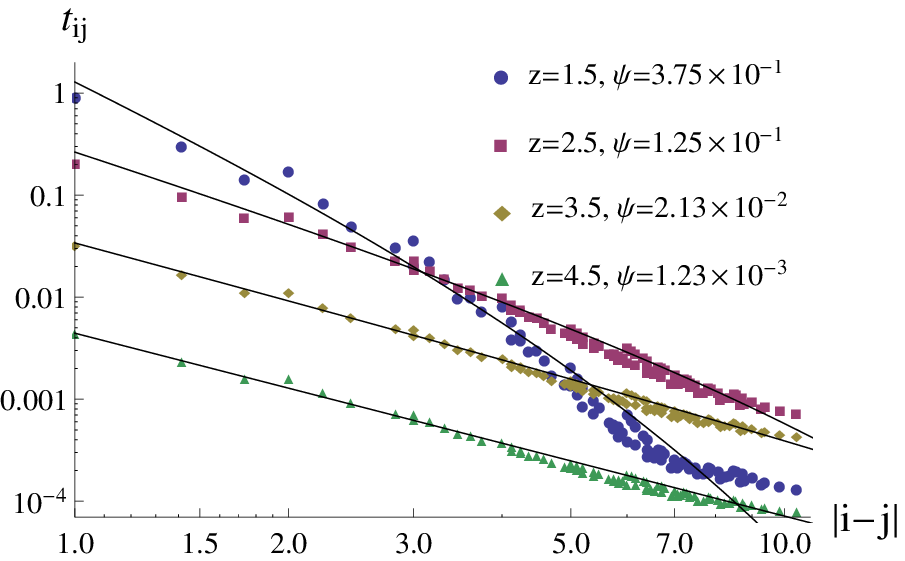}
\label{fig:t_alg_CP}
}
\subfigure[]{\includegraphics[width=3in]{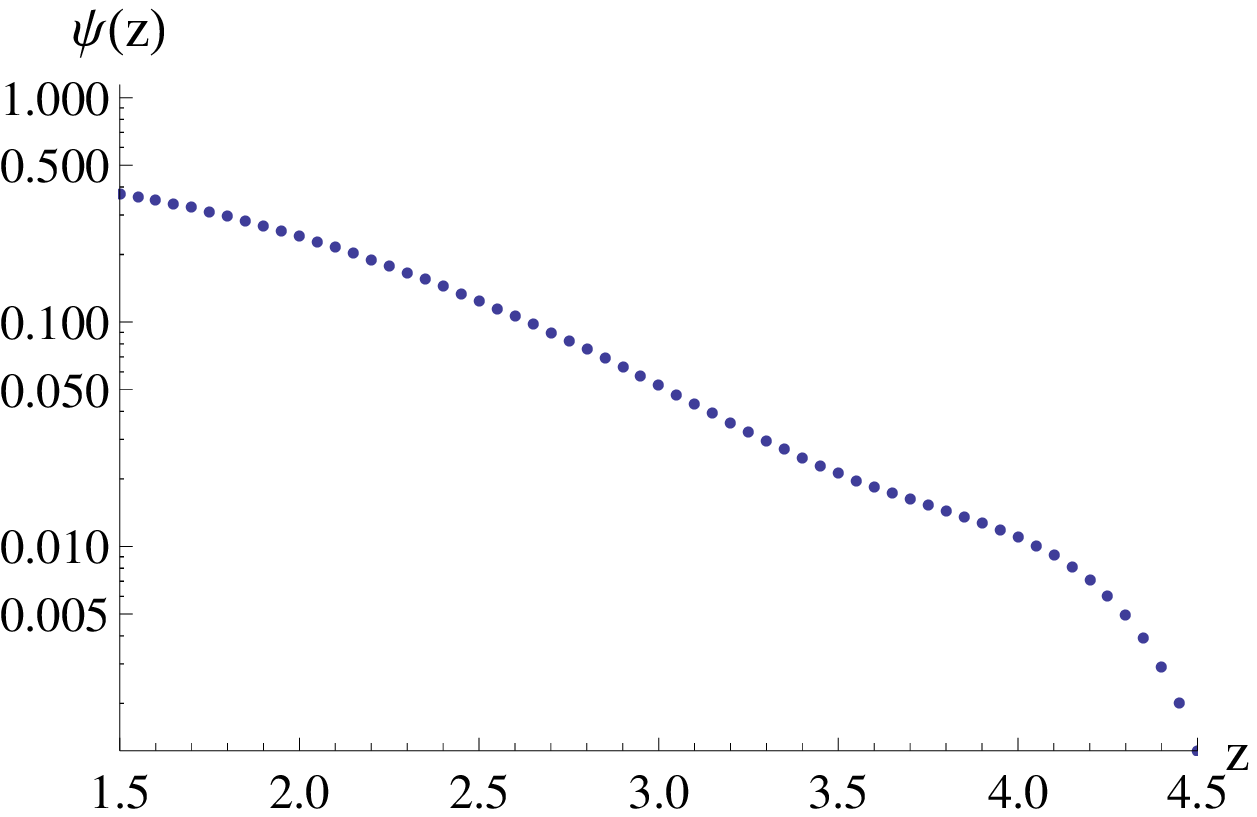}
\label{fig: psi of z around CP}
}
\caption{
(a) A log-log plot of the hopping fields $t_{ij}$ 
as a function of $|i-j|$ at several values of $z$ at the critical mass $\Delta \td m^2=0.2$ for $L=13$. 
(b) The logarithmic plot of $\psi(z)$ as a function of $z$ at $\Delta \td m^2 = 0.2$.
}
\end{figure}

\begin{figure}[h]
\includegraphics[width=3.5in]{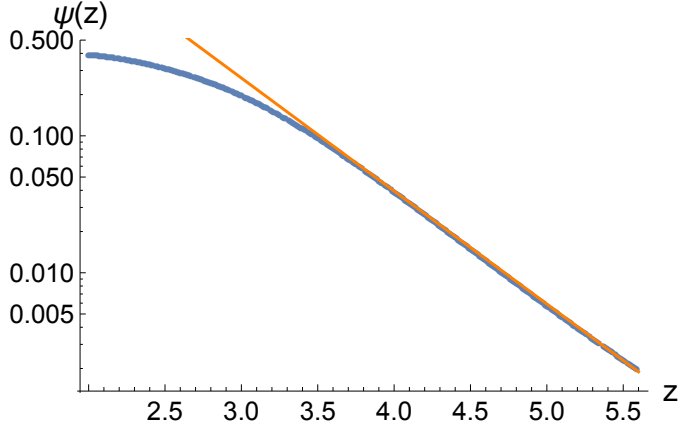}
\caption{
The logarithmic plot of $\psi(z)$ as a function of $z$ for a larger lattice with $L=21$ at the finite size critical point, $\Delta \td m^2=0.05$.
In the IR region, there is a clear exponential decay in the form of $\psi(z) \sim e^{-1.9z}$, which is shown as a straight line.
}
\label{fig:24}
\end{figure}

In Fig. \ref{fig:t_alg_CP}, we show the hopping fields $t_{ij}(z)$ as a function of $|i-j|$ 
for $L=13$ at the finite size critical point, 
$\Delta \td m^2 = \Delta \td m_c^2(13) = 0.2$.
The hopping fields are well fit by the form in Eq. (\ref{eq:expalg}).
As $z$ increases, the rate of exponential decay $\psi(z)$ decreases
as is shown in Fig. \ref{fig: psi of z around CP}.
Because of the invariance under the scale transformation at the critical point,
$\psi(z)$ is expected go to zero exponentially in the large $z$ limit (it is noted that $z$ is the logarithmic length scale).  
However, $\psi$ in the IR region deviates from the expected exponential behavior due to finite size effect for small $L$ (Fig. \ref{fig: psi of z around CP}).
To reduce the finite size effect, we push our system size to $L=21$ to read off $\psi(z)$ at the critical point. 
As is shown in Fig. \ref{fig:24}, $\psi$ indeed decays exponentially in the IR region.

At the critical point, the hopping fields retain locality with the characteristic length scale $1/\psi$.
In other words, the sites are more or less globally connected at length scales smaller than $1/\psi$,
and the theory is local only at larger length scales.
However, we don't expect there to be
a sense of flat geometry at the length scale of $1/\psi$
because the present theory is weakly coupled in the large N limit.

\begin{figure}[h]
\subfigure[]{\includegraphics[width=3in]{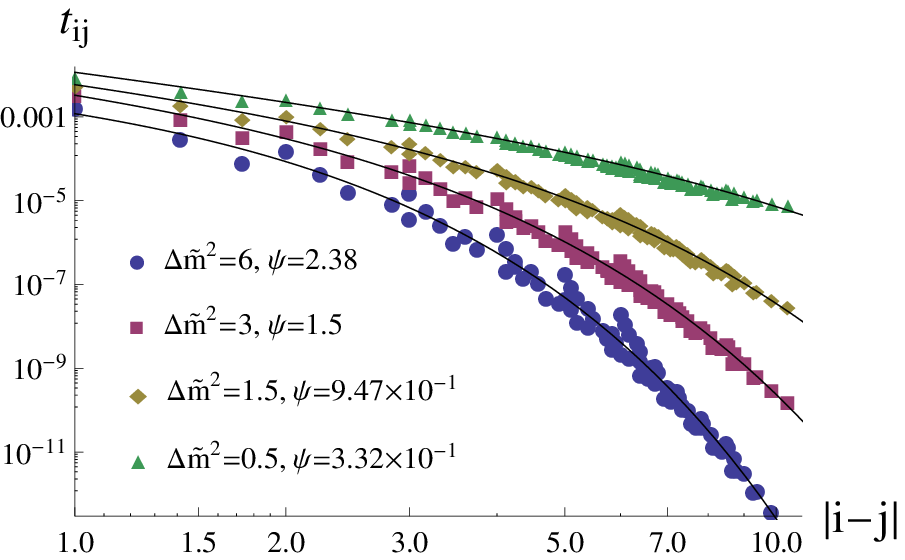}
\label{h_cut_325_1}
}
\subfigure[]{\includegraphics[width=3in]{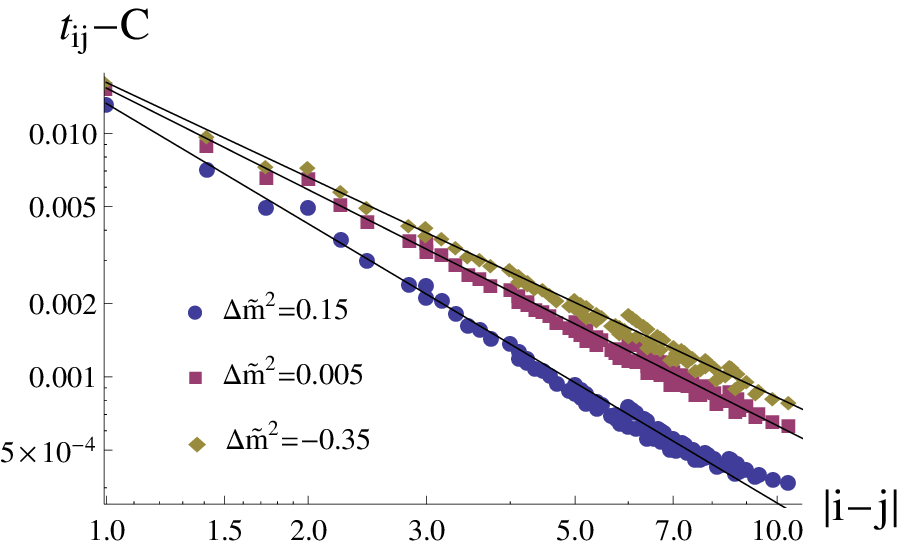}
\label{h_cut_325_2}
}
\caption{
(a) The hopping fields $t_{ij}$ at $z=4$ plotted on a log-log plot as a function of $|i-j|$ for $L=13$. 
Several values of $\Delta \td m^2 > \Delta \td m_c^2(13)$ in the insulating phase are shown, moving closer to the critical point line $\textbf{($\Delta \td m_c^2(13) = 0.2$)}$. 
The solid lines represent fits with the form in Eq. (\ref {eq:expalg}), and the exponential component $\psi$ is displayed for each curve.
(b) The same log-log plot as (a), but at $\Delta \td m^2 < \Delta \td m_c^2(13)$. 
Here the constant off-set in Eq. (\ref{tijCB}) has been subtracted out to show that the remainder is a pure algebraic decay.
}
\end{figure}

\begin{figure}[h!]
\includegraphics[width=4.5in]{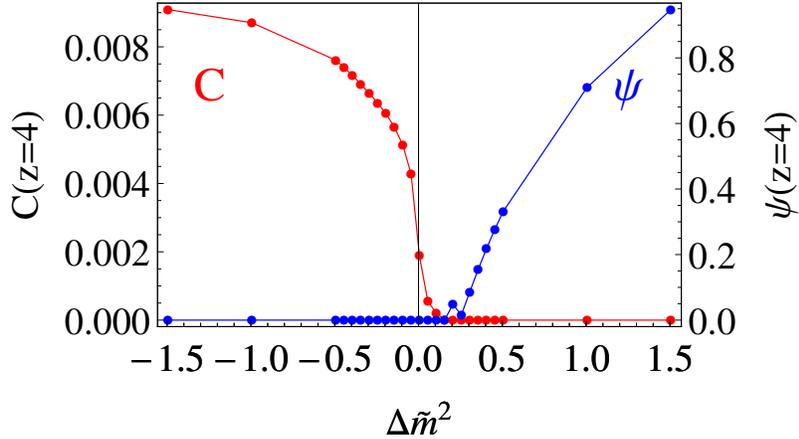}
\caption{
$C(z=4)$ and $\psi(z=4)$ for $L=13$ from Eqs. (\ref{tijCB}) and (\ref {eq:expalg}), respectively, plotted as functions of $\Delta \td m^2$ across the critical point inside the bulk.
}
\label{CPsi}
\end{figure}

\begin{figure}[h!]
\includegraphics[width=3.5in]{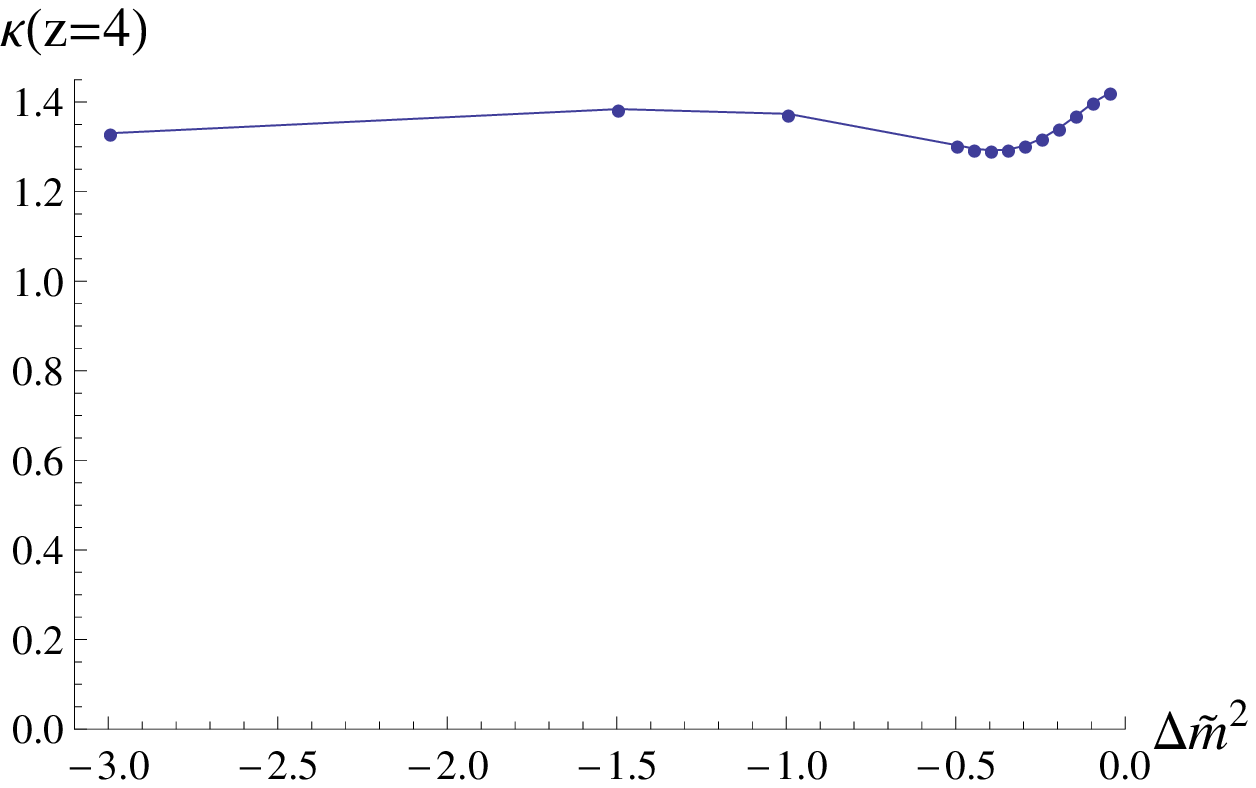}
\caption{
$\kappa(z=4)$ for $L=13$ from Eq. (\ref{tijCB}) plotted as a function of $\Delta \td m^2 < 0$ inside the bulk. 
}
\label{kappa_across}
\end{figure}

As $\Delta \td m^2$ is tuned across the horizon at a fixed z deep inside the bulk,
$t_{ij}$ shows the same critical behaviour 
as it does for increasing $z$ with fixed $\Delta \td m^2$ in the superfluid phase, 
which is discussed in Sec. IV. B. 2.
This is shown in Figs. \ref{h_cut_325_1} and \ref{h_cut_325_2}.
On the insulating side,
the hopping fields  decays exponentially, 
%%as $t_{ij}(z) \sim e^{- \psi(z) |i-j|}$,
where the rate of the exponential decay ($\psi$) continuously vanishes 
as the critical point is approached.
On the superfluid side,
the hopping fields decay algebraically with 
the off-set $C$ from Eq. (\ref{tijCB}) that continuously turns on across the horizon
for finite size lattices.
The dependence of $\psi$ and $C$ on $\Delta \td m^2$ 
across the horizon at a fixed $z$
is displayed in Fig. \ref{CPsi}.
Once the constant off-set is subtracted,
the hopping fields decay algebraically inside the horizon. 
This is displayed in Fig. \ref{h_cut_325_2}. 
While the hopping fields decay with the universal exponent at the horizon, 
the exponent inside the horizon changes continuously,
as is shown in Fig. \ref{kappa_across}. 

\begin{figure}[h!]
\includegraphics[width=3.5in]{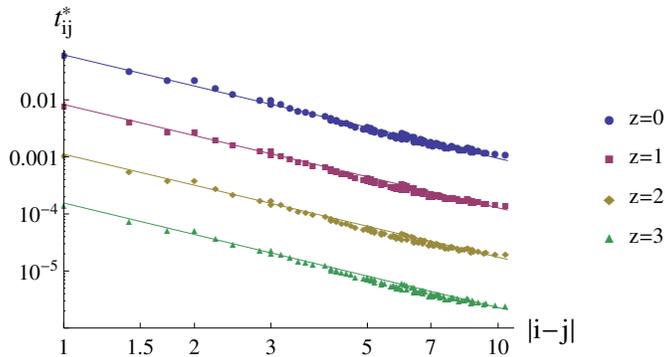}
\caption{
A log-log plot of $t^*_{ij}$ as a function of $|i-j|$ at different $z$ with fixed $\Delta \tilde{m}^2 = 0.2$ for $L=13$. 
The fields display an algebraic decay with the same power at all $z$.}
\label{fig: log log tstar bulk}
\end{figure}
In Fig. \ref{fig: log log tstar bulk}, we show the correlation functions $t^*_{ij}(z)$ 
at the finite size critical point $\Delta \tilde{m}^2 = 0.2$. 
They show purely algebraic decays of the form in Eq. (\ref{eq:32}) with $G=0$. 
As is the case in the superfluid phase, $\Delta$ is independent of $z$ and approaches $\Delta(\infty) \approx 0.55$ in the thermodynamic limit. 
%with the same exponent that 
%approaches $\Delta(\infty) \approx 0.55$ in the thermodynamic limit. 
%As is the case in the superfluid phase, $\Delta$ is independent of $z$.

\subsubsection{Holographic phase diagram}

Now we combine all the information 
to construct a holographic phase diagram
in the bulk.
In Fig. \ref{fig:holo_pd_L_7_9_11}, 
the  scale $z_F$ marks the crossover 
beyond which the space is fragmented.
In the insulating phase,
$z_F$ is largely independent of $L$:
the fragmentation of space
is a generic feature of the insulating phase.
On the other hand, 
$z_F$ diverges in the thermodynamic limit for $\Delta \td m^2 \leq 0$.
This is confirmed from the finite size scaling in Fig. \ref{fig:zF_L}.
Therefore, there is no fragmentation outside the insulating phase
in the thermodynamic limit.
The superfluid phase is
distinguished from the insulating phase
by the presence of the horizon in the bulk.
The horizon is characterized by the 
power-law decay of the hopping field, 
$t_{ij}(z_H) \sim \frac{1}{|i-j|^{\kappa_H}}$ 
with the universal  exponent that approaches $\kappa_H \approx 1$ 
in the thermodynamic limit.
At the critical point, 
the hopping fields decay as
$t_{ij}(z) \sim \frac{e^{-\psi(z)|i-j|}}{|i-j|^{\kappa(z)}}$,
where $\psi(z)$, $\kappa(z) > 0$ and $\psi(z) \rightarrow 0$, 
$\kappa(z) \rightarrow \kappa_H$ as $z \rightarrow \infty$.
%$t_{ij}(z_H) \sim \frac{e^{-\psi(z)|i-j|}}{|i-j|^{\kappa}}$, 
%where $\psi(z)$ is positive and asymptotically vanishes 
%only in the $z \rightarrow \infty$ limit.
Fig. \ref{fig:holo_pd_L_inf} summarizes 
the holographic phase diagram 
in the thermodynamic limit.

\section{Summary and discussion}

In summary, we derive the holographic action 
for the U(N) vector model regularized on a lattice.
The bulk equations of motion are solved 
numerically for finite lattices in three dimensions.
From the finite size scaling, we find that 
the insulating phase, the superfluid phase
and the critical point exhibit distinct geometric 
features in the bulk.
The IR geometry of the insulating phase
is characterized by ultra-locality with a decoupled lattice.
The superfluid phase exhibits a horizon 
at and beyond which the geometry becomes non-local.
The critical point shows a local geometry
with a characteristic length scale
that asymptotically diverges in the IR limit.

Although the U(N) vector model is exactly solvable in the large $N$ limit,
the present holographic description allows one to understand the concrete model
from a holographic perspective. 
This is a first step toward applying QRG to more non-trivial
models whose solutions at large $N$ are not known (e.g. matrix models). 
An advantage of using finite lattices is that 
the finite size effect can be used in solving the problem 
using a simplified IR boundary condition numerically. 
We finish by listing some open problems for future investigation.

Analytic solution : 
Although the numerical solutions provide a great deal of information
on the behaviour of the bulk, it is desirable to have an analytic solution,
especially at the critical point.
One approach is to reduce the infinite set of equations of motion
for the bi-local fields to a finite set by projecting the solutions
to the ones constrained by the numerical solution.
In general, it will be of interest to better understand theoretical structures of 
bi-local (or multi-local) field theories.

Finite temperature :
The Euclidean theory in $D$ dimensions can be viewed as 
the $(D-1)$-dimensional quantum theory at zero temperature in the imaginary time formalism.
One can turn on finite temperatures by making the size of the thermal circle finite.
In this case, the hopping fields in the temporal direction will be different from 
those in the spatial directions.
At the horizon, the non-locality is expected to develop only in the temporal direction but not in the spatial directions.  
In other words, the size of the thermal circle will shrink to zero while
the spatial area remains nonzero at the horizon.

Application to Fermi surfaces :
Having understood the insulator to superfluid phase transition in the bosonic model,
one can try to study the fermionic counterpart. 
In the fermionic system, 
the superfluid phase is replaced by an itinerant state with a Fermi surface.
It will be of great interest to understand how the Fermi surface manifests itself	
in the bulk geometry\cite{lee2009non,2011PhRvD..83f5029L,Čubrović24072009,2011PhRvD..83d6003H,2011JHEP...06..012F}.

Full $(D+1)$-dimensional diffeomorphism invariance in the continuum limit :
As is discussed in Sec. II,
the holographic action for the lattice model possesses 
a subset of the full diffeomorphism invariance 
of the continuum space.
At each $D$-dimensional slice with a constant $z$,   
the local permutation symmetry is present.
Given that the lattice provides a UV complete theory which flows
to the conformal field theory in the continuum (long-distance) limit,
it is expected that the holographic theory for the lattice model
recovers the continuum theory with the full diffeomorphism invariance
in the IR region.
However, this needs to be understood more explicitly. 

Connection to the higher spin theory :
In the continuum limit, it may be possible to relate
the equations of motion in the bulk 
with those of the higher spin gauge theories
proposed as the holographic dual for the vector models\cite{2002PhLB..550..213K,Vasiliev:1995dn,Vasiliev:1999ba,Giombi:2009wh,Vasiliev:2003ev,Maldacena:2011jn,Maldacena:2012sf}.
However, the connection is not clear a priori
because the form of the bulk theory is sensitive to the
specific regularization scheme.
In particular, we believe that the $1/N$ corrections cannot be included in the present formalism of the higher spin theory where the quartic interaction is implemented only through the UV boundary condition\cite{2002PhLB..550..213K}. 
This is because the quadratic theory with fluctuating sources is ill-defined 
unless one keeps the double-trace operator in the bulk, as is discussed in Sec. II.

\section{Acknowledgments}
We thank 
Yu Nakayama, Joe Polchinski, Subir Sachdev and Xiao-Liang Qi
for helpful discussions.
The research was supported in part by 
the Natural Sciences and Engineering Research Council of 
Canada,
the Early Research Award from the Ontario Ministry of 
Research and Innovation,
the Templeton Foundation
and CIFAR.
Research at the Perimeter Institute is supported 
in part by the Government of Canada 
through Industry Canada, 
and by the Province of Ontario through the
Ministry of Research and Information.

\bibliography{references}

%%\input{QRG_appendix_0303.tex}

%%%%%%%%%%%%%%%%%%%%%%%%%%%%%
\begin{appendix}
%%%%%%%%%%%%%%%%%%%%%%%%%%%%%

\section{Derivation of Eq. (\ref{S_1 plus delta S_1})}
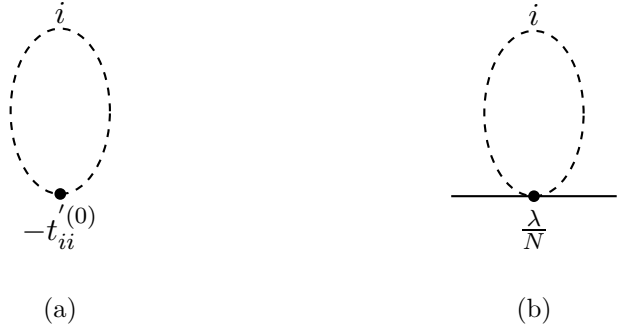
\begin{figure}[htp]
\centering
\subfigure[]{
\begin{tikzpicture}[scale=1.1, every node/.style={scale=1.1}]
\fill[black]  (-6,1.2) circle(2.0pt);
\node at (-6,0.8) {$-t^{'(0)}_{ii}$};
\draw[thick,dashed]  (-6,2.2) ellipse (0.6 and 1);
\node at (-6,3.4) {$i$};
\end{tikzpicture}
}
\hspace*{4cm}
\subfigure[]{
\begin{tikzpicture}[scale=1.1, every node/.style={scale=1.1}]
\fill[black]  (-3.2,1.2) circle(2.0pt);
\node at (-3.2,0.8) {$\frac{\lambda}{N}$};
\draw[thick,dashed]  (-3.2,2.2) ellipse (0.6 and 1);
\node at (-3.2,3.4) {$i$};
\draw[thick]  plot[smooth, tension=.7] coordinates {(-4.2,1.2) (-2.2,1.2)};
\end{tikzpicture}
}
\caption{The diagrams contributing to $\tilde{\mathscr{Z}}_1$ at order $dz$. (a) and (b) represent the first and second terms of Eq. (\ref{eq_zeta1}), respectively. The dashed propagator indicates the high energy modes while the solid propagator indicates the low energy modes.}
\label{fig_firstorder}
\end{figure}

\begin{figure}[h]
\centering
%\vspace{5mm}
\subfigure[]{
\begin{tikzpicture}[scale=1.1, every node/.style={scale=1.1}]
\node at (-6.8,-1.5) {$-t^{'(0)}_{ki}$};
\node at (-5,-1.5) {$-t^{'(0)}_{ij}$};
\node at (-8.2,-1) {$k$};
\node at (-5.9,-0.4) {$i$};
\node at (-3.6,-1) {$j$};
\draw[thick]  plot[smooth, tension=.7] coordinates {(-8,-1) (-6.8,-1)};
\draw[thick]  plot[smooth, tension=.7] coordinates {(-5,-1) (-3.8,-1)};
\draw[thick,dashed] (-6.8,-1) node (v2) {} arc (123.6891:55.5291:1.6);
\fill[black]  (-6.8,-1) circle(2.0pt);
\fill[black]  (-5,-1) circle(2.0pt);
\end{tikzpicture}
}
\hspace*{1cm}
\subfigure[]{
\begin{tikzpicture}[scale=1.1, every node/.style={scale=1.1}]
\draw[thick] (-2.2,-0.4) -- (-1,-0.4) node (v1) {} -- (-1,0.8) -- (-1,-1.6);
\draw[thick,dashed] (-1,-0.4) -- (0.6,-0.4) node (v3) {};
\node at (-1.2,-0.7) {$\frac{\lambda}{N}$};
\draw[thick] (1.8,-0.4) -- (0.6,-0.4) node (v4) {} -- (0.6,0.8);
\draw[thick] (0.6,-1.6) -- (0.6,-0.4);
\node at (0.8,-0.7) {$\frac{\lambda}{N}$};
\node at (-2.4,-0.4) {$i$};
\node at (0.6,-1.8) {$i$};
\node at (-1,-1.8) {$i$};
\node at (-1,1) {$i$};
\node at (2,-0.4) {$i$};
\node at (0.6,1) {$i$};
\node at (-0.2,-0.2) {$i$};
\end{tikzpicture}
}
\hspace*{1cm}
\subfigure[]{
\begin{tikzpicture}[scale=1.1, every node/.style={scale=1.1}]
\draw[thick] (-2.6,-4) -- (-1.4,-4) node (v5) {} -- (-1.4,-2.8) -- (-1.4,-5.2);
\draw[thick,dashed] (-1.4,-4) -- (0.2,-4) node (v6) {};
\draw[thick] (0.2,-4) -- (1.4,-4);
\fill[black]  (0.2,-4) circle(2.0pt);
\node at (-0.6,-3.8) {$i$};
\node at (1.6,-4) {$j$};
\node at (-2.8,-4) {$i$};
\node at (-1.4,-5.4) {$i$};
\node at (-1.4,-2.6) {$i$};
\node at (-1.6,-4.3) {$\frac{\lambda}{N}$};
\node at (0.2,-4.5) {$-t^{'(0)}_{ij}$};
\end{tikzpicture}
}
\caption{The diagrams contributing to $\tilde{\mathscr{Z}}_2$ at order $dz$. (a), (b), (c) represent the first, second and third terms of Eq. (\ref{eq_zeta2}), respectively. Dashed and solid propagators have the same meaning as in Fig. \ref{fig_firstorder}.}
\label{fig_secondorder}
\end{figure}
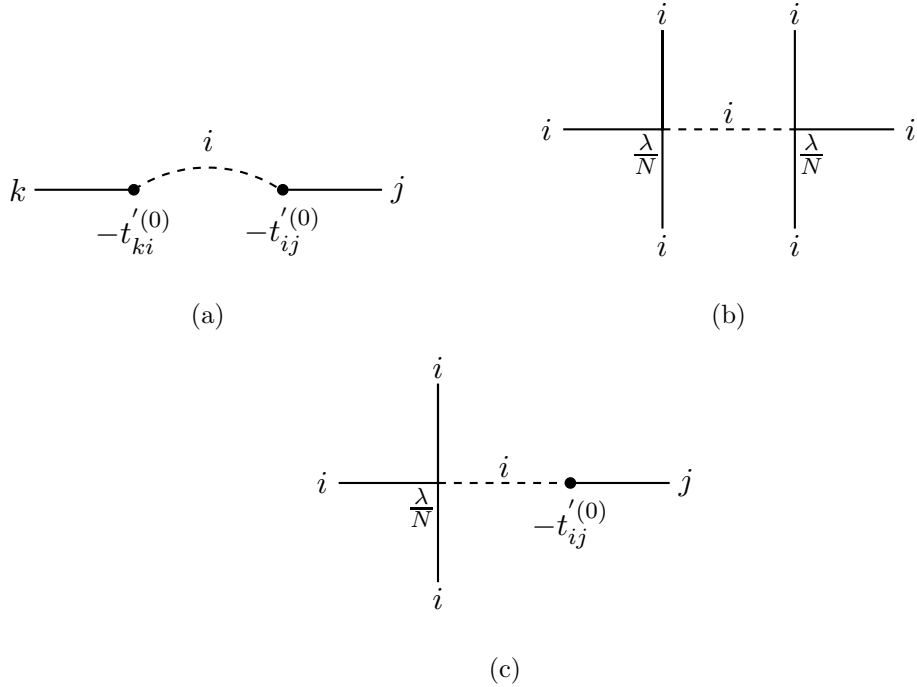

%%%%%%%%%%%%%%
Here we show the intermediate steps in going from Eq. (\ref{S_1}) to Eq. (\ref{S_1 plus delta S_1}). We can rearrange $\mathcal{S}^{(0)''}$ in the following way:
\beqq
\mathcal{S}^{(0)''} = N S_{UV}\left[t^{(0)}_{ij},t^{*(0)}_{ij}\right] + \mathcal{S}_L\left[\v{\phi},\v{\phi}^*\right] + \mathcal{S}_H\left[\tilde{\v{\phi}},\tilde{\v{\phi}}^*\right] + \mathcal{S}_{I}\left[\v{\phi},\v{\phi}^*, \tilde{\v{\phi}},\tilde{\v{\phi}}^*\right],
\eeqq
where
\begin{align*}
\mathcal{S}_L&=m^2\sum_i \left(\v{\phi}^*_i\cdot\v{\phi}_i\right)+ \frac{\lambda}{N} \sum_i e^{-4\alpha_i^{(1)} dz} \left(\v{\phi}^*_i\cdot\v{\phi}_i\right)^2-\sum_{ij}t^{'(0)}_{ij}e^{-( \alpha_i^{(1)} + \alpha_j^{(1)} ) dz}\left(\v{\phi}^*_i\cdot\v{\phi}_j\right),\nonumber\\
\mathcal{S}_H&=\sum_{i} \tilde{\mu}^2_i \left(\tilde{\v{\phi}}^*_i\cdot\tilde{\v{\phi}}_i\right),\nonumber\\
\mathcal{S}_{I}&=-\sum_{ij}t^{'(0)}_{ij}e^{-( \alpha_i^{(1)} + \alpha_j^{(1)} ) dz}\left[\left(\tilde{\v{\phi}}^*_i\cdot\tilde{\v{\phi}}_j\right)+\left(\tilde{\v{\phi}}^*_i\cdot\v{\phi}_j\right)+\left(\v{\phi}^*_i\cdot\tilde{\v{\phi}}_j\right)\right]\nonumber\\
&+\frac{\lambda}{N}\sum_i e^{-4\alpha_i^{(1)} dz} \left[\left(\tilde{\v{\phi}}^*_i\cdot\tilde{\v{\phi}}_i\right)^2+\left(\tilde{\v{\phi}}^*_i\cdot\v{\phi}_i\right)^2+\left(\v{\phi}^*_i\cdot\tilde{\v{\phi}}_i\right)^2\right.\nonumber\\
&~~~~~~~~~~~~~~~~~~~~~~~\left.+2\left(\v{\phi}^*_i\cdot\v{\phi}_i\right)\left\{\left(\tilde{\v{\phi}}^*_i\cdot\tilde{\v{\phi}}_i\right)
+\left(\tilde{\v{\phi}}^*_i\cdot\v{\phi}_i\right)+\left(\v{\phi}^*_i\cdot\tilde{\v{\phi}}_i\right)\right\}\right.\nonumber\\
&~~~~~~~~~~~~~~~~~~~~~~~\left.+2\left(\tilde{\v{\phi}}^*_i\cdot\tilde{\v{\phi}}_i\right)\left\{\left(\tilde{\v{\phi}}^*_i\cdot\v{\phi}_i\right)
+\left(\v{\phi}^*_i\cdot\tilde{\v{\phi}}_i\right)\right\}
+2\left(\tilde{\v{\phi}}^*_i\cdot\v{\phi}_i\right)\left(\v{\phi}^*_i\cdot\tilde{\v{\phi}}_i\right)\right].
\end{align*}
$S_L\left[\v{\phi},\v{\phi}^*\right]$ is the bare action for the low energy fields $\v{\phi}$ and $\v{\phi}^*$, $\mathcal{S}_H\left[\tilde{\v{\phi}},\tilde{\v{\phi}}^*\right]$ is the bare action for the high energy fields $\tilde{\v{\phi}}$ and $\tilde{\v{\phi}}^*$, and $\mathcal{S}_{I}$ gives the mixing between the low and the high energy fields.
Since $\alpha_i^{(1)}$ and $m$ are both $O(1)$ in $dz$, if we make $dz$ infinitesimal we have
\beqq
\frac{1}{\tilde{\mu}^2_i} = \frac{e^{2\alpha_i^{(1)} dz}-1}{m^2}\approx\frac{2\alpha_i^{(1)} dz}{m^2}\sim O(dz). 
\eeqq
Therefore, to derive the bulk action that is continuous in the holographic direction we need to take into account only terms that are linear in $1/\tilde{\mu}^2_i$. Keeping this in mind we integrate out the high energy fields $\tilde{\v\phi}$, $\tilde{\v\phi}^*$ using only their bare action. For this we write $\mathcal{Z}$ as
\beqq
\mathcal{Z}\propto\int\mathcal{D}\v\phi\mathcal{D}\v\phi^*\mathcal{D}t^{(0)} \mathcal{D}t^{*(0)}
e^{-\mathcal{S}^{(0)''}_1}~e^{-\Delta\mathcal{S}^{(0)''}_1},
%\label{Z with S_1 prime}
\eeqq
where
\begin{widetext}
\beqq
\mathcal{S}^{(0)''}_1=N \mathcal{S}_{UV}\left[t^{(0)}_{ij},t^{*(0)}_{ij}\right] + \mathcal{S}_L\left[\v{\phi},\v{\phi}^*\right]
\eeqq
\end{widetext}
and
\beqq
e^{-\Delta\mathcal{S}^{(0)''}_1}=\int \mathcal{D}\tilde{\v\phi}\mathcal{D}\tilde{\v\phi}^*~e^{-\mathcal{S}_H}
~e^{-\mathcal{S}_I}=
\mathscr{N}\langle e^{-\mathcal{S}_I}\rangle_{\mathcal{S}_H},
\eeqq
where
\beqq
\mathscr{N}= \int\mathcal{D}\tilde{\v\phi}\mathcal{D}\tilde{\v\phi}^*e^{-\mathcal{S}_H}
\eeqq
is the normalization factor.
We need to calculate the correction to the action $\Delta\mathcal{S}^{(0)''}_1$ to first order in $dz$. To make the calculation more tractable we rewrite
\beqq
\mathscr{N}\langle e^{-\mathcal{S}_I}\rangle_{\mathcal{S}_H}=\mathscr{N}\left(1+\sum_{p=1}^\infty \frac{(-1)^p}{p!}\langle \left(\mathcal{S}_I\right)^p\rangle_{\mathcal{S}_H}\right)
\eeqq
and rename the terms from the sum
\beqq
\tilde{\mathscr{Z}}_p=\frac{(-1)^p}{p!}\langle \left(\mathcal{S}_I\right)^p\rangle_{\mathcal{S}_H}.
\eeqq
The goal is to calculate every $\tilde{\mathscr{Z}}_p$ to order $dz$, then re-exponentiate to get $\Delta\mathcal{S}^{(0)''}_1$ to order $dz$. $\tilde{\mathscr{Z}}_1$ is given by 
\begin{widetext}
\beq
\tilde{\mathscr{Z}}_1=\frac{2 dz}{m^2}\left[N\sum_i \alpha_i^{(1)} t^{'(0)}_{ii} - 2\lambda\left(1+\f{1}{N}\right)\sum_i \alpha_i^{(1)} \left({\v \phi}_i^*\cdot{\v \phi}_i\right)\right].
\label{eq_zeta1}
\eeq
\end{widetext}
The first term of Eq. (\ref{eq_zeta1}) comes from tracing out the high energy modes from the hopping term, while the second term comes from the renormalization of the low energy mass due to integration of the high-low energy 4-boson interaction. The two contributions are illustrated in Figure \ref{fig_firstorder}. 
%%%%%%%%%%%
%%%%%%%%%%%%%
$\tilde{\mathscr{Z}}_2$ is given by
\begin{widetext}
\begin{align}
\tilde{\mathscr{Z}}_2=\frac{1}{2}\frac{2 dz}{m^2}\sum_{i} \alpha_i^{(1)} &\left[2\sum_{jk}  t^{'(0)}_{ki}t^{'(0)}_{ij}({\v \phi}^*_k\cdot{\v \phi}_j)+\frac{8\lambda^2}{N^2} ({\v \phi}_i^*\cdot{\v \phi}_i)^3 \right. \nonumber \\ 
&\left.- \frac{4\lambda}{N}\sum_{j} \left( t^{'(0)}_{ij}({\v \phi}_i^*\cdot{\v \phi}_j)
+ t^{'(0)}_{ji}({\v \phi}_j^*\cdot{\v \phi}_i) \right) ({\v \phi}_i^*\cdot{\v \phi}_i)\right].
\label{eq_zeta2}
\end{align}
\end{widetext}
The three contributions are illustrated in Fig \ref{fig_secondorder}. The first one represents fusion of hopping links to generate further hopping links (a). The second represents the fusion of two 4-boson vertices to generate a 6-boson vertex (b), while the final two terms refer to the fusion of the 4-vertex with the hopping term (c).
%%%%%%%%%%%%%%%
There are no more terms in any $\tilde{\mathscr{Z}}_p$ that are linear in $dz$, which implies that $\Delta\mathcal{S}^{(0)''}_1 = -(\tilde{\mathscr{Z}}_1 + \tilde{\mathscr{Z}}_2) + O(dz^2)$. This finishes our calculation of $\mathcal{S}^{(0)''}_1 + \Delta\mathcal{S}^{(0)''}_1$, which in Eq. (\ref{S_1 plus delta S_1}) we call  $\tilde{\mathcal{S}}^{(0)''}$.
%%%%%%%%%%%%%%

\section{Large $N$ solution for $U(N)$ vector model on a lattice}
Here we review the solution of the $U(N)$ vector lattice model at $N \rightarrow \infty$ \cite{Moshe:2003xn}. Our initial partition function at any given scale is
\beqaa
\mathcal{Z} = \int d \v{\phi}^* d \v{\phi} e^{-\left[-\sum_{ij} t_{ij}\v{\phi}_i^* \v{\phi}_j + \sum_{i} \left(m^2 |\v{\phi}_i|^2 + \f{\lambda}{N}|\v{\phi}_i|^4\right)\right]}.
\eeqaa
We introduce a Hubbard-Stratanovich field $\bar{\sigma}$ in the following way
\beqaa
\mathcal{Z} = \int d \v{\phi}^* d \v{\phi}\: d\bar{\sigma} e^{-\left[-\sum_{ij} t_{ij}\v{\phi}_i^* \v{\phi}_j + \sum_{i} \left(m^2 |\v{\phi}_i|^2 + \f{\lambda}{N}|\v{\phi}_i|^4
+ \f{N}{\lambda}(\bar{\sigma}_i - i \f{\lambda}{N}|\v{\phi}_i|^2)^2 \right)\right]}.
\eeqaa
Redefining $\sigma_i \equiv -i \bar{\sigma}_i$ gives
\beqaa
\mathcal{Z} \propto \int d \v{\phi}^* d \v{\phi}\: d\sigma e^{-\left[-\sum_{ij} t_{ij}\v{\phi}_i^* \v{\phi}_j + \sum_{i} (m^2+2\sigma_i) |\v{\phi}_i|^2
- \sum_i\f{N}{\lambda}\sigma_i^2\right]}.
\eeqaa
We want to replace $\sigma_i$ with its saddle point value at $N \rightarrow \infty$, which is assumed to be a site-independent constant $\sigma$. To find $\sigma$ we integrate out the $\v{\phi}$ and $\v{\phi}^*$ fields after going to momentum space 
\beqaa
\int d \v{\phi}^* d \v{\phi}\: e^{-\f1V \sum_k \left[(-t_k + m^2 + 2\sigma)|\phi_k|^2\right]} = 
\left(\prod_k \f{\pi V}{m^2 - t_k +2\sigma}\right)^{N}  = e^{N\: tr\: log\left(\f{\pi V}{m^2 - t_k +2\sigma}\right)}.
\eeqaa
%In the case of the $O(N)$ model the difference is that $N \rightarrow \f{N}{2}$. 
The effective action for $\sigma$ is now (ignoring constant terms)
\beqaa
S_{eff} = - V \f{N}{\lambda} \sigma^2 +	 N\;tr\;log\left(m^2 - t_k +2\sigma\right),
\eeqaa
and the saddle point equation becomes
\beqaa
\f{\partial S_{eff}}{\partial \sigma} &=& -2 V \f{N}{\lambda} \sigma + N\sum_k \f{2}{m^2 - t_k+2\sigma} = 0
\\ &\Rightarrow& \sigma = \f{1}{V} \sum_k \f{\lambda}{m^2 - t_k+2\sigma},
\eeqaa
which is the self-consistency condition for the mean field value $\sigma$, Eq. (\ref{eq_self_constist_momentspace_discrete}).
%%%%%%%%%%%%%%

\section{Numerical Method}
Here we discuss in detail the method we use to solve the equations of motion (EOM) of Eq. (\ref{EOMs}). The equations themselves are first order and seemingly simple, however each equation for $t_{ij}$ and $t^*_{ij}$ involves a sum over all other fields $t_{qp}$ and/or $t^*_{qp}$. The number of EOM grows as the volume $V$, and the number of terms in the sum also grows as $V$, leading to a very large number of terms. The equations are also non-linear. We use a tau spectral collocation method \cite{boyd2001chebyshev,canuto2006spectral,fornberg1998practical} combined with a Newton-Raphson algorithm \cite{press2007numerical} to solve this system. 

Each bilocal field is decomposed into a truncated polynomial series
\begin{align} 
t_{ij}(z)=\sum_{\al=0}^n a_{\al}^{ij}T_{\al}(x(z)),~~~~~t^*_{ij}(z)=\sum_{\al=0}^n b_{\al}^{ij}T_{\al}(x(z)),
\label{spectral_decomp}
\end{align}
where $T_{\al}(x(z))$ is the $\al$-th Chebyshev polynomial of the first kind. Here $x(z)$ must live on $[-1,1]$, which is the domain on which the Chebyshev polynomials are orthonormal. The domain of the RG scale $z$ is $[0,z^*]$, which implies that $x(z) = \f{2 z}{z^*} - 1$. The number $n$ is called the order of the tau spectral approximation, and $n+1$ is also the number of collocation points, i.e. the number of points on the $x$/$z$-axis at which the spectral Chebyshev representation agrees exactly with the original function. $n$ is also a measure of the accuracy of the solution; for smooth functions, one expects the error in the solution to decrease exponentially with $n$. In our case we take the collocation points to be the Chebyshev Gauss-Lobatto (CGL) points $x_{\al}=cos(\f{\pi \al}{n})$. For each lattice size we take $n=20$. 

We solve our equations for the coefficients $a,b$ instead of the full functions $t,t^*$. This allows us to solve one large (non-differential) matrix equation rather than many differential equations. For this we rewrite the EOM and boundary conditions in terms of the new ``vectors'' $\vec{a}_{ij}$ and $\vec{b}_{ij}$, where the $\{ij\}$ label the lattice link and the vector label is the Chebyshev label (labeled as ``$\al$'' in Eq.(\ref{spectral_decomp})).  

The rewritten EOM have 4 types of terms. The first is a constant, which is represented as a vector with that constant as the first component ($\al$ = 0) and all other components zero.
The second type is a linear term, which is just $c_{ij} t_{ij}$ (or $c_{ij} t^*_{ij}$) and it becomes $c_{ij}\: \vec{a}_{ij}$ (or $c_{ij}\: \vec{b}_{ij}$). The third type is the differential operator. We denote it by the matrix $L$ and for the CGL points it is given by 
\[
L_{\al \be} =
\begin{cases} 
\f{2}{z^*} \f{2 \be}{c(\al+1)} & \text{if} \;\; \be > \al \;\; \text{and}\;\;  \be+\al = \mathrm{odd} \\ 
0 & \mathrm{otherwise} 
\end{cases},
\] 
where $c(\ga)= 1+ \delta_{\ga,0}$. The last type of term is the quadratic term, e.g. $t_{ij}\:t_{pq}$. We choose to express it in spectral form via a procedure described in Sec. 3.4.4 of \cite{canuto2006spectral}. The first step of the procedure is to expand the sum to order $m=2n$, leaving all the coefficients zero that are higher order than $n$, i.e. $a^{ij}_{\al} = a^{pq}_{\al} = 0 \quad \text{for} \quad n<\al\leq2n$. Then we perform an inverse discrete Chebyshev transform to obtain the real space coefficients for every field, which are given by 
\begin{align*}
\tilde{a}^{ij}_{\be} = \ds_{\al=0}^m a^{ij}_{\al} cos\left(\f{\pi \be \al}{m}\right). 
\end{align*}
We multiply the real space coefficients to obtain the coefficients of the product in real space $\tilde{s}^{ij,pq}_{\be} = \tilde{a}^{ij}_{\be} \cdot \tilde{a}^{pq}_{\be}$, for $\be\in \{0,m\}$. We now perform an inverse discrete Chebyshev transform back to spectral space, which is given by 
\begin{align*}
s^{ij,pq}_{\al} = \ds_{\be=0}^m \tilde{s}^{ij,pq}_{\be} \f{2}{m \:\bar{c}_{\al} \:\bar{c}_{\be}} cos\left(\f{\pi \be \al}{m}\right) ,
\end{align*}
where 
\[
\bar{c}_{\ga} =
\begin{cases} 
2, & \ga=0,m \\ 
1 & \ga \in \{1,m-1\} 
\end{cases}.
\] 
Finally, 
\begin{align*} 
t_{ij}(z) t_{pq}(z)=\sum_{\al=0}^n s^{ij,pq}_{\al} \: T_{\al}(x(z))
\end{align*}
is the spectral representation of the product of $t_{ij}(z) \: t_{pq}(z)$. 

Now it is straightforward to express the EOM and boundary conditions in spectral form as a set of algebraic equations. The number of equations is equal to the total number of fields $t_{ij}$ and $t^*_{ij}$ times $n+1$ (the number of collocation points), which we enumerate and put into vector form  
$\vec{R}(\vec{a}_{ij},\vec{b}_{ij})=0$. Finally, we combine all of the $\{\vec{a}_{ij},\vec{b}_{ij}\}$ into one big vector $\vec{c}$
\begin{align}
\vec{R}(\vec{c})=0.
\label{R_equal_0}
\end{align}
As $\vec{R}$ is quadratic in $\vec{c}$, we cannot solve it directly via a matrix inversion.  
We employ an approximation method, namely a Newton-Raphson method, to solve Eq. (\ref{R_equal_0}). For this we linearize the equations
\begin{align*}
\vec{R}(\vec{c}_0)+(\vec{c}_1-\vec{c}_0)\frac{\partial\vec{R}}{\partial\vec{c}}\bigg|_{\vec{c}_0}=0,
\end{align*}
and solve for $\vec{c}_1$. This process is iterated $w$ times to arrive at a solution $\vec{c}_w$, which satisfies $L^2_{w}\equiv\sqrt{\ds_{\al} (\vec{c}_w-\vec{c}_{w-1})_{\al}} < \epsilon$. We choose $\epsilon = 10^{-11}$, which is small enough to indicate that the Newton-Raphson method has converged up to floating point round-off. 

An important question is that of convergence: when does the Newton-Raphson method converge, depending on the initial starting vector $\vec{c}_0$, and how many iterations does it take? 
The convergence domain that we find is summarized below (the reader should keep in mind that in tuning $\tilde{m}^2 = m^2 - \tilde{t}_{ii}$ we change $\tilde{t}_{ii}$ and keep $m^2$ constant (in this case $m^2 = 25$)). 

The solution in the deep insulating phase is available from the perturbative analytical solution of Eq. (\ref{analytical_pert_solution}). From this starting point one can obtain the solution at a certain distance away from the critical point on the insulating side, say $\Delta \tilde{m}^2 = 10$. From this solution the Newton-Raphson method converges quickly (within 10 iterations) for any point $\Delta \tilde{m}^2 > 0$. Also, it converges at $\Delta \tilde{m}^2 \lesssim 0$, however more slowly ($\approx$ within 20 iterations). This gives us the solutions at and just beyond the critical point on the superfluid side, which are of the most interest physically. Starting from a solution at  $10 \gg \Delta \tilde{m}^2 > 0$, we see a faster convergence to these desired points. Accessing points in the deep superfluid state however is more difficult. To definitively converge to a point $\Delta \tilde{m}_1^2 \ll 0$ one has to start with $\Delta \tilde{m}_0^2 = \Delta \tilde{m}_1^2 + \delta \tilde{t}_{ii}
$, where $\delta \tilde{t}_{ii}$ depends on the lattice size $L$. For $L=13$ we find that $\delta \tilde{t}_{ii} \approx 1$ is a good choice. The slower convergence in the superfluid phase is due to the sensitivity of $\vec{R}(\vec{c})$ in Eq. (\ref{R_equal_0}) on $\vec{c}$.

In summary, the numerical code can be parallelized when finding solutions on the insulating side, as well as at and near the critical point on the superfuild side, but finding solutions in the deeper superfluid phase needs to be done in serial.

\section{Comparison between the general and on-site boundary conditions}
\begin{figure}[h]
	\centering
\subfigure[]{\includegraphics[width=3in]{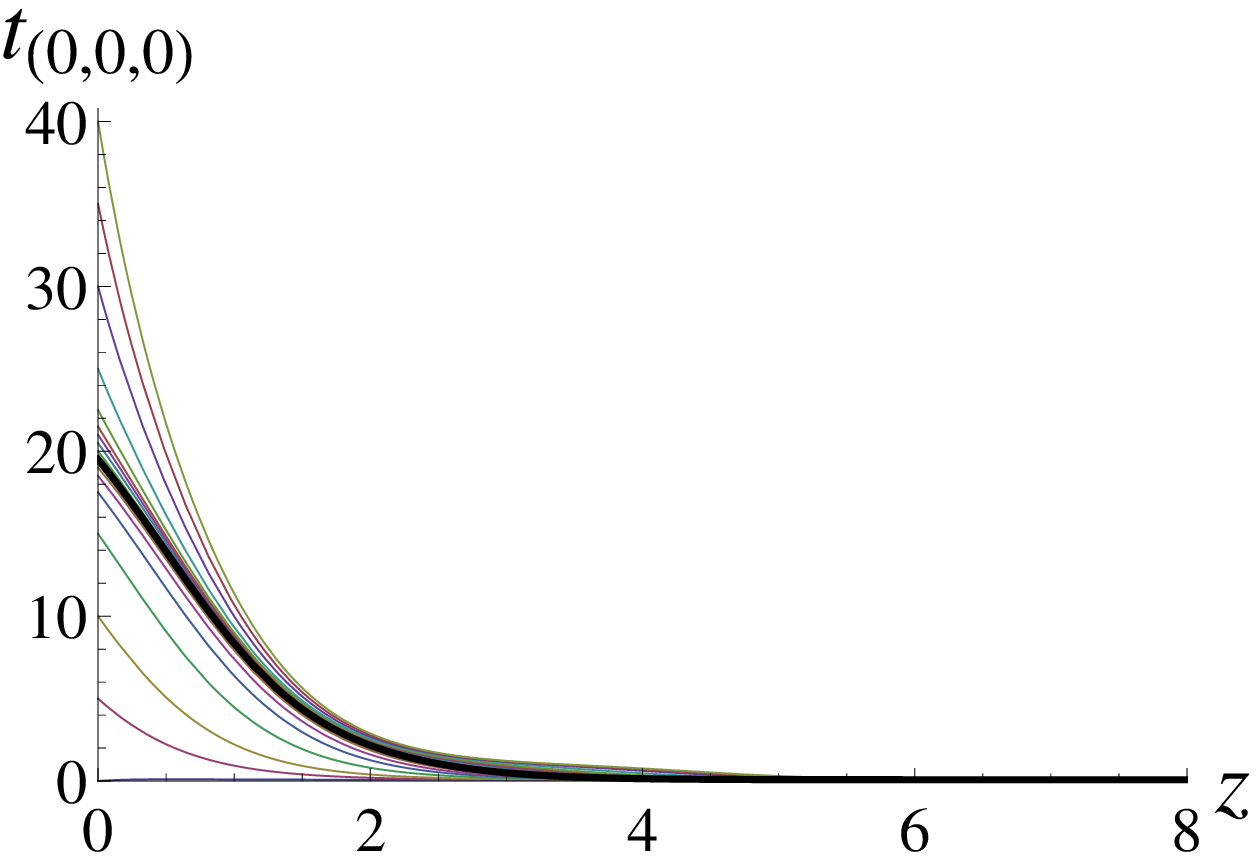}}
\subfigure[]{\includegraphics[width=3in]{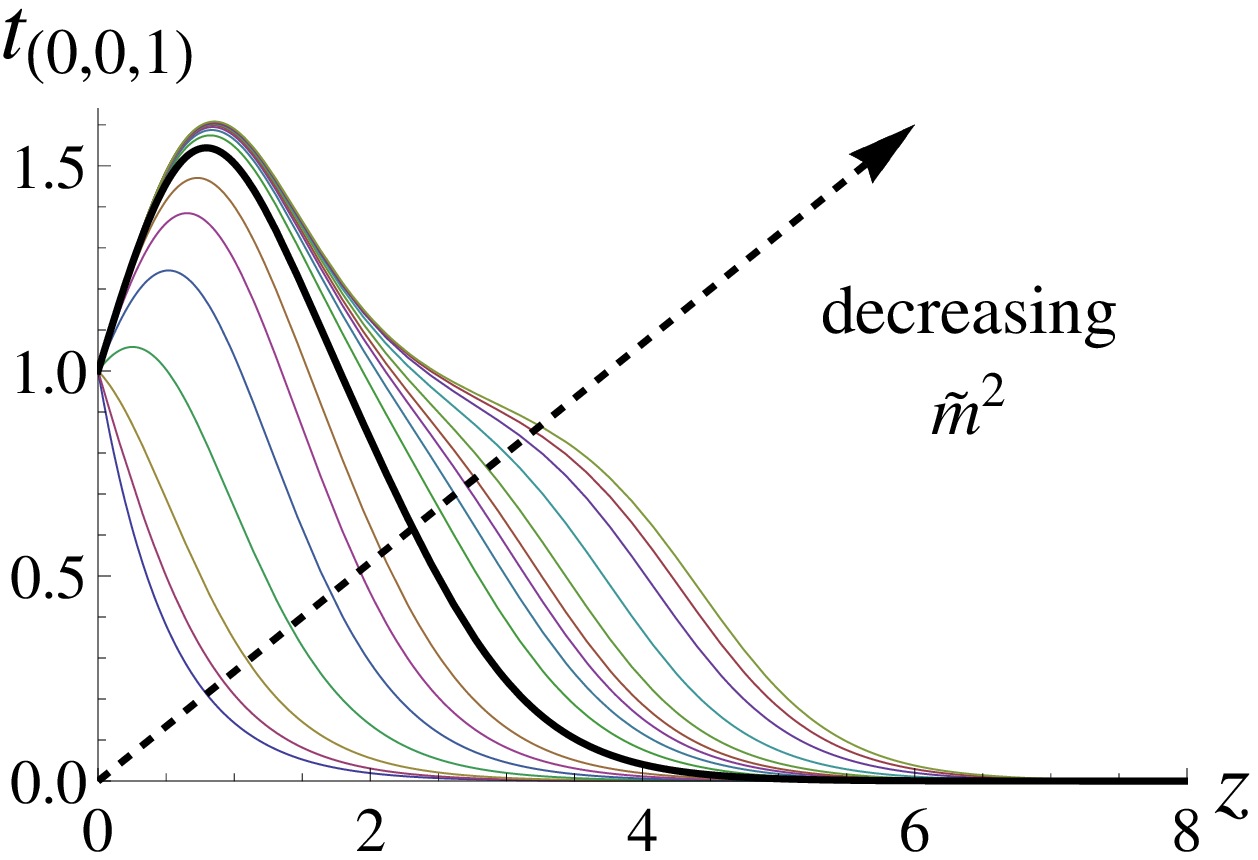}}\\
\subfigure[]{\includegraphics[width=3in]{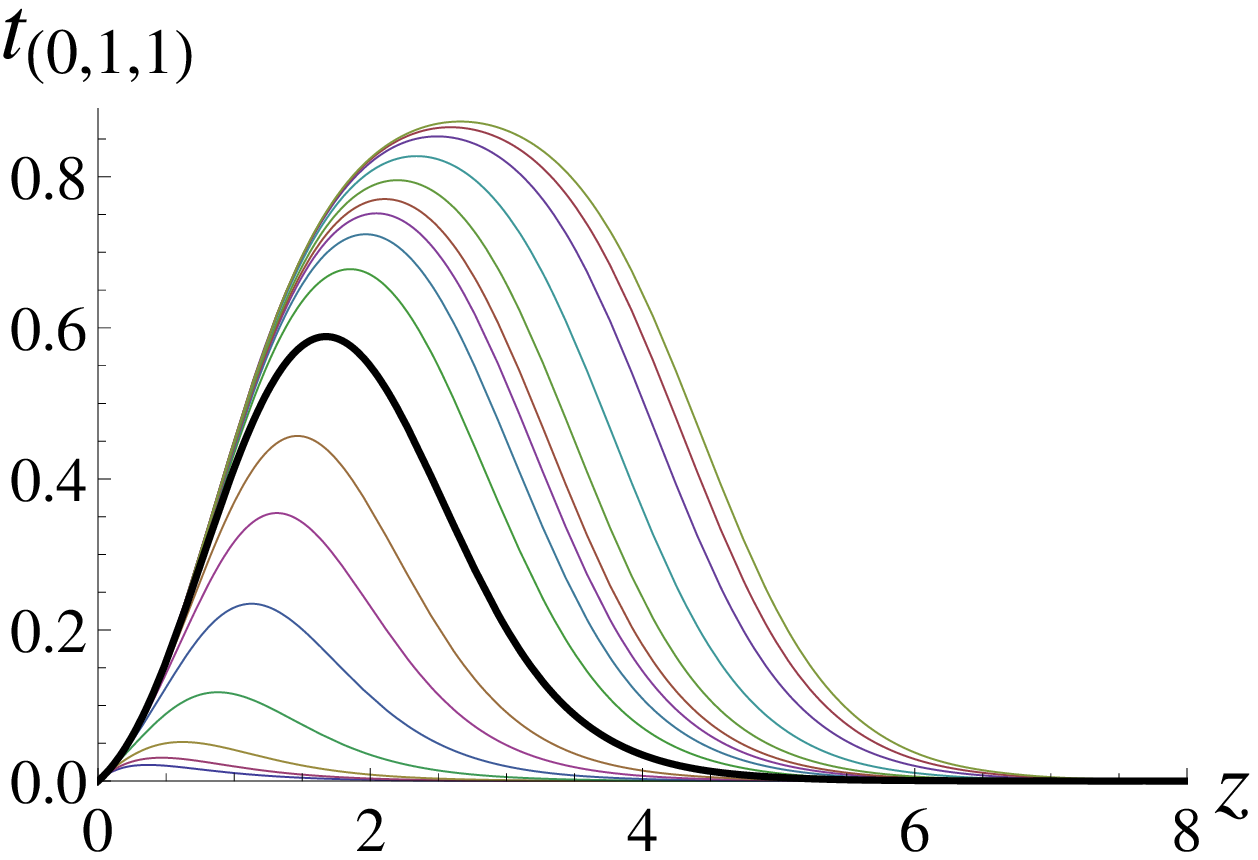}}
\subfigure[]{\includegraphics[width=3in]{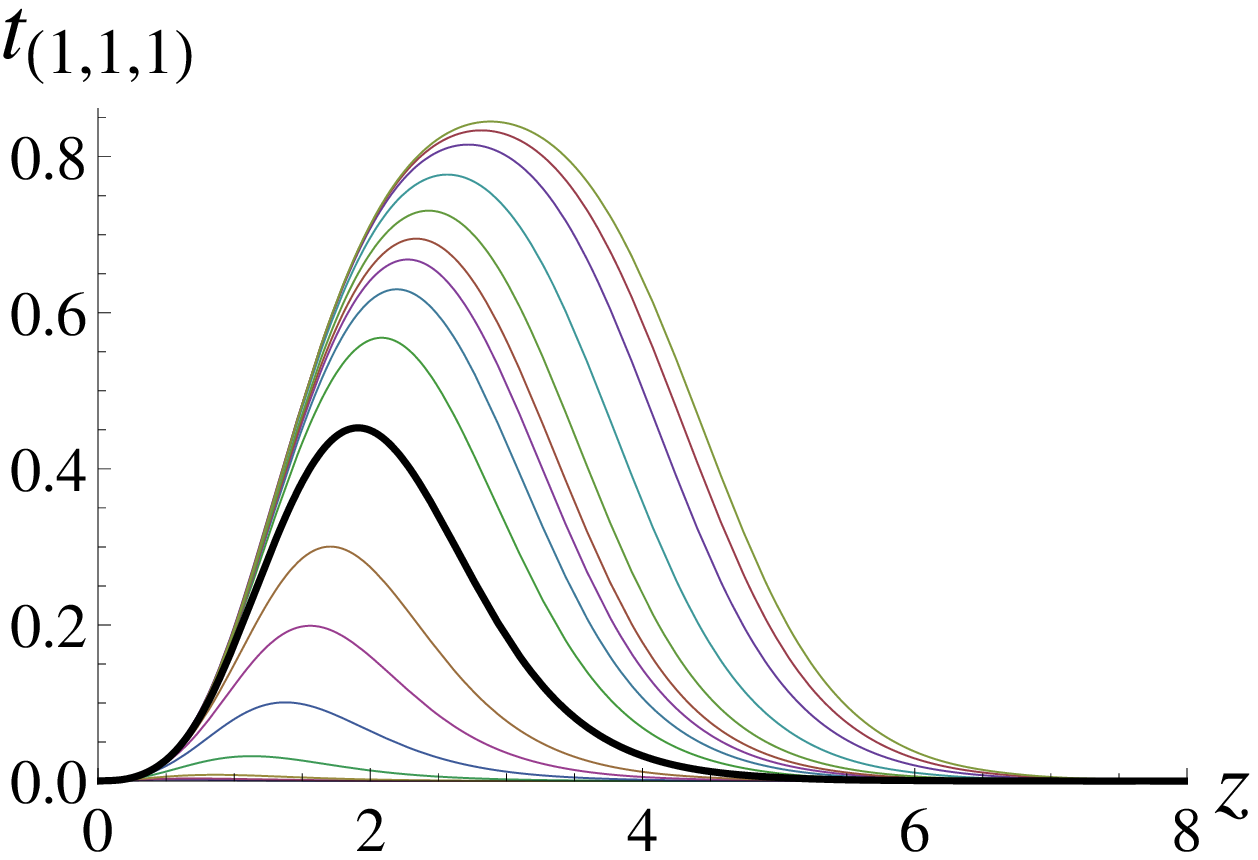}}
\caption{The same plots as in Fig. \ref{3x3x3 t fields GBC zstar 8} with the difference being that the IR BC used is the on-site one of Eq. (\ref{eq_tstar_IR_on_site}), instead of the general BC of Eq. (\ref{eq_tstar_IR_general}).}
\label{3x3x3 t fields MIBC zstar 8}
\end{figure}

\begin{figure}[h]
	\centering
\subfigure[]{\includegraphics[width=3in]{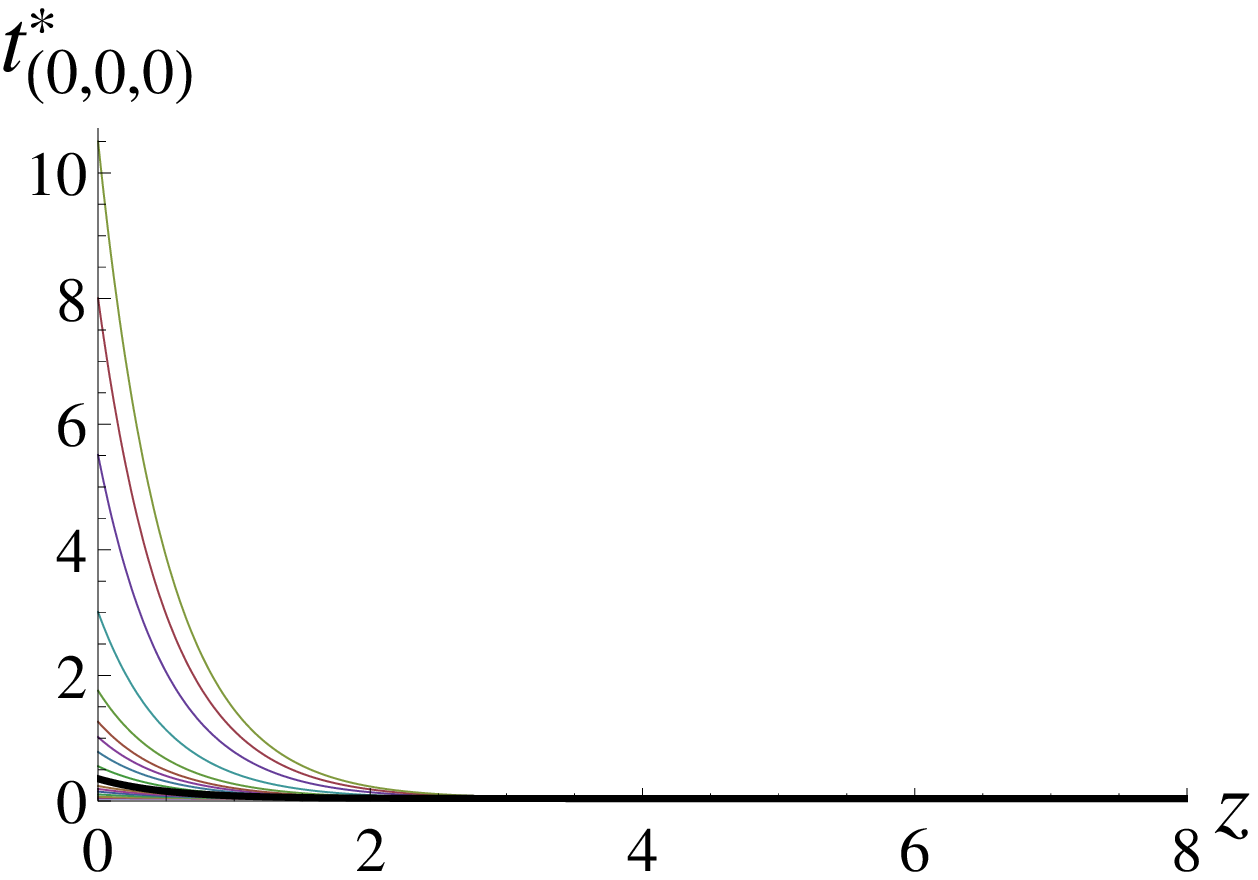}}
\subfigure[]{\includegraphics[width=3in]{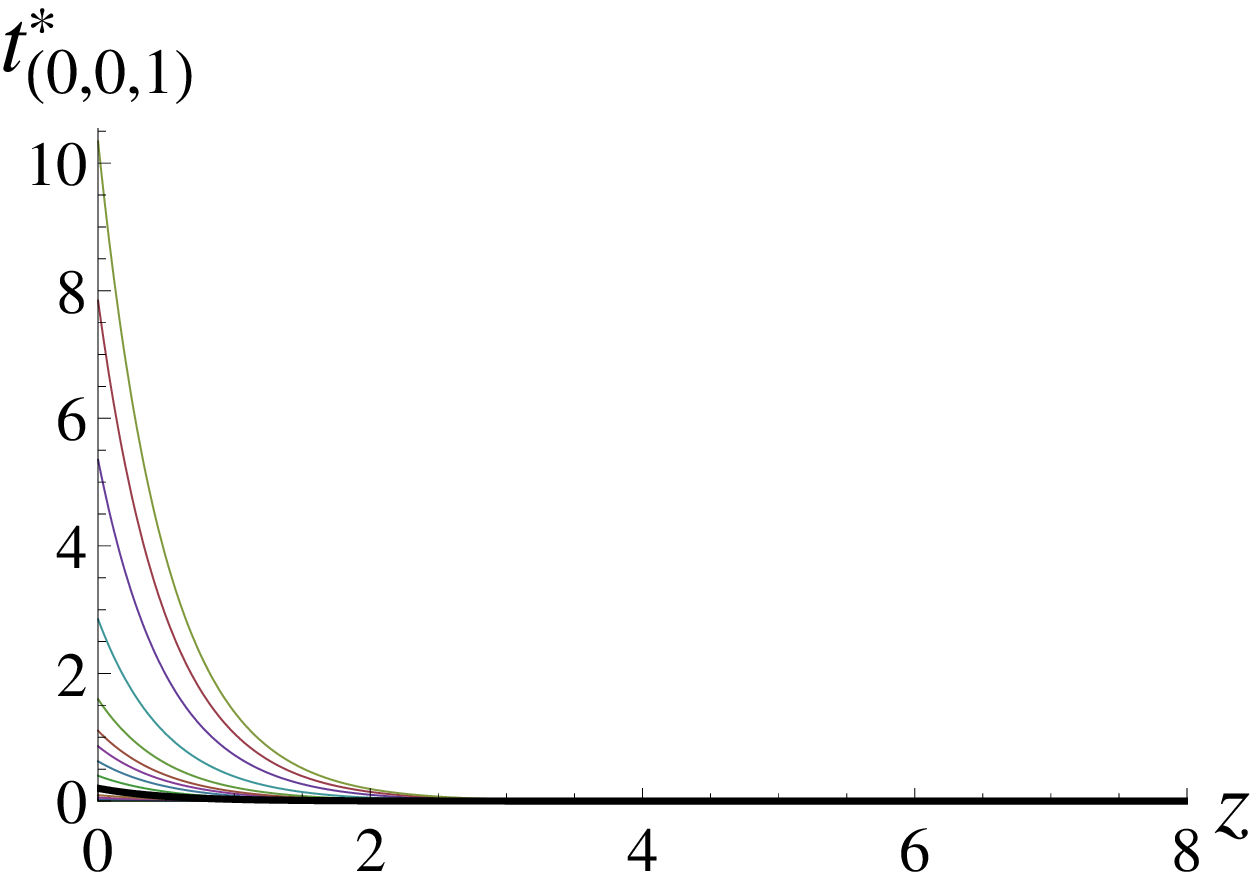}}\\
\subfigure[]{\includegraphics[width=3in]{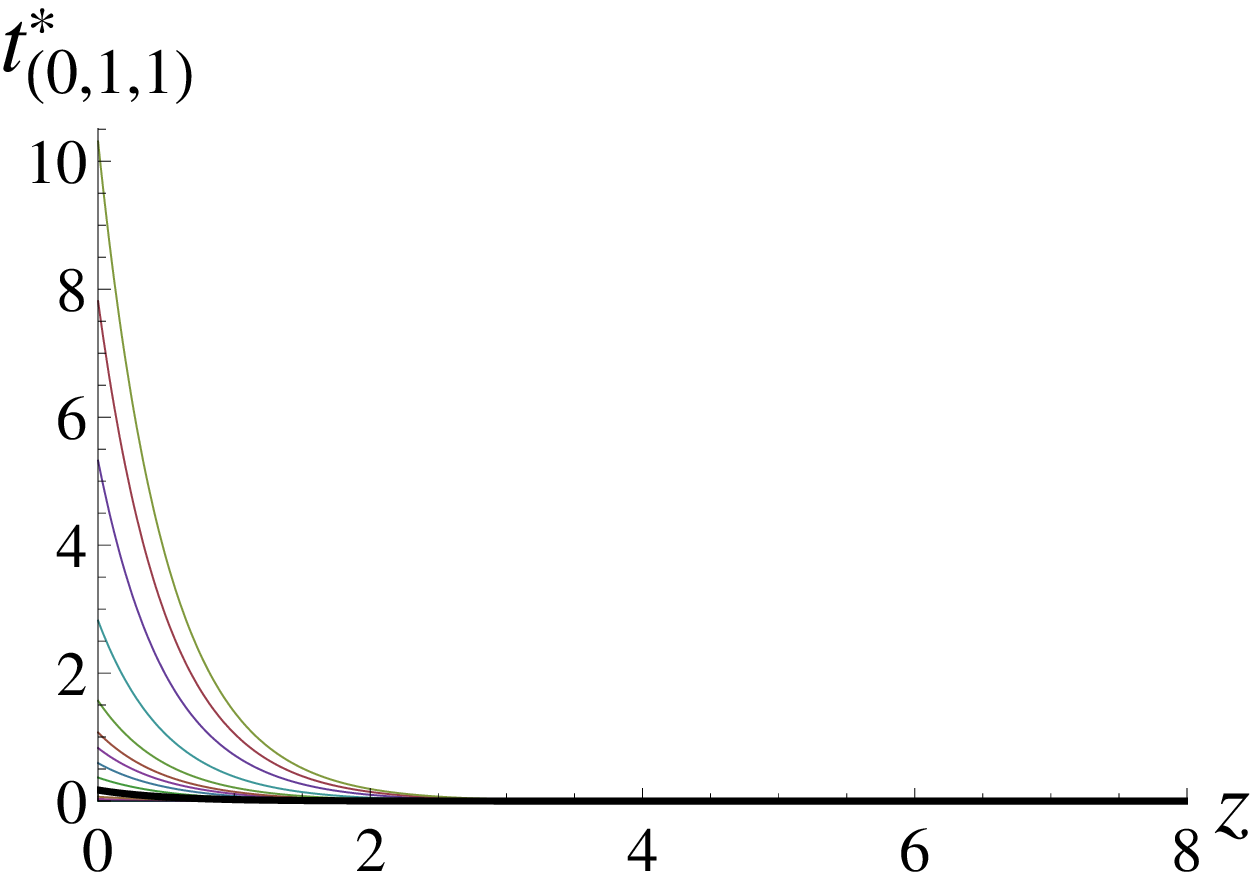}}
\subfigure[]{\includegraphics[width=3in]{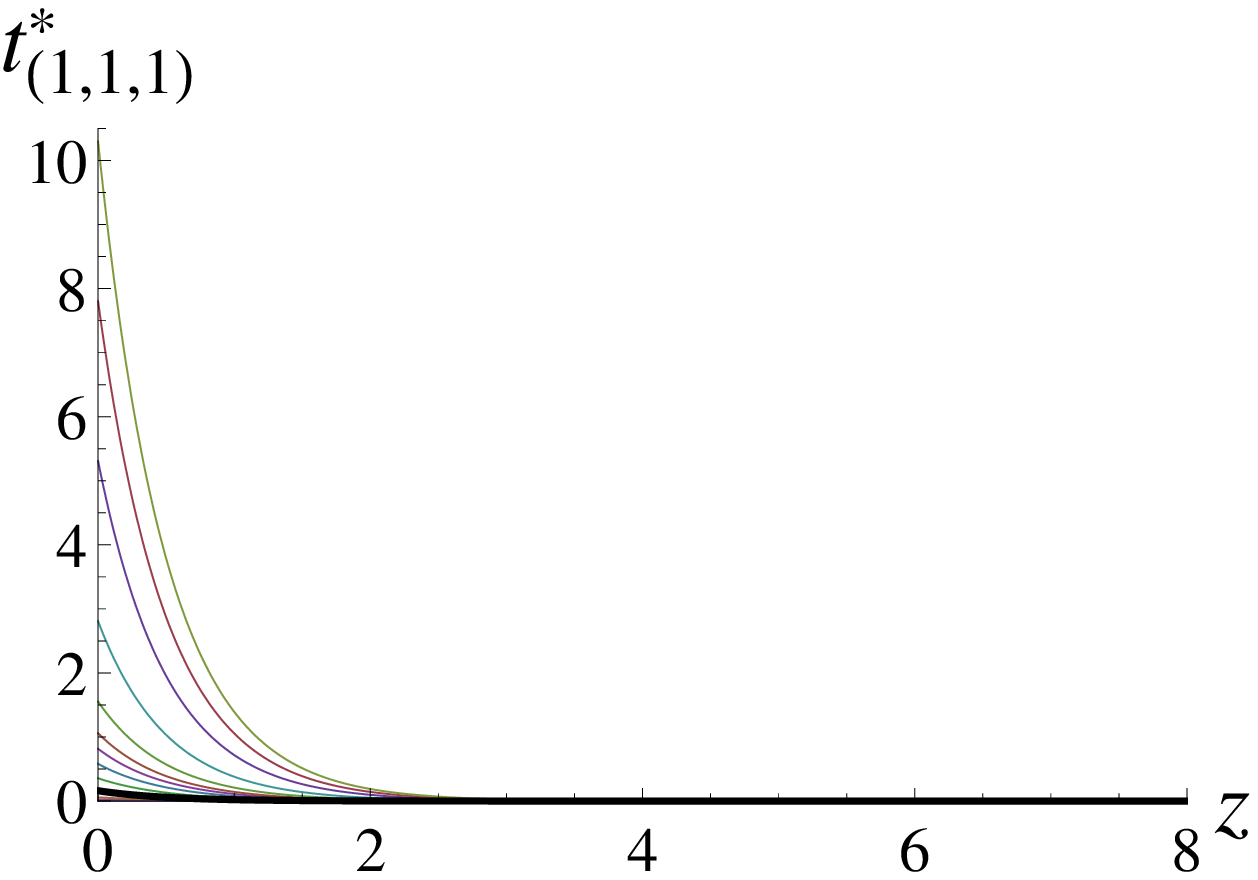}}
\caption{The same plots as in Fig. \ref{3x3x3 tstar fields GBC zstar 8} with the difference being that the IR BC used is the on-site one of Eq. (\ref{eq_tstar_IR_on_site}), instead of the general BC of Eq. (\ref{eq_tstar_IR_general}).}
\label{3x3x3 tstar fields MIBC zstar 8}
\end{figure}

In this appendix we check that the on-site boundary condition (BC) gives the same solutions to the EOM as the general BC.  
The solutions for the $L=3$ lattice with the on-site BC are plotted in Figs. \ref{3x3x3 t fields MIBC zstar 8} and \ref{3x3x3 tstar fields MIBC zstar 8}. We can see that the fields look identical to those of Figs. \ref{3x3x3 t fields GBC zstar 8} and \ref{3x3x3 tstar fields GBC zstar 8}, and after comparison we find that they indeed lie on top of each other. 
%%%%%%%%%
%%%%%%%%%

%%%%%%%%%
However, besides the bulk shape it is also important to compare the exact IR values to those obtained using the general BC, especially when the theory starts out in the superfluid phase at the UV. Tables \ref{table 1} and \ref{table 2} summarize this comparison for two values of $\tilde{m}^2 = 25$ and $-15$, corresponding to the deep insulating and deep superfluid phases, respectively. 
%%%%%%%%%
\begin{table}[h]
\begin{tabular}{|l|l|l|l|l|}
\hline
$\tilde{m}^2 = 25$ & $t_{ii}(z^*)$ & $|t_{i \neq j}(z^*)|$    & $t^*_{ii}(z^*)$ & $|t^*_{i \neq j}(z^*)|$ \\ \hline
general BC                                & $0.08$                  & $\le 1.3 \times 10^{-7}$ & $0.04$                                    & $\lesssim 2 \times 10^{-10}$            \\ \hline
on-site BC                               & $0.08$                  & $\le 1.3 \times 10^{-7}$ & $0.04$                                    & $< 10^{-19}$                                                        \\ \hline
\end{tabular}
\caption{Comparison at $z=z^*$ of general and on-site BCs for $\tilde{m}^2 = 25$ (deep insulating phase).}
\label{table 1}
\end{table}
%%%%%%%%%
\begin{table}[h]
\begin{tabular}{|l|l|l|l|l|}
\hline
$\tilde{m}^2 = -15$ & $t_{ii}(z^*)$ & $|t_{i \neq j}(z^*)|$    & $t^*_{ii}(z^*)$ & $|t^*_{i \neq j}(z^*)|$ \\ \hline
general BC                                & $0.079812$                  & $\approx 0.00071$ & $0.0399997$                                    & $\approx 1.14 \times10^{-6}$            \\ \hline
on-site BC                               & $0.079812$                  & $\approx 0.00071$ & $0.0399997$                                    & $< 10^{-15}$                                                        \\ \hline
\end{tabular}
\caption{Comparison at $z=z^*$ of general and on-site BCs for $\tilde{m}^2 = -15$ (deep superfluid phase).}
\label{table 2}
\end{table}
%%%%%%%%%
In both phases the $t^*_{i \neq j}(z^*)$ values are the only ones that differ at all between the two BCs. For the on-site BC they are zero always, which is what we set them to ($10^{-15}$ is the machine precision), and for the general BC this is not the case. However, $t^*_{i \neq j}(z^*)$ for the general BC are still extremely small when compared to $t^*_{ii}(z^*)$ \textit{in both phases}, which is why it works to approximate them by zero. This explains the validity of this approximation. 

\begin{figure}[h]
	\centering
\subfigure[]{\includegraphics[width=3in]{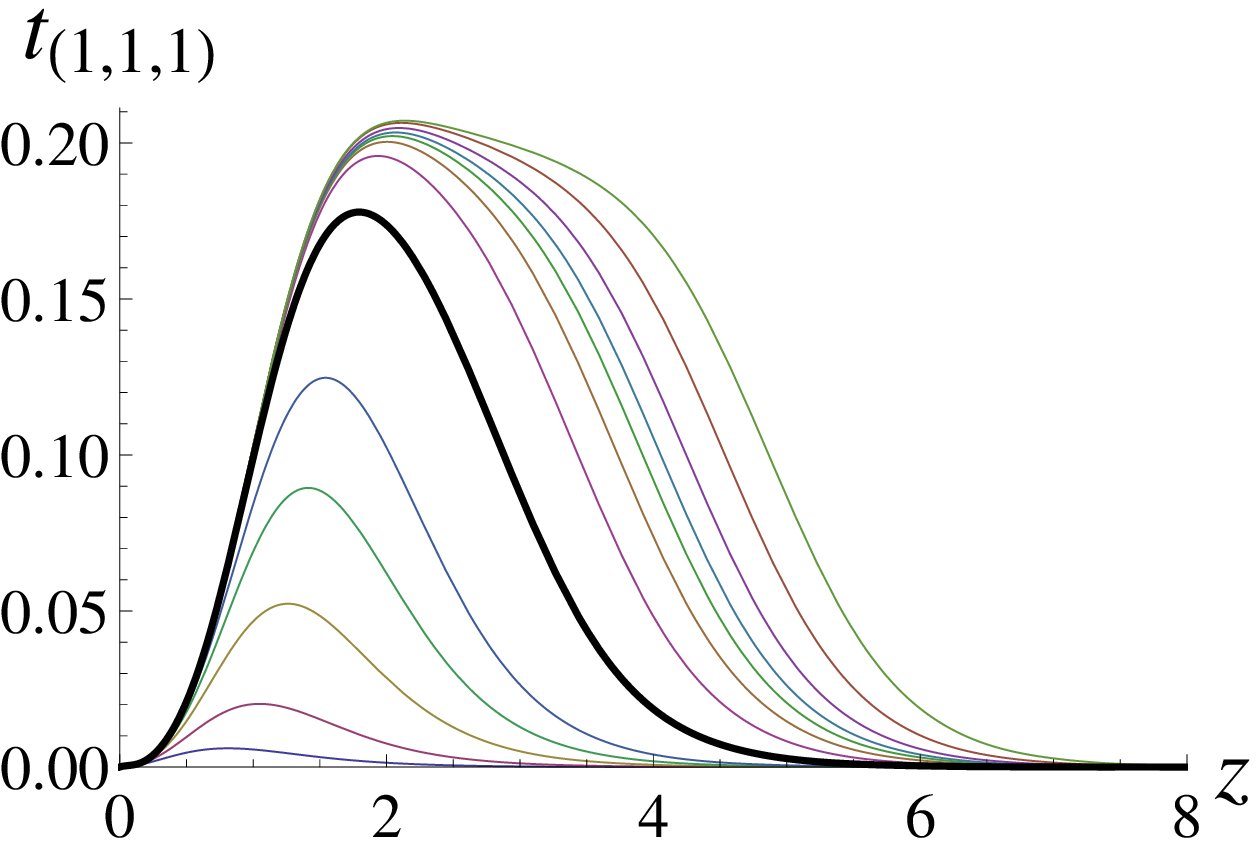}}
\subfigure[]{\includegraphics[width=3in]{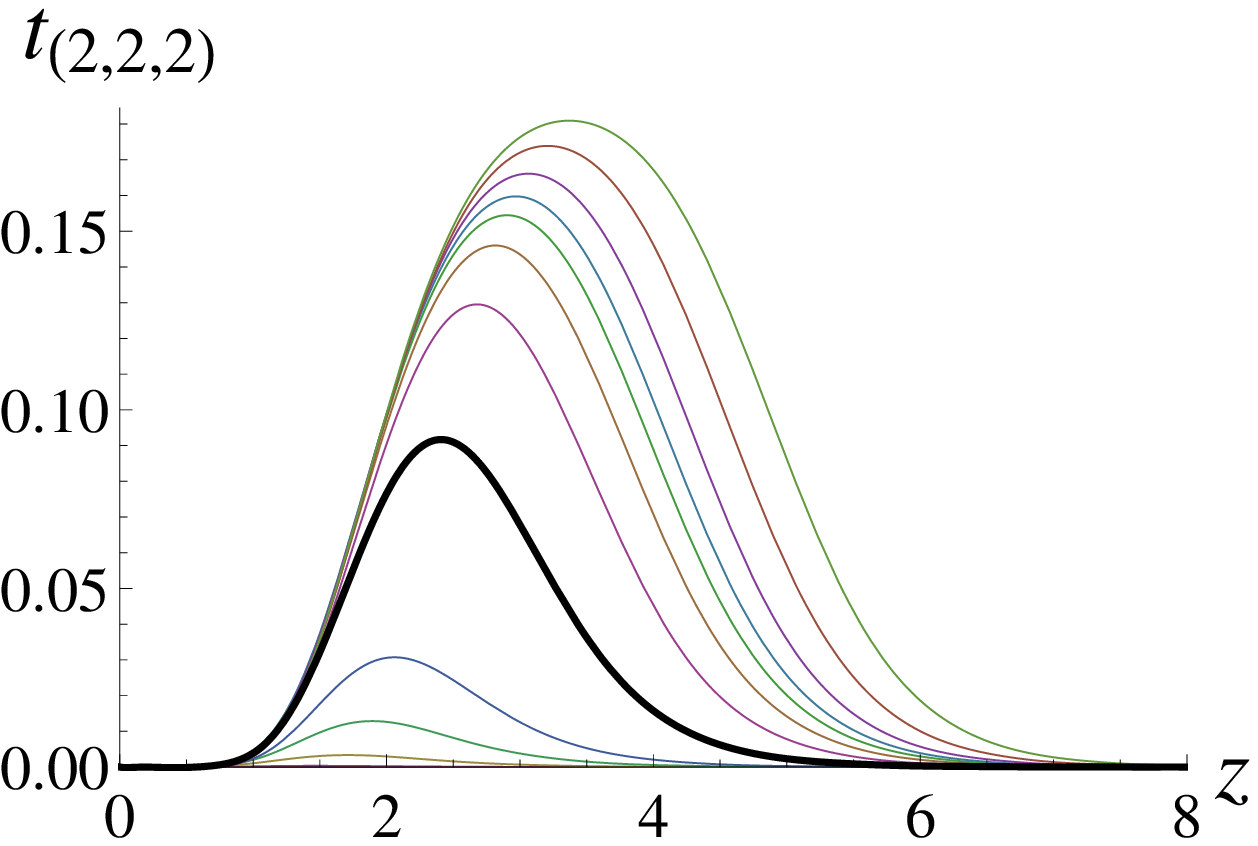}}
\subfigure[]{\includegraphics[width=3in]{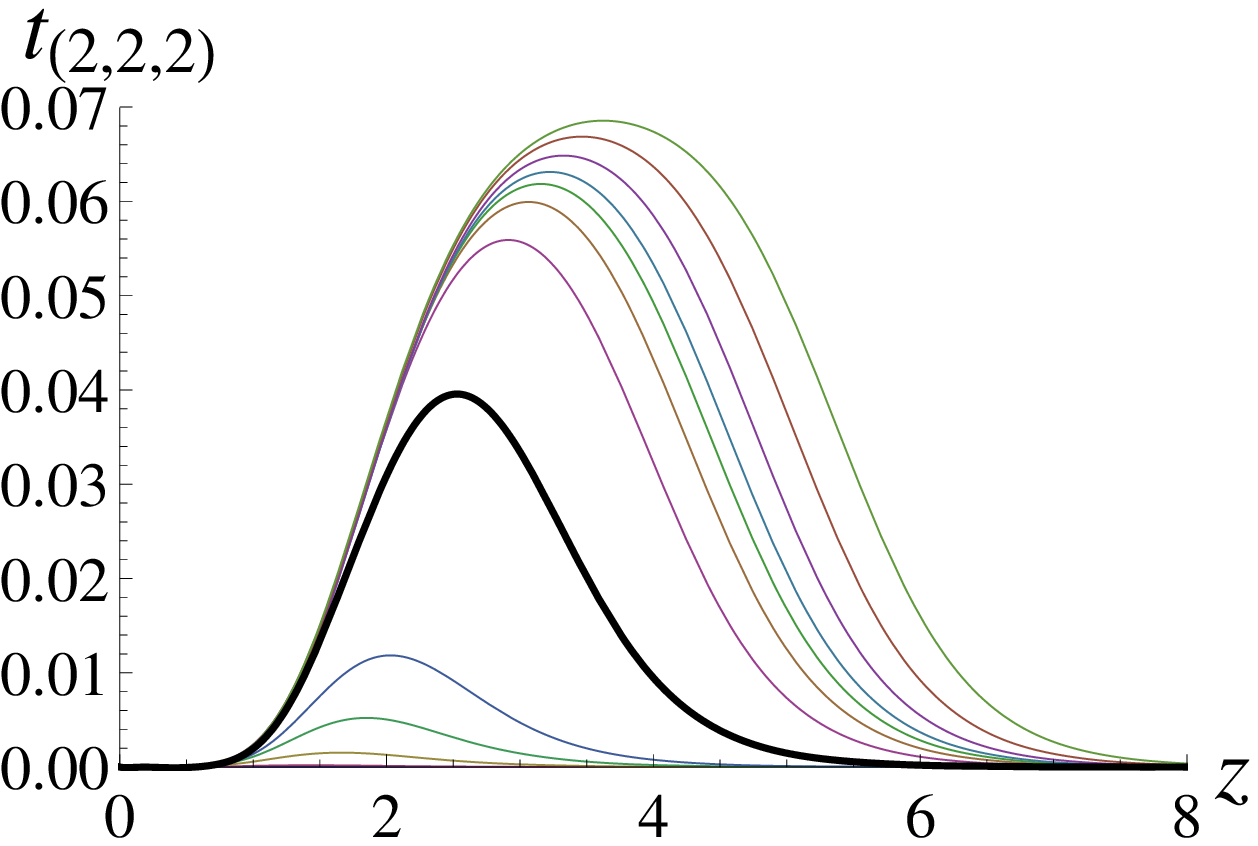}}
\subfigure[]{\includegraphics[width=3in]{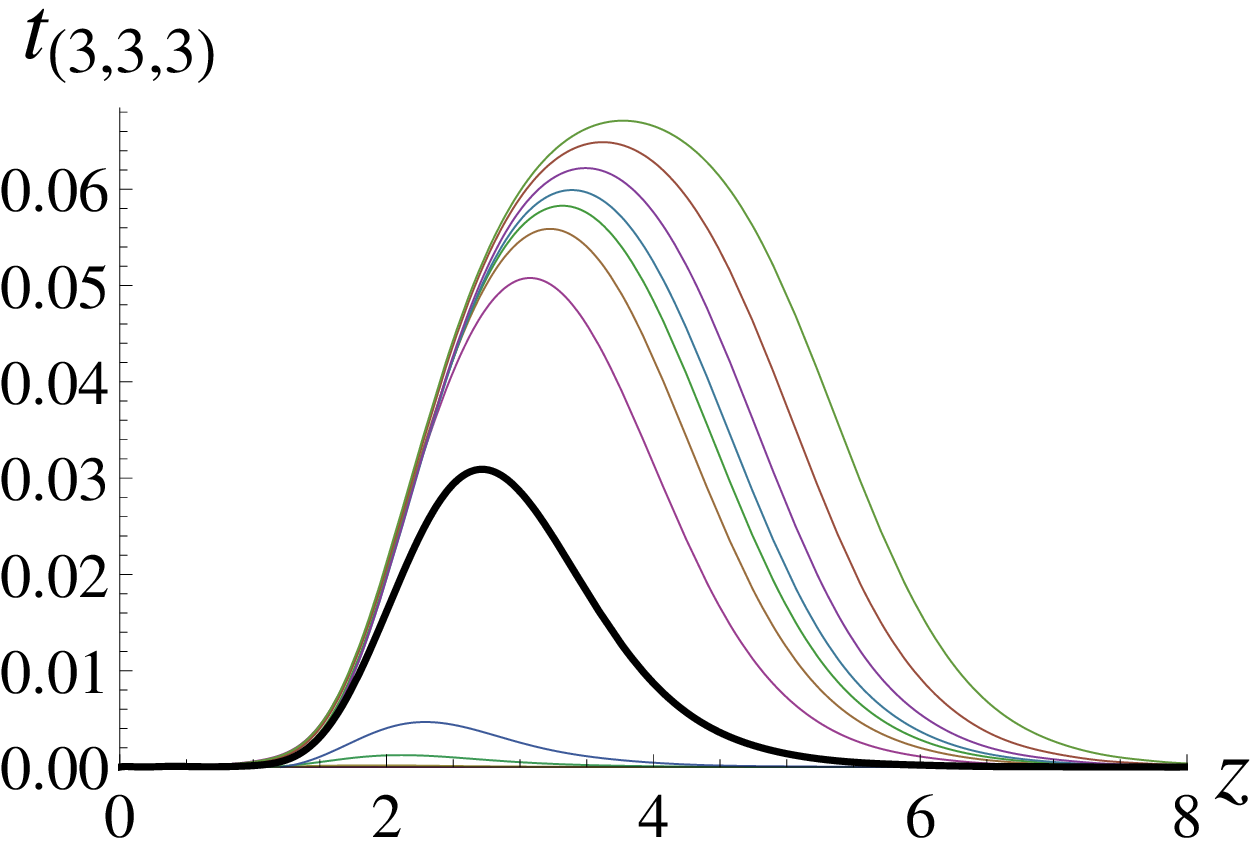}}
\subfigure[]{\includegraphics[width=3in]{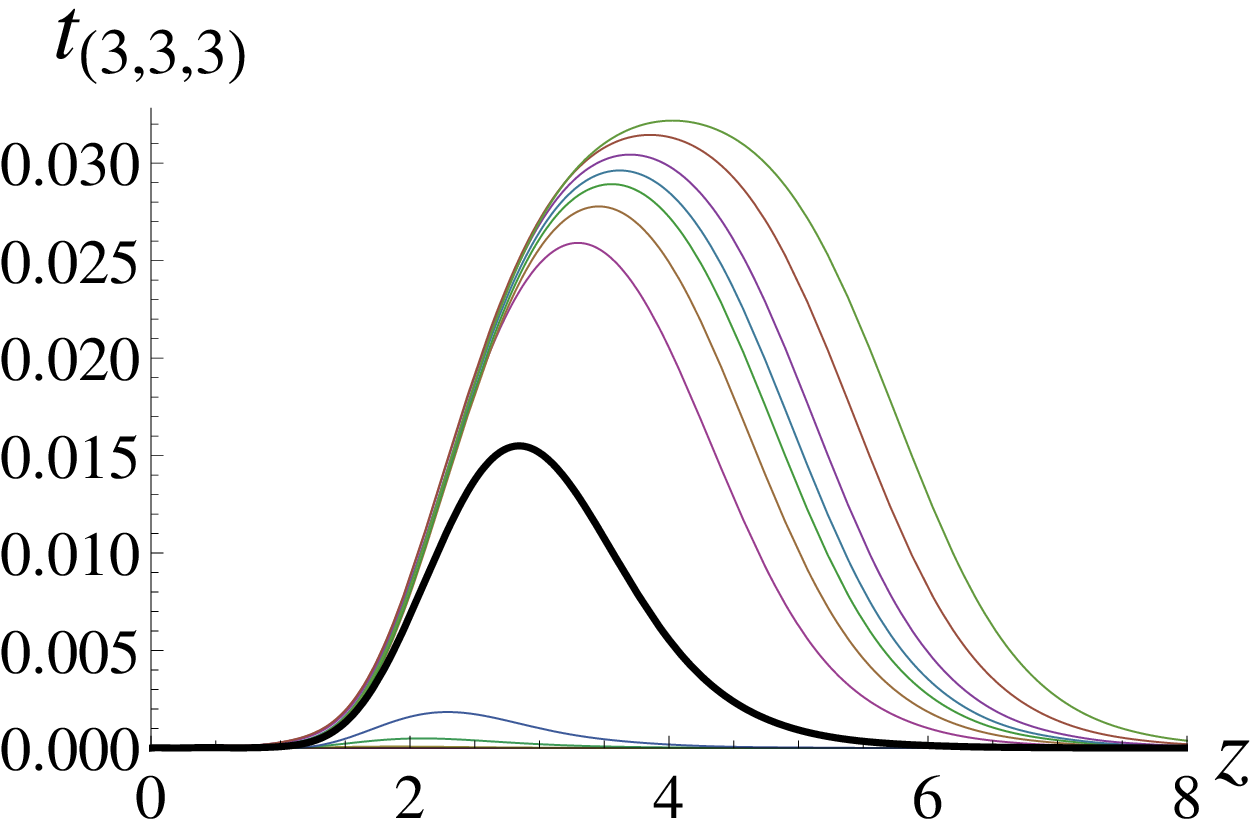}}
\subfigure[]{\includegraphics[width=3in]{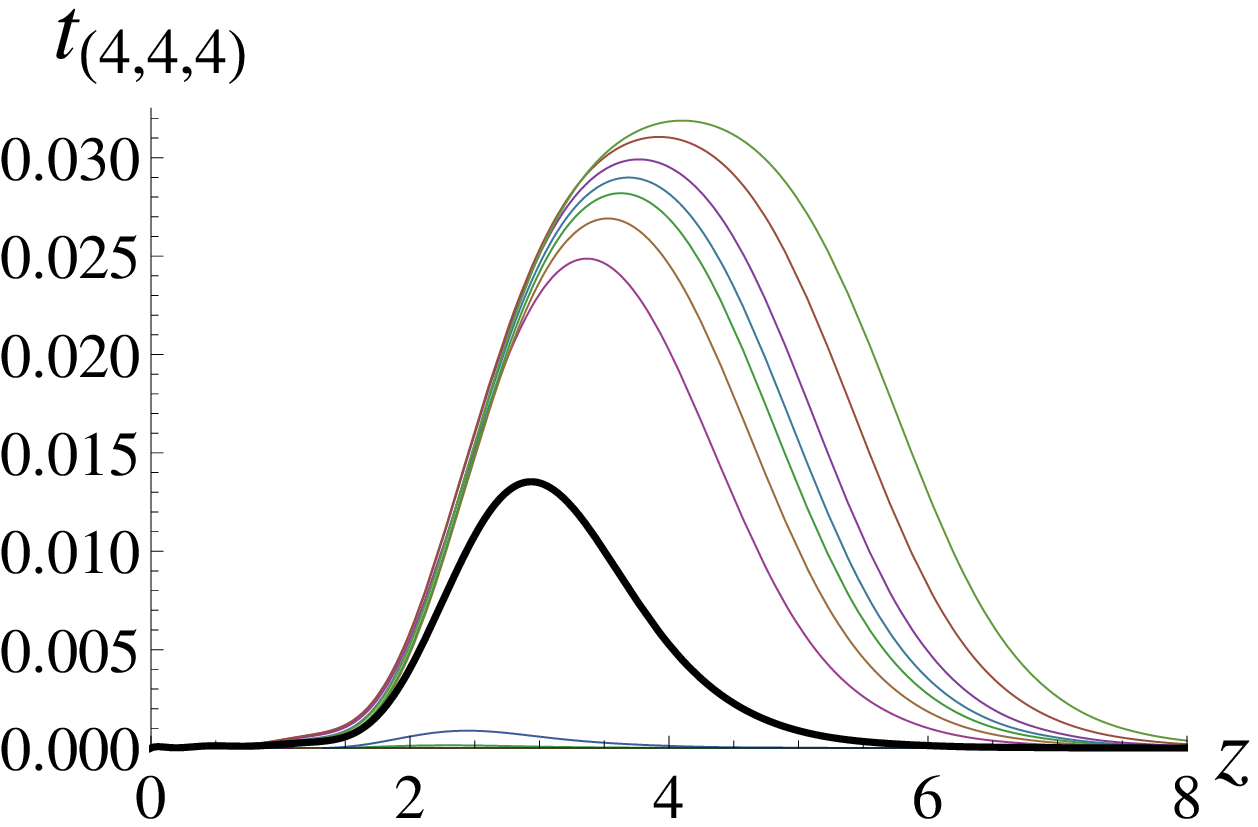}}
\caption{
Same plots as Fig. \ref{3x3x3 t fields MIBC zstar 8} for larger lattices 
$L=5$ (a,b),
$L=7$ (c,d),
$L=9$ (e,f).
In all plots, $\tilde{m}^2$ are chosen to be $15$, $10$, $7.5$, $6.5$, $6$, $5.49454$, $5$, $4.5$, $4$, $3.5$, $2.5$, $0$, $-5$, and the arrow from 
Fig. \ref{3x3x3 t fields MIBC zstar 8} applies here as well. The critical point ($\tilde{m}^2 = 5.49454$) is again denoted by think curves.
}
\label{L579 fields MIBC zstar 8}
\end{figure}

In Fig. \ref{L579 fields MIBC zstar 8}, we show the hopping fields $t_{ij}(z)$ for larger lattices $L=5,7,9$ obtained using the on-site BC. They look qualitatively similar to those in Fig. \ref{3x3x3 t fields MIBC zstar 8}.

%%%%%%%%%
%%%%%%%%%%%%%%%%%%%%%%
\section{Value of $\tilde{m}^2_c$ in the $L\rightarrow \infty$ limit.}
The discrete version of our action in Eq. (\ref{eq_original_discrete_action}) has a critical point in the thermodynamic limit, $\tilde{m}^2_c$, which can be computed exactly in the large $N$ limit. Using Eq. (\ref{eq_self_constist_momentspace_discrete}) at $z=0$ (in the thermodynamic limit the sum becomes and integral), we compute $\tilde{m}^2_c$ for the values of $\td t_{ij}$ and $\td J_{ijpq}$ which we set at the UV boundary: $\td t_{ij}$ is zero beyond nearest neighbors and $\td J_{ijpq} = 0$. At $z=0$ the hopping fields in momentum space are
\beqaa
\td t_{\vec{k}} = \sum_{(l,m,n)} e^{i\vec{k} \cdot (l,m,n)} \td t_{(l,m,n)} = \td t_{(0,0,0)} + 2\td t_{nn}\left(cos(k_x) + cos(k_y) + cos(k_z)\right),
\eeqaa
where $\td t_{nn} =1$ are the nearest neighbor hoppings. At the critical point the denominator of the integrand in Eq. (\ref{eq_self_constist_momentspace_discrete}) is zero exactly for $\vec{k}=0$, since $\td t_{\vec{k}=\vec{0}} > \td t_{\vec{k}\neq \vec{0}}$. We can plug in the value of $\sigma_c$ for which this is true on both sides of Eq. (\ref{eq_self_constist_momentspace_discrete}), which gives
\beqaa
\f12 \left((\td t_{(0,0,0)}-m^2)_c + 6\right) = \int_0^{2\pi}\int_0^{2\pi}\int_0^{2\pi} \f{d^3k}{(2\pi)^3} \f{\lambda}{6 - 2\left(cos(k_x) + cos(k_y) + cos(k_z)\right)}.
\eeqaa
Reverting back to the notation of $\td t_{(0,0,0)} = \td t_{ii}$, we obtain 
\beqaa
\tilde{m}^2_c = (m^2- \td t_{ii})_c = 5.49454.
\eeqaa

%%%%%%%%%%%%%%%%%
\end{appendix}
%%%%%%%%%%%%%%%%%

\end{document}